\documentclass{aa}
\usepackage[]{epsfig, graphicx}
\def\ltsima{$\; \buildrel < \over \sim \;$}
\def\simlt{\lower.5ex\hbox{\ltsima}}
\def\gtsima{$\; \buildrel > \over \sim \;$}
\def\simgt{\lower.5ex\hbox{\gtsima}}
\def\cgs{{erg cm$^{-2}$ s$^{-1}$}}
\def\ergs{{erg s$^{-1}$}}
\def\cm2{{cm$^{-2}$}}

\def\xnu{{$\chi^{2}_\nu$}}

\def\lum{{$L_{2-10}$}}
\def\p1{{Paper I}}

\def\xmm{{\em XMM--Newton}}
\def\chandra{{\em Chandra}}
\def\exosat{{\em EXOSAT}}
\def\beppo{{\em BeppoSAX}}
\def\ginga{{\em Ginga}}
\def\asca{{\em ASCA}}
\def\nhgal{{N$_{\rm H}^{\rm Gal}$}}
\def\nh{{N$_{\rm H}$}}
\def\chandra{{\em Chandra}}
\def\rosat{{\em ROSAT}}
\def\xmm{{\em XMM--Newton}}
\def\nhgal{{N$_{\rm H}^{\rm Gal}$}}
\def\nh{{N$_{\rm H}$}}  
\def\epic{{\em EPIC}}
\def\mosuno{{\em MOS1}}
\def\mosdue{{\em MOS2}}
\def\pn{{\em pn}}
\def\mos{{\em MOS}}

\def\f14{{10$^{-14}$}}
\def\f13{{10$^{-13}$}}
\def\sun{{$_{\odot}$}}
\begin{document}
   \title{The XMM-Newton view of PG quasars}

   \subtitle{I. X--ray continuum and absorption.}
\author{ E.~Piconcelli, E.~Jimenez-Bail\'{o}n, M.~Guainazzi, N.~Schartel, 
P.M.~Rodr\'{i}guez-Pascual, M.~Santos-Lle\'{o}}
 \institute{XMM--Newton Science Operation Center (ESAC), Apartado
 50727, E--28080 Madrid, Spain} \authorrunning{E.~Piconcelli et al.}
 \titlerunning{The XMM--Newton view of PG quasars. I.}
 \offprints{epiconce@xmm.vilspa.esa.es}
 \date{Received / accepted}

\abstract{

We present results of a systematic analysis of the \xmm~spectra of 40
quasars  (QSOs)  (z$\leq1.72$)   from the  Palomar--Green  (PG) Bright
Quasar Survey sample (M$_B<-$23).  The  sample includes 35 radio-quiet
quasars (RQQs) and  5 radio-loud quasars  (RLQs).  The analysis of the
spectra above 2  keV reveals that the hard  X--ray continuum emission
can   be    modeled   with      a     power law      component   with
$\langle\Gamma_{2-12~keV}\rangle = 1.89\pm0.11$
and $\langle\Gamma_{2-12~keV}\rangle$ = 1.63$^{+0.02}_{-0.01}$ for
the RQQs  and RLQs, respectively.  Below 2~keV, a strong,  broad
excess is present in most QSO spectra.  This feature has been  fitted
with four different models assuming  several  physical    scenarios.
All tested   models (blackbody,  multicolor  blackbody,
bremsstrahlung and   power  law) satisfactorily fitted the majority of
the  spectra. However, none of them is  able to provide an adequate
parameterization for the  soft excess  emission  in  all QSOs,
indicating the  absence of an universal shape for this spectral
feature.
An additional cold  absorption  component was  required only in three
sources.  On the other hand,  as recently pointed out by Porquet et al. (2004) for a smaller sample of PG QSOs,
warm absorber features are present in
~50\% of  the QSO spectra in contrast with their rare occurrence
($\sim$   5--10\%) found in previous studies.  The    \xmm~view  of
optically--selected bright QSOs  therefore suggests  that there are
no significant  difference in the X--ray  spectral  properties  once
compared  with  the  low--luminosity  Seyfert  1 galaxies.
Properties of the Fe K$\alpha$ emission lines are presented in a
companion paper.

   }

   \maketitle
%

\section{Introduction}

Quasars (QSOs)  emit a large  amount of their  emission in  the X--ray
band, where  luminosities can  reach  10$^{46-47}$~\ergs~(Zamorani et al. 1981).  
Variability
studies  indicate  that the X--rays  originated in  a region very
close to  the central object, probably  an accreting supermassive black hole.
Therefore,  the analysis  of the emission   in this
energy range provides  strong constraints for models of the mechanism
powering  the QSOs.  X--ray   emission   properties of QSOs have   been
largely studied  during the last  decades.  Different samples  of QSOs
have  been analyzed with previous  X--ray satellites such as {\it Einstein}
(Zamorani   et    al.  1981),   \exosat~(Comastri     et  al.   1992),
\ginga~(Lawson \& Turner 1997), \asca~(Cappi  et  al. 1997; George  et
al. 2000; Reeves \& Turner 2000), \rosat~(Schartel  et al. 1997; Laor et
al. 1997) and \beppo~(Mineo et al. 2000).

From these  studies it emerges that the typical  spectrum of a QSO in the
hard ($>$2~keV) band is dominated by a power law--like emission with a  photon index
$\Gamma\sim2$.  For Seyfert galaxies, the most popular physical scenario
(Haardt \& Maraschi 1991)~explains this  hard spectral component  as the emission
originating in  a hot corona   placed  above the accretion  disk  which
comptonizes the UV-soft X--ray thermal   emission  from the disk   and
up-scatters it   into the hard   X--ray band.   Part of  the continuum
emission  is  then reprocessed  in  the disk  and/or in  other nuclear
physical components (i.e.   the molecular torus, clouds) producing the
typical reflection   spectrum   (George \& Fabian   1991).    The most
striking feature  in  the reflection spectrum   is  the fluorescent Fe
K$\alpha$  emission line around 6.4~keV.  This scenario has been also extended to the QSOs,
although the observational evidences of a reflection component in their X--ray spectra are poor. 
Signatures of  the presence of large amounts of
ionized and/or cold gas were also detected in the soft X-ray portion of
QSO spectra (Cappi et al. 1997; George et al. 1997). It  has been  reported as well that a fraction of  QSOs
present a  gradual  upturn  in the
emission emerging below  2-3~keV, the so-called  {\it soft   excess} (Comastri et al. 1992), whose nature is still debated.


Systematic analysis of QSOs with  high spectral resolution provide
a very useful tool in order to investigate their nature. In this
paper we  analyze a sample of QSOs  observed with \xmm, which provides
the highest throughput and sensitivity to date.  We investigate the spectral
characteristics of the objects in the sample and search for  common
features in order to elucidate the physical
mechanisms  responsible for the X--ray emission.  The sample  includes both radio loud
and    radio quiet quasars     (RLQs   and RQQs, respectively).    The
classification is based on the strength of the radio emission relative
to  the optical one.  Many studies  show systematic differences in the
spectral properties in the X--ray band of both types of objects, as for
instance the slope of the power law (Laor et al. 1997; Reeves \& Turner 2000)  which is evidence for a
possible difference in the physical processes which give rise the X--ray
emission.   Therefore  the  chosen  sample   allows  to  provide useful information about 
possible differences between RLQs and RQQs.
Results based on \xmm~observations of a smaller subsample of PG QSOs have been recently presented by Porquet et al. (2004; P04 hereafter).
Thanks to the larger number of objects, our survey allows to obtain a complete description of the X--ray spectral properties of 
optically--selected QSOs based on a sounder statistical ground.

In Sect.~\ref{sec:observations} we present the general characteristics
of the objects included in the sample, as well as the
\xmm~observations and the applied reduction technique. We performed a 
systematic analysis of all objects assuming several physical scenarios
for the emission of the QSO. The corresponding results are presented
in Sect.~\ref{sec:analysis}, which  also includes a  detailed analysis
of the most peculiar objects. In  Sect.~\ref{sec:discussion} we
discuss 
 observational constraints obtained by this
study to  the physical scenario for the X--ray emission/absorption
mechanisms at work in QSOs and we also compare them with previous similar works.
Finally, the main results of this study  are summarized   in Sect.~\ref{sec:conclusions}.


\section{\xmm~observations}\label{sec:observations}
\subsection{The sample}
The present  sample is constituted by  all Palomar--Green (PG) quasars (Schmidt \& Green 1983) with available
\epic~observations in the  \xmm~Science  Archive (XSA) as  of February
2004 plus   PG~1001$+$054, yielding   a preliminary
sample of 42 objects.  Two  QSOs (0003$+$199 and 1426$+$015) were 
excluded from the present study since all their
\epic~exposures are  affected by   pile--up. Basic
data for  the  final  sample of  40 objects   (i.e. name, coordinates,
redshifts and the  line--of--sight Galactic column density) are  given
in Table~\ref{tab:sample}.  The  redshifts span  values from 0.036  to
1.718.  All but 5 QSOs (0007$+$106, 1100$+$772, 1226$+$023, 1309$+$355,
1512$+$370, i.e.  $\sim$13\%  of the sample)  are radio--quiet sources
according to the definition   proposed  in Kellermann et  al.   (1994)
i.e. exhibit a ratio of radio to optical emission  $R_L$ $\leq$ 10. It
is  worth  noting that  our  sample includes all but two\footnote{Both
1425$+$267 and 1543$+$489 have been observed  by \xmm~but there were
not public data.} (1425$+$267 and 1543$+$489) of the sources in
the complete sample studied in  Laor et al. (1997) (i.e. all the PG QSOs with $M_B<-$23, $z \leq$ 0.4
and   \nhgal~$<$  1.9  $\times$ 10$^{22}$   \cm2).

\begin{table*}
\caption{The Sample}
\label{tab:sample}
\begin{footnotesize}
\begin{center}
\begin{tabular}{l c c c c c}
\hline \hline \\
{\bf PG Name} & {\bf Other name} & {\bf RA} & {\bf Dec} & {\bf z} & {\bf N$^{\rm Gal}_{H}$} \\
 & & (J2000) & (J2000) & & (10$^{20}\,{\rm cm^{-2}}$)\\
\hline\hline
0007$+$106$^\star$&III~Zw~2    & 00 10 31.0  &$+$10 58 30   &0.089  &6.09$^a$\\
0050$+$124& I~Zw~1 & 00 53 34.9 &$+$12 41 36 & 0.061 & 4.99$^a$ \\ 
0157$+$001& MKN~1014 & 01 59 50.2 & $+$00 23 41  & 0.163&2.46$^b$\\
0804$+$761&1H 0758$+$762& 08 10 58.6  &$+$76 02 42&0.100  &3.26$^c$ \\
0844$+$349&  TON~914        & 08 47 42.4 &$+$34 45 04 & 0.064 & 3.32$^c$  \\
0947$+$396$^L$&          & 09 50 48.4 &$+$39 26 50 & 0.206 & 1.92$^b$ \\
0953$+$414$^L$&          & 09 56 52.1  &$+$41 15 34   &0.239  &1.12$^c$\\
1001$+$054$^L$&          & 10 04 20.1  &$+$05 13 00   &0.161  &1.88$^a$\\
1048$+$342$^L$&          & 10 51 43.8  &$+$33 59 26   &0.167  &1.75$^b$\\
1100$+$772$^\star$&3C 249.1  & 11 04 13.7  &$+$76 58 58   &0.312  &3.17$^e$\\
1114$+$445$^L$&          & 11 17 06.4  &$+$44 13 33   &0.144  &1.93$^b$\\
1115$+$080&          & 11 18 16.0  &$+$07 45 59   & 1.718 & 3.62$^b$  \\
1115$+$407$^L$&          & 11 18 30.2  &$+$40 25 53   &0.154  &1.74$^b$\\
1116$+$215$^L$& TON~1388 &11 19 08.6  &$+$21 19 18   & 0.177 & 1.44$^a$ \\
1202$+$281$^L$&          & 12 04 42.1  &$+$27 54 11   & 0.165 & 1.72$^d$ \\ 
1206$+$459&          & 12 08 58.0  &$+$45 40 35   &1.158  &1.31$^d$\\
1211$+$143&          & 12 14 17.7  &$+$14 03 13   & 0.081 & 2.76$^c$ \\
1216$+$069$^L$&          & 12 19 20.9  &$+$06 38 38   &0.334  &1.57$^c$\\
1226$+$023$^{L\star}$&3C~273     & 12 29 06.7  &$+$02 03 08   &0.158  &1.89$^d$\\ 
1244$+$026&          & 12 46 35.3  &$+$02 22 09   & 0.048 & 1.93$^a$\\
1307$+$085&          & 13 09 47.0  &$+$08 19 48 & 0.155 & 2.11$^d$ \\
1309$+$355$^{L\star}$& TON~1565 & 13 12 17.8  &$+$35 15 21   &0.184  &1.00$^b$\\
1322$+$659$^L$&          & 13 23 49.5  &$+$65 41 48   &0.168  &1.89$^b$\\
1352$+$183$^L$&          & 13 54 35.6  &$+$18 05 17 & 0.158 & 1.84$^a$\\
1402$+$261$^L$&          & 14 05 16.2  &$+$25 55 35  & 0.164 & 1.42$^a$\\
1404$+$226&          & 14 06 21.8  &$+$22 23 35   &0.095  &3.22$^a$  \\
1407$+$265&          & 14 07 07.8  &$+$26 32 30   &0.940  &1.38$^a$\\
1411$+$442$^L$&          & 14 13 48.3  &$+$44 00 14   &0.0896 &1.05$^c$\\
1415$+$451$^L$&          & 14 17 01.24 &$+$44 56 16 &0.114 & 1.10$^e$ \\
1427$+$480$^L$&          &14 29 43.0&$+$47 47 26& 0.221 & 1.69$^b$\\
1440$+$356$^L$& MKN~478  & 14 42 07.4 &$+$35 26 23 & 0.079 & 0.97$^b$ \\
1444$+$407$^L$&          & 14 46 45.9 &$+$40 35 05& 0.267 & 1.27$^e$ \\
1501$+$106& MKN~841 & 15 04 01.2 &$+$10 26 16 & 0.036 & 2.19$^b$\\
1512$+$370$^{L\star}$&4C $+$37.43&15 14 43.0 &$+$36 50 50   &0.371  &1.36$^e$\\
1613$+$658&MKN 876   & 16 13 57.2 &$+$65 43 09   &0.129  &2.66$^a$ \\
1626$+$554$^L$&          & 16 27 56.0 &$+$55 22 31   &0.133  &1.55$^b$\\
1630$+$377&          & 16 32 01.2 &$+$37 37 49   &1.466  &0.90$^a$\\
1634$+$706&          & 16 34 31.4 &$+$70 31 34 & 1.334 & 5.74$^a$ \\
2214$+$139&MKN 304   & 22 17 11.5 &$+$14 14 28   &0.066  &4.96$^d$\\
2302$+$029&          & 23 04 45.0 &$+$03 11 46& 1.044 & 5.27$^d$  \\
\hline
\end{tabular}\end{center}
$^\star$ Radio--loud objects. $^L$ Objects included in Laor et al. (1997)
References for N$_{\rm H}$~values: $^a$ Elvis, Lockman \& Wilkes (1989);  $^b$ Murphy et
al. (1996); $^c$ Lockman \& Savage (1995); $^d$ Dickney \& Lockman
(1990); $^e$ Stark et al. (1992).
\end{footnotesize}
\end{table*}
\subsection{Data reduction}
The raw data from the \epic~instruments were processed using the standard Science Analysis System (SAS)
v5.4.1 (Loiseau 2003) to produce the linearized event files for \pn, \mosuno~ and \mosdue.
Only events with single and double patterns for the \pn~(PATTERN
$\leq$ 4) and single, double, triple and quadruple events for the
\mos~(PATTERN $\leq$ 12), were used for the spectral analysis. The subsequent event selection was
performed  taking into account  the  most updated calibration files at
the time of the reduction (September 2003).   All known flickering and
bad pixels were removed. Furthermore, periods of background flaring in
the \epic~data were excluded using  the method described in Piconcelli
et al.  (2004).   Useful exposure times  after  cleaning are listed in
Table~\ref{tab:journal} for each camera together with the date and the
revolution  of each reduced observation.   In the  case of 0844$+$349,
1244$+$026  and  1440$+$356,  the  \pn~observations  are  affected  by
pile--up and, therefore, only    the \mos~spectra were used    for the
analysis.  On the other hand,  for 1226$+$023 and 1501$+$106 only
\pn~data were  analyzed  due  to  impossibility to select a background region since the \mos~
observations were carried out in small window mode.   Source
spectra were extracted  from circular regions centered  on the peak of
the X-ray   counts. Backgrounds were estimated  in a source--free circle of equal
radius close to the source on the  same CCD.  Appropriate response and
ancillary files for both the \pn~and the
\mos~cameras were created using the RMFGEN and ARFGEN tools in the 
$SAS$ software, respectively.
As the difference between the {\em MOS1} and {\em MOS2} response matrices are a few percent, 
we created a combined \mos~spectrum and
response matrix. The background--subtracted spectra for the \pn~and combined
\mos~cameras were then simultaneously fitted. According to  the current calibration uncertainties 
we performed the spectral analysis in the 0.3-12~keV band for
\pn~and in the 0.6--10 keV band for the \mos~cameras, respectively.


\begin{table*}
\caption{Journal of Observations.}
\begin{center}
\label{tab:journal}
\begin{footnotesize}
\begin{tabular}{c c c c c c||c c c c c c}
\hline \hline \\
PG Name & Date &Rev. &\multicolumn{3}{c}{Exposure Time (ks)} & PG Name
&   Date   &Rev.  &  \multicolumn{3}{c}{Exposure  Time    (ks)} \\ & &
&\pn&\mosuno&\mosdue & & & &\pn&\mosuno&\mosdue\\
\hline\hline
&&&& & & &&&\\
0007$+$106&2000--07--04&104 &9.9&7.5&7.5 & 		    1307$+$085 & 2002--06--13 & 460 & 11.2 &  13.3   & 13.3 \\     
0050$+$124 & 2002--06--22 & 464 & 18.6 &  20.7   & 19.6   &    1309$+$355&2002--06--10&458&25.2&28.4&27.3\\                
0157$+$001 & 2000--07--29 & 117 & 10.1 &  10.4   & 10.4  &     1322$+$659&2002--05--11&443&8.6&11.4&11.6\\                 
0804$+$761&2000--11--04&166&0.5&6.6&6.5 & 		    1352$+$183 & 2002--07--20 & 478 & 12.3 &  14.7   & 14.9 \\     
0844$+$349 & 2000--11--04 & 166 & -    &  23.3   & 23.3  &     1402$+$261 & 2002--01--27 & 391 & 9.1  &  11.9   & 11.9 \\     
0947$+$396 & 2001--11--03 & 349 & 17.5 &  20.9   & 21.1  &     1404$+$226&2001--06--18&279&16.3&19.4&18.1\\                
0953$+$414&2001--11--22&358&11.5&12.4&14.7 & 		    1407$+$265&2001--12--22&373&35.1&36.8&36.9\\                
1001$+$054&2003--05--04&623&8.8&10.1&11.0 & 		    1411$+$442&2002--07--10&473&23.1&35.4&35.4\\                
1048$+$342&2002--05--13&444&28.1&32.1&32.1 & 		    1415$+$451 & 2002--12--08 & 549 & 21.1 &  24.1   & 24.1 \\     
1100$+$772&2002--11--01&530&19.2&22.5 &22.5 & 		    1427$+$480 & 2002--05--31 & 453 & 35.2 &  38.9   & 38.5 \\     
1114$+$445&2002--05--14&445&37.7&42.2&42.3 & 		    1440$+$356 & 2001--12--23 & 373 & - &  28.2   & 24.8 \\     
1115$+$080 & 2001--11--25 & 360 & 53.8 &  61.8   & 61.8  &     1444$+$451 & 2002--08--11 & 489 & 18.5 &  21.0   & 21.0 \\     
1115$+$407&2002--05--17&446&15&20.2&19.9 & 		    1501$+$106 & 2001--01--14 & 201 & 9.4  &  -        & -     \\  
1116$+$215 & 2001--12--02 & 363 & 5.5  &  8.3    & 8.4  & 	    1512$+$370&2002--08--25&496&17.6&20.4&20.4\\                
1202$+$281 & 2002--05--30 & 453 & 12.9 &  17.3   & 17.3  &     1613$+$658&2001--04--13&246&3.5&3.3 &3.5 \\                 
1206$+$459 & 2002--05--11 & 443	& 9.1  & 12.2& 12.2 &		    1626$+$554&2002--05-05&440&5.5&8.7&7.2\\                    
1211$+$143 & 2001--06--15 & 278 & 49.5 &  53.4   & 53.3  &     1630$+$377&2002--01--06&380&12.3&15.7&16\\                  
1216$+$069&2002--12--18&554&14&16.5&16.5 &  		    1634$+$706 & 2002--11--22 &  541& 15.7 &  19.0   & 18.8 \\     
1226$+$023&2000--06--15&95&20.8&$-$&$-$ & 		    2214$+$139&2002--05--12&444&29&33.4&35.5\\                  
1244$+$026 & 2001--06--17 & 279 & -    &  12.1   & 12.1  &     2302$+$029 & 2001--11--29 & 362 & 9.0  &  12.4  & 12.2 \\      
\hline
\end{tabular}
\end{footnotesize}
\end{center}
\end{table*}

\section{Analysis of the spectra}\label{sec:analysis}
The source spectra were grouped such that each spectral bin contains
at least 35 (or more in the case of the brightest sources) 
counts in order to apply the modified $\chi^2$ minimization
technique and they were analyzed using XSPEC v.11.2 (Arnaud 1996).
Galactic absorption (see Table~\ref{tab:sample}) is always implicitly included in all the spectral
models presented hereafter.  The photoelectric absorption cross
sections of Morrison \& McCammon (1983) and the solar abundances of
Anders \& Grevesse (1989) were used. The quoted errors refer to  the
90\% confidence level for one interesting parameter
(i.e. $\Delta\chi^2$ = 2.71; Avni 1976). Throughout this paper we
assume a flat {\em $\Lambda$CDM} cosmology with ($\Omega_{\rm
M}$,$\Omega_{\rm \Lambda}$)  = (0.3,0.7) and a Hubble constant of 70
km s$^{-1}$ Mpc$^{-1}$ (Bennett et al. 2003). All fit parameters are
given in the quasar rest frame.

\begin{table*}
\caption{Spectral fitting results. I: Power law model applied to the
  hard band (2--12~keV). 1001$+$054 is excluded due to the very poor
  statistics in this energy range.}
\begin{center}
\label{tab:pl_212}
\begin{tabular}{c c c l ||c c c l}
\hline \hline  &&&&&&&\\ PG Name & $\Gamma_{\rm 2-12}$ &$\chi^2$ &
dof & PG Name & $\Gamma_{\rm 2-12}$ &$\chi^2$ &   dof \\ \hline\hline
0007$+$106&1.61$^{+0.05}_{-0.04}$&135  & 179&1307$+$085 &
1.46$^{+0.07}_{-0.08}$	&117 & 92 \\ 0050$+$124& 2.28$^{+0.03}_{-0.03}$
& 276 & 266&1309$+$355&1.72$^{+0.07}_{-0.06}$&95& 87   \\ 0157$+$001&
2.1$^{+0.2}_{-0.2}$ & 31 &
48&1322$+$659&2.2$^{+0.2}_{-0.2}$&91&88\\
0804$+$761&2.05$^{+0.07}_{-0.07}$&   148          &  135&1352$+$183  &
1.91$^{+0.09}_{-0.09}$&	126 & 97 \\ 0844$+$349 &
2.11$^{+0.05}_{-0.05}$ & 106 & 109&1402$+$261& 2.06$^{+0.09}_{-0.08}$
& 77 & 92\\ 0947$+$396& 1.81$^{+0.06}_{-0.06}$ & 121 &
137&1404$+$226&2.3$^{+0.8}_{-0.8}$&  5           &9       \\
0953$+$414&2.01$^{+0.07}_{-0.06}$&64&69&1407$+$265&2.19$^{+0.07}_{-0.07}$&83
& 118      \\
1001$+$054&$-$&$-$&$-$&1411$+$442&0.38$^{+0.11}_{-0.13}$&111& 67    \\
1048$+$342&1.77$^{+0.05}_{-0.05}$&113&149& 1415$+$451 &
2.03$^{+0.08}_{-0.08}$ & 65 & 105\\ 1100$+$772&
1.65$^{+0.05}_{-0.05}$&180  & 220&1427$+$480& 1.90$^{+0.07}_{-0.05}$
& 115 & 155\\ 1114$+$445&1.48$^{+0.04}_{-0.03}$&  255&241&1440+356 &
2.03$^{+0.06}_{-0.07}$  & 173 & 104\\ 1115$+$080 &
2.00$^{+0.11}_{-0.11}$	&87 & 87& 1444$+$407 & 2.12$^{+0.12}_{-0.12}$  &
30 & 48\\ 1115$+$407&2.16$^{+0.07}_{-0.07}$&94 &100&1501$+$106 &
1.88$^{+0.03}_{-0.03}$  & 164 & 154\\ 1116$+$215& 2.14$^{+0.08}_{-0.08}$
& 105 & 106&1512$+$370&1.82$^{+0.06}_{-0.07}$&97&130       \\
1202$+$281 &  1.69$^{+0.05}_{-0.05}$ & 189 & 195&
1613$+$658&1.70$^{+0.10}_{-0.10}$&     48 &66 \\
1206$+$459&2.0$^{+0.4}_{-0.5}$&15&12&1626$+$554&1.95$^{+0.10}_{-0.10}$&130
& 137      \\ 1211$+$143& 1.76$^{+0.03}_{-0.03}$ & 380 &
278&1630$+$377&2.2$^{+0.2}_{-0.4}$&  17           &    10   \\
1216$+$069&1.73$^{+0.08}_{-0.10}$&54&79&1634$+$706&2.04$^{+0.09}_{-0.09}$  & 79 &84\\ 
1226$+$023&1.634$^{+0.009}_{-0.009}$&260             &193
&2214$+$139&0.94$^{+0.04}_{-0.04}$&   469&256\\ 
1244$+$026&2.55$^{+0.11}_{-0.11}$ & 69 & 55& 2302$+$029  & $2.2^{+0.3}_{-0.2}$ &13 & 23\\ 
\hline\\
\end{tabular}
\end{center}
\end{table*} 

\subsection{Continuum emission above 2 keV}\label{sec:pwlw}

\begin{figure}
\begin{center}
\epsfig{figure=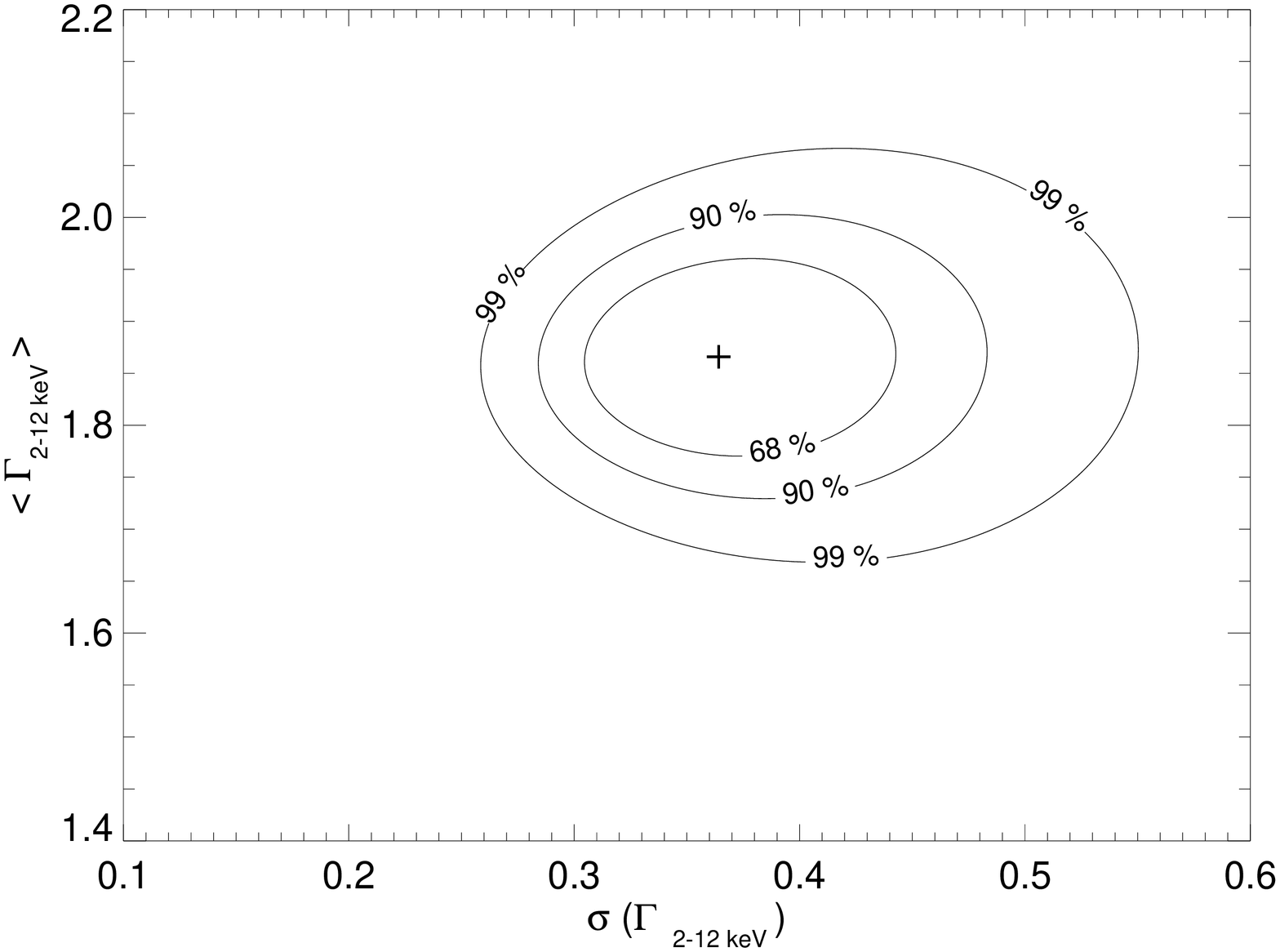,width=4.3cm,angle=90}
\caption{Confidence contours for the simultaneous determination of the
photon index
$\langle\Gamma_{2-12}\rangle$ and the intrinsic dispersion
$\sigma(\Gamma_{2-12})$ together with the best-fit values found.}
\label{fig:gamma_2_12_contours}
\end{center}
\end{figure}

A simple redshifted power law model  has been fitted  to the hard X-ray
band, excluding  the  data below   2  keV  where additional   spectral
components like soft excess and absorbing  features can heavily modify
the primary source continuum.  Such a fit resulted acceptable for most
sources,  with   only  five   objects (i.e.   1211$+$143,  1226$+$023,
1411$+$442, 1630$+$377 and 2214$+$139) yielding an associate \xnu~$>$
1.2. Some     of  these sources  show a    very   flat  spectrum  with
$\Gamma_{2-12}$  $<$    1 which suggests    the  presence of intrinsic
absorption  obscuring the primary  continuum.  The resulting best--fit
parameters are displayed in Table~\ref{tab:pl_212}.  Using the maximum
likelihood technique (see  Maccacaro et al. 1988), we have  calculated the
best  simultaneous   estimate    of  the  average   photon    index of
$\langle\Gamma_{2-12}\rangle$      and         the  intrinsic   spread
$\sigma(\Gamma_{2-12})$.  Figure~\ref{fig:gamma_2_12_contours} shows  the
68\%, 90\% and 99\% confidence contours for the two parameters together
with  the best-fit values obtained, i.e. $\langle\Gamma_{2-12}\rangle$
=     $1.87^{+0.09}_{-0.10}$        and  $\sigma(\Gamma_{2-12})$     =
0.36$^{+0.08}_{-0.06}$ (errors  have been calculated using  the
68\% contour level).

\begin{table*}
\caption{Spectral fitting results. II: Power law model (Model PL) applied to the 0.3--12 keV band.}
\label{tab:pl_0312}
\begin{center}
\begin{tabular}{c c c c c|c c c c c}
\hline \hline \\ PG Name & $\Gamma$& A$_{pl}$ & $\chi^2$ &  d.o.f.& PG
 Name & $\Gamma$& A$_{pl}$ & $\chi^2$ &  d.o.f. \\ \hline\hline
0007$+$106&1.82$^{+0.01}_{-0.02}$&2.32$^{+0.02}_{-0.02}$ $\times$
10$^{-2}$ &407 &307 & 1307$+$085 & $1.76^{+0.04}_{-0.02}$ & 5.5
$^{+0.2}_{-0.2}\times 10^{-4}$ & 781 & 214\\ 0050$+$124 &
$1.77^{+0.02}_{-0.02}$ & $1.60^{+0.02}_{-0.02}\times10^{-3}$ & 1359&
394&1309$+$355&1.94$^{+0.02}_{-0.03}$&3.01$^{+0.05}_{-0.05}$ $\times$
10$^{-4}$ &615 &220\\ 0157$+$001 & $2.50^{+0.03}_{-0.03}$ &
$8.51^{+0.13}_{-0.13}\times10^{-4}$ & 183 & 168&
1322$+$659&2.81$^{+0.02}_{-0.02}$&1.64$^{+0.01}_{-0.01}$ $\times$
10$^{-3}$ &515&216\\
0804$+$761&2.28$^{+0.01}_{-0.02}$&7.21$^{+0.07}_{-0.07}$ $\times$
10$^{-3}$&364&266& 1352$+$183 & $2.50^{+0.02}_{-0.02}$ &
$1.596^{+0.015}_{-0.015}\times10^{-3}$ & 472 & 248 \\ 0844$+$349 & $2.41^{+0.01}_{-0.01}$ &
$3.80^{+0.04}_{-0.04}\times10^{-3}$ & 451 & 166& 1402$+$261 &
$2.71^{+0.02}_{-0.02}$ & 2.10 $^{+0.02}_{-0.02}\times 10^{-3}$ & 449 &
219\\ 0947$+$396 &  $2.307^{+0.016}_{-0.016}$&$1.341^{+0.012}_{-0.012}\times 10^{-3}$ & 496 &
265& 1404$+$226&4.18$^{+0.05}_{-0.05}$&2.20$^{+0.07}_{-0.07}$ $\times$
10$^{-4}$ &570 &79\\
0953$+$414&2.46$^{+0.01}_{-0.01}$&3.03$^{+0.01}_{-0.01}$ $\times$
10$^{-3}$ &823&290&
1407$+$265&2.28$^{+0.02}_{-0.02}$&2.00$^{+0.03}_{-0.03}$ $\times$
10$^{-3}$ &243 &246\\
1001$+$054&4.0$^{+0.4}_{-0.4}$&1.7$^{+0.2}_{-0.2}$ $\times$
10$^{-5}$ &156 &10&
1411$+$442&2.6$^{+0.1}_{-0.1}$&5.0$^{+0.2}_{-0.2}$ $\times$
10$^{-5}$ &730 &108\\
1048$+$342&2.24$^{+0.02}_{-0.02}$&8.50$^{+0.08}_{-0.08}$ $\times$
10$^{-4}$ &559&275& 1415$+$451 & $2.55^{+0.02}_{-0.02}$ &
$8.49^{+0.08}_{-0.08}\times10^{-4}$ & 458 & 232 \\
1100$+$772&2.01$^{+0.01}_{-0.01}$&2.16$^{+0.01}_{-0.02}$ $\times$
10$^{-3}$ &606 &348& 1427$+$480 &$2.36^{+0.014}_{-0.014}$ &
$8.13^{+0.07}_{-0.07}\times10^{-4}$ & 555 & 283\\
1114$+$445&1.14$^{+0.01}_{-0.01}$&2.90$^{+0.05}_{-0.05}$ $\times$
10$^{-4}$ &3607 &368& 1440$+$356 & $2.65^{+0.02}_{-0.02}$ &
3.09$^{+0.03}_{-0.02}\times10^{-3}$ & 731 & 161\\ 1115$+$080 &
$1.77^{+0.02}_{-0.02}$ & $6.3^{+0.2}_{-0.2}\times10^{-4}$ & 354&  214& 1444$+$407 &
$2.72^{+0.03}_{-0.03}$ & 8.14 $^{+0.10}_{-0.10}\times 10^{-4}$ & 357 &
168 \\ 1115$+$407 &2.72$^{+0.02}_{-0.02}$&1.49$^{+0.02}_{-0.02}$
$\times$ 10$^{-3}$ &530 &234&  1501$+$106 & $2.47^{+0.02}_{-0.02}$ &
9.38 $^{+0.04}_{-0.04}\times 10^{-3}$ & 2281 & 225\\ 1116$+$215 &
$2.60^{+0.02}_{-0.02}$ & $3.45^{+0.03}_{-0.03}\times10^{-3}$ &456 & 234& 1512$+$370&
2.15$^{+0.01}_{-0.02}$&1.49$^{+0.01}_{-0.01}$ $\times$ 10$^{-3}$ &452
&257\\ 1202$+$281 & $1.69^{+0.05}_{-0.05}$ & $1.16^{+0.08}_{-0.08}\times10^{-3}$ & 730 &
323& 1613$+$658&2.13$^{+0.04}_{-0.04}$&1.82$^{+0.04}_{-0.04}$ $\times$
10$^{-3}$ &354 &184\\
1206$+$459&1.74$^{+0.09}_{-0.09}$&2.4$^{+0.2}_{-0.2}$ $\times$
10$^{-4}$ &40 &42&
1626$+$554&2.25$^{+0.02}_{-0.02}$&1.95$^{+0.05}_{-0.05}$ $\times$
10$^{-3}$ &345 &271\\ 1211$+$143& $3.14^{+0.02}_{-0.02}$ &
1.72$^{+0.02}_{-0.03}\times10^{-3}$ & 15193 & 404&
1630$+$377&2.25$^{+0.07}_{-0.07}$&5.2$^{+0.4}_{-0.4}$ $\times$
10$^{-4}$ &70&61\\
1216$+$069&2.19$^{+0.03}_{-0.03}$&1.0$^{+0.1}_{-0.2}$ $\times$
10$^{-3}$ &336 &204& 1634$+$706 & $2.22^{+0.08}_{-0.02}$ &
$3.57^{+0.09}_{-0.09}\times10^{-3}$ & 238 & 210\\
1226$+$023&1.86$^{+0.01}_{-0.01}$&2.85$^{+0.01}_{-0.01}$ $\times$
10$^{-2}$ &6455 &264&
2214$+$139&0.58$^{+0.02}_{-0.02}$&1.34$^{+0.03}_{-0.03}$ $\times$
10$^{-4}$ &2483 &367\\ 1244$+$026 & $2.90^{+0.02}_{-0.02}$ &
3.15$^{+0.04}_{-0.04}\times10^{-3}$ & 233 & 112& 2302$+$029 &
$2.29^{+0.06}_{-0.06}$ & $6.7^{+0.4}_{-0.4}\times10^{-4}$ & 94 & 68
\\ \hline
\end{tabular}
\end{center}
($^{\dag}$)~Flux at 1 keV in units of photons/keV/cm$^{2}$/s.
\end{table*}

Although the quality of the fit in the 2-12~keV  band is good for most
quasars, the residuals show an excess  around 6~keV which  suggests iron fluorescence emission.
A detailed analysis of the Fe K$\alpha$ emission line is deferred to a second paper (see Jimenez-Bailon et al. 2004; Paper II hereafter).

\subsection{Systematic modeling of the 0.3--12 keV continuum}
\label{sec:soft}

The  extrapolation  of the power  law   to energies  lower  than 2 keV
clearly  revealed the  presence of  large  deviations  in all  but one
(i.e. 1206$+$459) spectra, with the most common residual feature being
a smooth  excess of soft X-ray  flux. On the  other hand, a handful of
objects (i.e.  1001$+$054, 1114$+$445, 1115$+$080 and  2214$+$139)  exhibit a deep
and sharp   deficit in  the 0.5--2  keV   band likely due   to complex
absorption.

\subsubsection{Power law fit}

The 0.3--12 keV spectrum of each source was fitted with a single
power law model (PL) in order to provide an overall indication of the
X--ray broad--band spectral shape ($\langle\Gamma_{0.3-12}\rangle$    =  2.30$\pm0.15$ and 
dispersion of $\sigma(\Gamma_{0.3-12})$ = 0.60$^{+0.12}_{-0.09}$).
The vast  majority of   \xnu~values associated  with  this
parameterization show that it yields a very poor description of the
\epic~data (see Table~\ref{tab:pl_0312} for the
results).   
This fact is an evident consequence of the inclusion of the 0.3--2 keV range in the
fit,  which is dominated by the emission of the steeper ``soft
excess'' component.
Therefore,  we fitted  more complex  models
accounting for the   soft  excess  (Sect.~\ref{sec:soft}) and the   
absorption features (Sect.~\ref{sec:complex}) if present in the spectrum. 
The presence of iron  emission lines is investigated in Paper II.

\subsubsection{Soft excess modeling}
\begin{figure*}
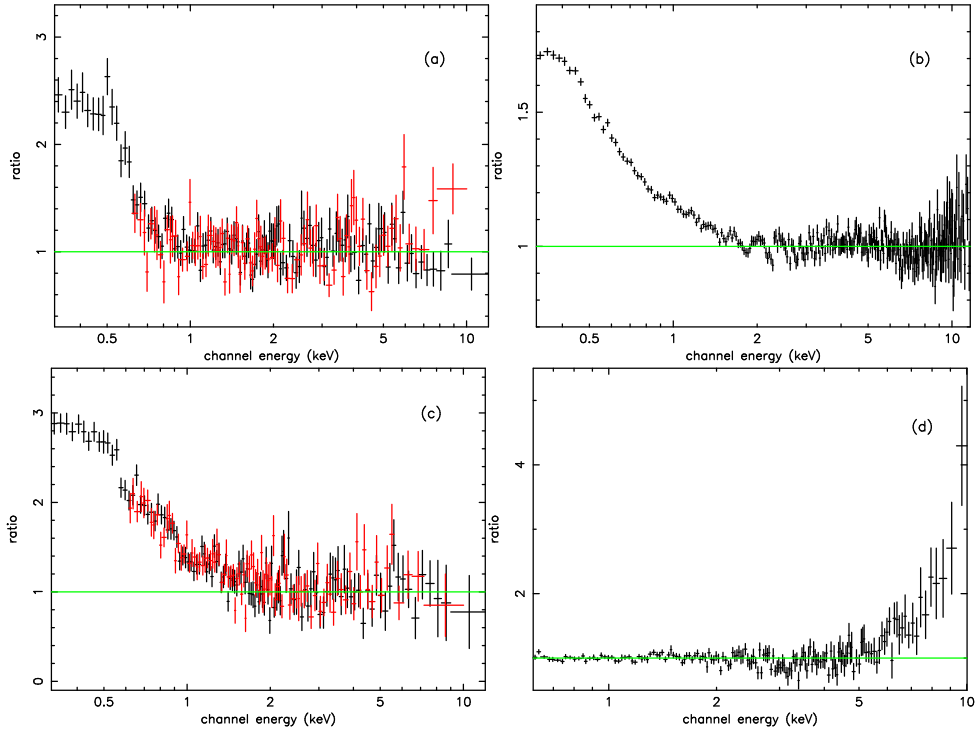

\begin{center}
\epsfig{figure=ep1621_f2a.ps,width=4.8cm,angle=-90}\epsfig{figure=ep1621_f2b.ps,width=4.8cm,angle=-90}
\epsfig{figure=ep1621_f2c.ps,width=4.8cm,angle=-90}\epsfig{figure=ep1621_f2d.ps,width=4.8cm,angle=-90}
\caption{The ratios (data/model) resulting from fitting  a power law
to the 2-12 keV \epic~data (\pn~data in black, \mos~in red) and extrapolating to lower energies ({\it
from top-left)} in the case of the QSO ({\it from
top-left)} (a) 1307+085 ($z$ = 0.155): the soft excess emission in this source results
to be best--fitted with a blackbody (i.e. model A; see Table~\ref{tab:best}); (b) 1226+023 ($z$ = 0.158):
 note the broader excess than in  1307+085 (the two QSOs have a similar $z$) which requires two blackbodies (i.e. model E; 
see Table~\ref{tab:best}) to be parametrized; (c) 1352+183 ($z$ = 0.158) for which model C (i.e. bremsstrahlung $+$ power law) gives
the best fit. {\it Bottom-right:} (d) the ratio resulted fitting with model A the \epic~data of 1440+356 ($z$ = 0.079). Note the large
residuals at high energies. This QSO is best--fitted with 
a double power law model (i.e. model D; see Table~\ref{tab:best}).}
\label{fig:se}
\end{center}
\end{figure*}

We  accounted for the soft excess emission by means of four different two--component continua:
(A) blackbody ({\tt BBODY} in XSPEC)$+$ power law; (B)
multicolor blackbody ({\tt DISKBB} in XSPEC) $+$ power law;  (C)
bremsstrahlung ({\tt BREMSS} in XSPEC) $+$ power law; (D) power law
$+$ power law. In Fig.~\ref{fig:se} are plotted some examples of
  source spectra fitted with the above models which can be considered representative of the different
 shapes of soft excess emission found in the sample.

Furthermore, photoelectric absorption edges were added to the fits
whenever appropriate. We
included these additional components by
the $F-test$ criterion at a significance level $\geq$95\%. 
For the high--$z$ QSO 1206$+$459, whose emission did not show any deviation from a simple
power law continuum, we did not perform any further spectral fits.
Eight sources (i.e. 0050$+$124, 1001$+$054, 1114$+$445, 1115$+$080, 1226$+$023, 1404$+$226,
1411$+$442, and 2214$+$139) exhibited spectra more complex than a
two--component continuum (plus absorption edge) and 
yielded a \xnu~$\geq$1.2 for each of the tested models. They are not listed in the relative Tables, but rather they
are individually discussed in Sect.~\ref{sec:complex}.

The best--fitting parameters resulted by the application of model A,
B, C and D are shown in Table~\ref{tab:A},~\ref{tab:B},~\ref{tab:C}
and ~\ref{tab:D}, respectively.
\begin{table*}
\caption{Spectral fitting results. III: Power law plus blackbody model
  for the soft excess component (Model A).}
\label{tab:A}
\begin{center}
\begin{tabular}{c c c c c c c c c c}
\hline  {\bf PG Name} &  A$^{\dag}_{pl}$ & $\Gamma$ &  A$^{\dag}_{BB}$ &
kT & Edge & $\tau$&$P_{F-test}$ & $\chi^2_{\nu}$ & dof \\ 
&&          &($\times 10^{-5}$)&(keV)&(keV)& &(\%)    & &\\
\hline\hline\\ 
0007$+$106&1.97$^{+0.07}_{-0.07}$
$\times$10$^{-3}$&1.67$^{+0.02}_{-0.02}$
&1.5$^{+0.2}_{-0.2}$&0.16$^{+0.09}_{-0.09}$&$-$&$-$&$-$&318 &305
\\ 
0157+001 & $7.1^{+0.5}_{-0.5}\times10^{-4}$ & $2.24^{+0.06}_{-0.07}$
&1.0$^{=}_{-}0.2$ & $0.127^{+0.010}_{-0.010}$ & $-$ &
$-$ & $-$ & 149 & 166 \\

0804$+$761&6.6$^{+0.2}_{-0.1}$ $\times$
10$^{-3}$&2.18$^{+0.02}_{-0.04}$&5.6$^{+1.6}_{-1.3}$
&0.112$^{+0.013}_{-0.013}$&$-$&$-$&$-$&309&264\\ 

0844+349  &   $3.31^{+0.08}_{-0.08}\times10^{-3}$  &  2.24$^{+0.03}_{-0.03}$  &
7.0$^{+2}_{-1}$ & $0.108^{+0.008}_{-0.008}$ & $-$ & $- $& $-$ & 189& 164 \\

0947+396 & $1.01^{+0.04}_{-0.05}\times10^{-3}$ & 1.99$^{+0.04}_{-0.04}$
&2.0$^{+0.2}_{-0.3}$ & $0.160^{+0.007}_{-0.008}$ &
$0.75^{+0.02}_{-0.03}$ & $0.28^{+0.07}_{-0.08}$ & $>$99.9 &254 & 261 \\

0953$+$414&2.21$^{+0.07}_{-0.07}$$\times$ 10$^{-3}$
&2.12$^{+0.02}_{-0.04}$&5.5$^{+0.4}_{-0.4}$&0.16
$^{+0.04}_{-0.04}$&0.73$^{+0.02}_{-0.02}$&0.33$^{+0.05}_{-0.05}$&$>$99.9&340
&286\\ 

1048$+$342&6.5$^{+0.2}_{-0.2}$ $\times$
10$^{-4}$&1.93$^{+0.03}_{-0.03}$&1.15$^{+0.09}_{-0.09}$ &0.133
$^{+0.05}_{-0.05}$&$-$ &$-$&$-$&283&273\\

1100$+$772&1.83$^{+0.02}_{-0.05}$ $\times$ 10$^{-3}$&
1.84$^{+0.03}_{-0.02}$ & 1.87$^{+0.20}_{-0.20}$&
0.141$^{+0.008}_{-0.010}$&$-$&$-$&$-$&401 &346 \\

1116+215 &  $2.8^{+0.1}_{-0.1}\times10^{-4}$ & $2.29^{+0.04}_{-0.03}$ &
$6.0^{+0.5}_{-0.6}$ & $0.113^{+0.005}_{-0.006}$ & $-$
&$-$ & $-$ & 220 & 232 \\ 

1202+281 & $1.37^{+0.05}_{-0.05}\times10^{-3}$  & $1.80^{+0.03}_{-0.03}$ & $2.7^{+0.2}_{-0.2}$ &
$0.159^{+0.007}_{-0.007}$ & $0.72^{+0.02}_{-0.02}$ &$0.33^{+0.07}_{-0.07}$ & $>$99.9 & 328 & 319 \\

1211+143 & $9.6^{+0.3}_{-0.2}\times10^{-4}$ &$1.715^{+0.010}_{-0.012}$ &
$9.36^{+0.12}_{-0.17}$ & $0.1157^{+0.0013}_{-0.0015}$ &
$0.771^{+0.012}_{-0.005}$ & $0.35^{+0.04}_{-0.03}$ & $>$99.9 & 771 & 400
\\ & & & & & $0.953^{+0.02}_{-0.008}$ & $0.28^{+0.03}_{-0.03}$ & $>$99.9 & 571 & 398\\ & & & & &
$7.28^{+0.16}_{-0.12}$ & $0.54^{+0.09}_{-0.09}$ & $>$99.9 & 458 &396\\

1216$+$069&7.8$^{+0.3}_{-0.4}$ $\times$ 10$^{-4}$&
1.91$^{+0.05}_{-0.05}$ & 1.4$^{+0.2}_{-0.2}$&
0.13$^{+0.01}_{-0.01}$&$-$&$-$&$-$&181 &202 \\

1244+026  &     $2.17^{+0.09}_{-0.08}\times10^{-3}$   &  2.48$^{+0.03}_{-0.03}$    &
4.93$^{+0.04}_{-0.04}$ &   $0.16^{+0.01}_{-0.01}$ &$-$  &
$-$ & $-$   &    125      &   110    \\    

1307+085      &
$4.4^{+0.2}_{-0.2}\times10^{-4}$   &      $1.50^{+0.04}_{-0.04}$     &
$1.2^{+0.1}_{-0.1}$    &    0.13$^{+0.01}_{-0.01}$   &   $0.74^{+0.02}_{-0.02}$   &   $
0.54^{+0.10}_{-0.14}$    &       99.9    &   232    &     210     \\

1309$+$355&2.76$^{+0.12}_{-0.12}$                             $\times$
10$^{-4}$&1.80$^{+0.04}_{-0.05}$
&0.53$^{+0.14}_{-0.10}$&0.104$^{+0.008}_{-0.008}$&0.74$^{+0.04}_{-0.02}$&0.4$^{+0.2}_{-0.2}$&$>$99.9&238&216\\

1402+261  &   $1.58^{+0.08}_{-0.07}\times10^{-3}$   &  $2.33^{+0.05}_{-0.05}$ &
$4.2^{+0.5}_{-0.4}$           &     $0.135^{+0.003}_{-0.004}$        &
$0.68^{+0.04}_{-0.05}$ & $0.24^{+0.08}_{-0.08}$ & 99.8 & 243 & 215 \\

1407$+$265&1.82$^{+0.07}_{-0.07}$                             $\times$
10$^{-3}$&2.21$^{+0.03}_{-0.03}$
&0.7$^{+0.3}_{-0.3}$&0.20$^{+0.03}_{-0.03}$&$-$&$-$&$-$&223&244\\

1415+451 & $6.4^{+0.3}_{-0.2}\times10^{-4}$ & $2.15^{+0.05}_{-0.05}$ &
$1.7^{+0.2}_{-0.2}$        &      $0.143^{+0.008}_{-0.009}$          &
$0.78^{+0.04}_{-0.04}$ & $0.19^{+0.09}_{-0.11}$ & 99.7 & 196 & 228 \\ &
&  & & &   $0.62^{+0.02}_{-0.03}$ & $0.27^{+0.09}_{-0.1}$& 97.9 & 189
& 226\\ 

1427+480  & $6.9^{+0.3}_{-0.3}\times10^{-4}$  &  $2.04^{+0.03}_{-0.03}$ & $1.16^{+0.10}_{-0.10}$ &
$0.155^{+0.007}_{-0.005}$ & $0.72^{+0.02}_{-0.02}$& $0.25^{+0.06}_{-0.07}$ & $>$99.9
& 269 & 279 \\



1512$+$370&1.20$^{+0.05}_{-0.05}$         $\times$          10$^{-3}$&
1.92$^{+0.03}_{-0.03}$            &            1.7$^{+0.3}_{-0.2}$&
0.14$^{+0.01}_{-0.01}$&$-$&$-$&$-$&258              &255            \\

1613$+$658&1.50$^{+0.09}_{-0.09}$           $\times$        10$^{-3}$&
1.80$^{+0.07}_{-0.03}$             &           3.1$^{+0.4}_{-0.4}$&
0.13$^{+0.01}_{-0.01}$&0.73$^{+0.02}_{-0.02}$&0.5$^{+0.1}_{-0.2}$& 99.9 &148 &180 \\

1626$+$554&1.73$^{+0.06}_{-0.06}$                             $\times$
10$^{-3}$&2.07$^{+0.05}_{-0.05}$       &2.2$^{+0.4}_{-0.4}$       &
0.100$^{+0.001}_{-0.001}$&$-$&$-$&$-$&270 &268\\

1630$+$377&3.9$^{+0.8}_{-0.3}\times$10$^{-4}$                     &
2.05$^{+0.06}_{-0.12}$&0.7$^{+0.8}_{-0.3}$&0.21$^{+0.04}_{-0.04}$&
$-$&$-$&$-$&55&59\\

1634+706 & $2.8^{+0.3}_{-0.3}\times10^{-3}$ & $2.07^{+0.02}_{-0.05}$ &
$2.1^{+0.5}_{-0.5}$ & $0.31^{+0.03}_{-0.03}$ &  $-$ & $-$  & $-$ & 209 & 208
\\

2302+029     &    $4.9^{+0.7}_{-0.6}\times10^{-4}$   &     2.05$^{+0.09}_{-0.09}$       &
$1.2^{+0.6}_{-0.3}$ & $0.18^{+0.02}_{-0.02}$ & $-$ & $-$ & $-$ & 59 & 66 \\\hline
\end{tabular}\end{center}
($^{\dag}$)~Flux at 1 keV in units of photons/keV/cm$^{2}$/s.
\end{table*}

\begin{table*}
\caption{Spectral Fitting results: IV. Power law plus multi--color
  blackbody model for the soft excess component (Model B).}
\label{tab:B}
\begin{center}
\begin{tabular}{c c c c c c c c c c}
\hline  {\bf PG Name} &  A$^{\dag}_{pl}$ & $\Gamma$ &  A$_{DiskBB}$ & kT &
Edge & $\tau$&$P_{F-test}$& $\chi^2_{\nu}$ & dof \\ 
   &       & &&(keV) &(keV)&&(\%)&&\\ \hline\hline\\
0007$+$106&1.9$^{+0.1}_{-0.1}$ $\times$
10$^{-3}$&1.63$^{+0.04}_{-0.04}$
&29$^{+10}_{-8}$&0.23$^{+0.02}_{-0.02}$&$-$&$-$&$-$&318 &305\\

0157+001 & $ 6.6^{+0.7}_{-0.6} \times10^{-4}$ & $2.18^{+0.10}_{-0.10}$ &
$50^{+30}_{-20}$ & $0.17^{+0.02}_{-0.02}$ & $-$ &
$-$ & $-$ & 149 & 166\\

0804$+$761&6.5$^{+0.2}_{-0.2}$ $\times$
10$^{-3}$&2.16$^{+0.04}_{-0.02}$&916$^{+1363}_{-508}$&0.14$^{+0.02}_{-0.02}$&$-$&$-$&$-$&307
&264 \\ 

0844$+$349 & $ 3.29^{+0.08}_{-0.08} \times10^{-3}$ & $
2.23^{+0.03}_{-0.04}$ & $2355^{+2390}_{-1185}$ &
$0.12^{+0.01}_{-0.01}$ & $-$ & $-$ & $-$& 189 & 164\\ 

0947$+$396 & $
9.3^{+0.5}_{-0.7} \times10^{-4}$ & $1.93^{+0.04}_{-0.06}$ &
$69^{+17}_{-14}$ & $0.189^{+0.013}_{-0.011}$ & $0.77^{+0.03}_{-0.05}$ &
$0.24^{+0.08}_{-0.06}$ & $>$99.9 & 241 & 261 \\

0953$+$414&2.1$^{+0.1}_{-0.1}$ $\times$
10$^{-3}$&2.06$^{+0.03}_{-0.03}$&259$^{+41}_{-40}$&0.17$^{+0.08}_{-0.06}$&0.74$^{+0.02}_{-0.02}$&0.26$^{+0.05}_{-0.05}$&$>$99.9&320&286\\

1048$+$342&6.1$^{+0.3}_{-0.3}$ $\times$
10$^{-4}$&1.88$^{+0.04}_{-0.04}$&81$^{+18}_{-15}$&0.16$^{+0.08}_{-0.04}$&$-$&$-$&$-$&270&273\\
1100$+$772&1.77$^{+0.06}_{-0.06}$ $\times$ 10$^{-3}$&
1.82$^{+0.03}_{-0.03}$ &135$^{+64}_{-38}$&
0.15$^{+0.01}_{-0.01}$&$-$&$-$&$-$&398 &346 \\

1115$+$407&1.08$^{+0.03}_{-0.05}$ $\times$
10$^{-3}$&2.33$^{+0.04}_{-0.05}$&268$^{+76}_{-52}$&0.153$^{+0.008}_{-0.010}$&0.67$^{+0.02}_{-0.04}$&0.21$^{+0.06}_{-0.06}$&99.99&249&230\\

1116+215 & $ 2.6^{+0.1}_{-0.1}\times10^{-3}$ & $2.24^{+0.05}_{-0.05}$ &
$600^{+300}_{-200}$ & $0.15^{+0.01}_{-0.01}$ & $0.72^{+0.04}_{-0.04}$ &
$0.13^{+0.08}_{-0.09}$ & 96.7 & 208 & 230 \\ 

1202+281 & $
1.28^{+0.06}_{-0.07}\times10^{-3}$ & $1.75^{+0.03}_{-0.04}$ &
$80^{+20}_{-20}$ & $0.194^{+0.013}_{-0.010}$ & $0.73^{+0.02}_{-0.02}$ & $0.27^{+0.07}_{-0.06}$ &
 $>$99.9 & 312 & 319 \\ 

1211+143  & $   9.4^{+0.3}_{-0.3}\times10^{-4}$  & $1.696^{+0.012}_{-0.03}$  &
$900^{+60}_{-100}$       &          $0.152^{+0.004}_{-0.002}$        &
$0.772^{+0.009}_{-0.003}$ & $0.48^{+0.05}_{-0.03}$ & $>$99.9 & 1004 & 400
\\ & & & & &  $0.959^{+0.009}_{-0.007}$ & $0.45^{+0.05}_{-0.04}$ & $>$99.9
& 592 & 398\\ & &
& & & $7.28^{+0.14}_{-0.13}$ & $0.54^{+0.09}_{-0.09}$ & $>$99.9 & 462 & 396 \\

1216$+$069&7.5$^{+0.4}_{-0.4}$          $\times$         10$^{-4}$&
1.89$^{+0.02}_{-0.03}$                &            183$^{+114}_{-64}$&
0.13$^{+0.01}_{-0.01}$&$-$&$-$&$-$&180 &202 \\

1244+026 & $ 2.07^{+0.10}_{-0.07} \times10^{-3}$ & $2.44^{+0.03}_{-0.03}$
&        $234^{+18}_{-13}$   &       $0.19^{+0.07}_{-0.05}$      &
$-$  & $-$  & $-$ &  130  & 110 \\

1307$+$085   & $   4.3^{+0.3}_{-0.3} \times10^{-4}$  &  $1.48^{+0.04}_{-0.04}$ &
$112^{+74}_{-51}$ & $0.15^{+0.02}_{-0.02}$ &  $0.75^{+0.02}_{-0.02}$ & $0.5^{+0.2}_{-0.2}$ & 99.9
&    238   &  210    \\  

1309$+$355&2.7$^{+0.1}_{-0.1}$  $\times$
10$^{-4}$&1.79$^{+0.04}_{-0.04}$
&88$^{+254}_{-66}$&0.13$^{+0.03}_{-0.03}$&0.74$^{+0.03}_{-0.02}$&0.5$^{+0.2}_{-0.2}$&99.9&238
&216       \\           

1322$+$659&1.15$^{+0.06}_{-0.06}$     $\times$
10$^{-3}$&2.32$^{+0.05}_{-0.05}$&661$^{+155}_{-119}$&0.132$^{+0.006}_{-0.006}$&
$-$ & $-$ & $-$ & 250 & 214\\

1352+183   &  $    9.8^{+1.0}_{-1.0} \times10^{-4}$   &  $1.97^{+0.08}_{-0.08}$    &
$94^{+20}_{-14}$    &  $0.203^{+0.010}_{-0.012}$    &   $0.68^{+0.02}_{-0.03}$   &
$0.38^{+0.08}_{-0.10}$   & 100   &    276 &   244   \\   &  &  &   &  &
$0.87^{+0.08}_{-0.06}$      &  $0.17^{+0.09}_{-0.07}$ & 99.5 & 264 &
 242 \\ 

1402+261  &  $
1.46^{+0.08}_{-0.11}\times10^{-3}$      &    $2.26^{+0.05}_{-0.07}$   &
$320^{+80}_{-80}$ &   $0.159^{+0.012}_{-0.009}$  &  $0.72^{+0.04}_{-0.02}$   &
$0.20^{+0.09}_{-0.07}$    &    $>$99.9   &     227      &   215     \\

1407$+$265&1.80$^{+0.09}_{-0.09}$                             $\times$
10$^{-3}$&2.20$^{+0.04}_{-0.04}$
&32$^{+43}_{-17}$&0.14$^{+0.03}_{-0.03}$&$-$&$-$&$-$&224&244 \\


1415+451 & $ 5.9^{+0.4}_{-0.4}\times10^{-4}$ & $2.09^{+0.06}_{-0.06}$
& $90^{+30}_{-20}$ & $0.176^{+0.016}_{-0.014}$ & $0.80^{+0.05}_{-0.05}$ &
$0.15^{+0.09}_{-0.09}$ & 99.9 & 186 & 228\\ & & & & &
$0.64^{+0.03}_{-0.04}$ & $0.21^{+0.06}_{-0.09}$ & 98.8 & 179 & 226 \\ 

1427+480 &  $ 6.5^{+0.3}_{-0.4}\times10^{-4}$ & $2.00^{+0.04}_{-0.05}$
& $60^{+10}_{-10}$ & $0.155^{+0.007}_{-0.006}$ & $0.73^{+0.03}_{-0.03}$ & $0.19^{+0.07}_{-0.06}$ &
$>$99.9 & 261  & 279 \\  


1444+407 & $ 5.5^{+0.4}_{-0.4}\times10^{-4}$ & $2.26^{+0.07}_{-0.07}$ &
$134^{+23}_{-23}$  & $0.16^{+0.01}_{-0.01}$ & $0.72^{+0.03}_{-0.03}$
&  $0.28^{+0.08}_{-0.11}$  &  99.9  &   193  &  164\\

1512$+$370&1.15$^{+0.05}_{-0.06}$         $\times$          10$^{-3}$&
1.89$^{+0.04}_{-0.04}$                &             147$^{+72}_{-43}$&
0.142$^{+0.012}_{-0.006}$&$-$&$-$&$-$&249          &255             \\

1613$+$658&1.5$^{+0.1}_{-0.1}$          $\times$         10$^{-3}$&
1.78$^{+0.07}_{-0.07}$              &            263$^{+305}_{-113}$&
0.15$^{+0.02}_{-0.02}$&0.74$^{+0.02}_{-0.02}$&0.4$^{+0.2}_{-0.1}$&99.9&151      &180 \\

1626$+$554&1.71$^{+0.07}_{-0.07}$                             $\times$
10$^{-3}$&2.06$^{+0.06}_{-0.06}$                                     &
635$^{+691}_{-300}$&0.12$^{+0.02}_{-0.02}$&$-$&$-$&$-$&268 &268  \\

1630$+$377&3.8$^{+0.7}_{-0.8}$       $\times$
10$^{-4}$&2.0$^{+0.1}_{-0.1}$&70$^{+750}_{-0.50}$ &0.10$^{+0.04}_{-0.02}$& $-$&$-$&$-$&55&59\\

1634+706  & $ 2.5^{+0.3}_{-0.4}\times10^{-3}$ & $2.03^{+0.06}_{-0.07}$
& $21^{+13}_{-8}$ & $0.19^{+0.02}_{-0.02}$ & $-$  & $-$ & $-$  & 206& 208 \\

2302+029 &  $    4.9^{+0.6}_{-0.7}\times10^{-4}$ &  $2.05^{+0.09}_{-0.11}$   &  $
170^{+500}_{-130} $ & $0.11^{+0.03}_{-0.02}$ & $-$ & $-$ & $-$ & 59 & 66 \\
\hline
\end{tabular}
\end{center}
($^\dag$)~Flux at 1 keV in units of photons/keV/cm$^{2}$/s.
\end{table*}



\begin{table*}
\caption{Spectral Fitting results: V.  Power law plus bremsstrahlung model for the
  soft excess (Model C).}
\label{tab:C}
\begin{center}
\begin{tabular}{c c c c c c c c c c}
\hline PG Name &  A$^{\dag}_{PL}$ & $\Gamma$ &  A$^{\dag}_{Brem}$ & kT & Edge
&$\tau$& $P_{F-test}$ & $\chi^2_{\nu}$ & dof \\ & 
& & &(keV) & (keV)&&(\%)&&\\ \hline \hline
& & & & &&&&\\
0007$+$106&1.6$^{+0.2}_{-0.2}$ $\times$
10$^{-3}$&1.56$^{+0.04}_{-0.08}$ &2.1$^{+0.3}_{-0.3}$ $\times$
10$^{-3}$&0.69$^{+0.09}_{-0.09}$&$-$&$-$&$-$&321 &305 \\ 

0157+001     &     $5.7^{+1.1}_{-1.1}\times10^{-4}$        &  $2.08^{+0.16}_{-0.14}$  &
$1.4^{+0.2}_{-0.3}\times10^{-3}$    &       $0.44^{+0.08}_{-0.09}$      &
$-$ & $-$ & $-$ & 150 & 166\\

0804$+$761&6.3$^{+0.5}_{-0.5}$                             $\times$
10$^{-3}$&2.13$^{+0.04}_{-0.04}$&1.2
$^{+0.6}_{-0.3}$$\times$10$^{-2}$&0.31$^{+0.08}_{-0.08}$&$-$&$-$&$-$&305&264\\

0844$+$349  &        $3.2^{+0.1}_{-0.1}\times10^{-3}$   &     $2.22^{+0.03}_{-0.03}$    &
$3.0^{+2}_{-1}\times10^{-2}$ & $0.20^{+0.01}_{-0.01}$ & $-$  & $-$ & $-$ &  189
&164\\

0947$+$396  &  $7.9^{+0.6}_{-1.3}\times10^{-4}$ &  $1.84^{+0.05}_{-0.11}$ &
$2.9^{+0.2}_{-0.2}\times10^{-3}$ &  $0.49^{+0.07}_{-0.03}$  & $0.79^{+0.04}_{-0.04}$  &
$0.17^{+0.08}_{-0.05}$ & $>$99.9 & 229 & 261 \\

0953$+$414&8.5$^{+0.5}_{-0.5}$
$\times$10$^{-3}$&1.99$^{+0.04}_{-0.06}$&1.8$^{+0.1}_{-0.1}$
$\times$
10$^{-3}$&0.42$^{+0.03}_{-0.03}$&0.76$^{+0.03}_{-0.03}$&0.18$^{+0.05}_{-0.05}$&$>$99.9&312&286\\

1048$+$342 &5.4$^{+0.4}_{-0.4}$ $\times$
10$^{-4}$&1.79$^{+0.05}_{-0.05}$&1.9$^{+0.1}_{-0.1}$ $\times$
10$^{-3}$&0.41$^{+0.03}_{-0.03}$& $-$ & $-$ &$-$&259&273\\

1100$+$772&1.62$^{+0.09}_{-0.10}$ $\times$ 10$^{-3}$&
1.76$^{+0.04}_{-0.05}$ & 3.0$^{+0.4}_{-0.3}$ $\times$ 10$^{-3}$&
0.45$^{+0.07}_{-0.06}$&$-$&$-$&$-$&390 &346 \\

1115$+$407&9.7$^{+0.7}_{-0.9}$ $\times$
10$^{-4}$&2.25$^{+0.04}_{-0.05}$ &5.70$^{+0.04}_{-0.04}$ $\times$
10$^{-3}$&0.34$^{+0.03}_{-0.03}$&0.69$^{+0.04}_{-0.04}$&0.15$^{+0.07}_{-0.06}$&99.9&239
&230 \\ 

1116$+$215    &    $2.4^{+0.2}_{-0.2}\times10^{-3}$       &   $2.18^{+0.06}_{-0.06}$    &
$1.2^{+0.2}_{-0.1}\times10^{-2}$            &       0.33$^{+0.04}_{-0.04}$     &
$0.75^{+0.05}_{-0.06}$ & $0.11^{+0.07}_{-0.06}$ & 95.6 & 203 & 230\\ 

1202$+$281 & $1.13^{+0.09}_{-0.09}\times10^{-3}$ & $1.68^{+0.05}_{-0.05}$ &
$3.5^{+0.2}_{-0.2}\times10^{-3}$ & $0.49^{+0.05}_{-0.05}$ & $0.74^{+0.03}_{-0.03}$ &
$0.20^{+0.06}_{-0.06}$ & $>$99.9 & 306 & 319 \\ 

1216$+$069&7.0$^{+0.6}_{-0.6}$
$\times$ 10$^{-4}$& 1.84$^{+0.06}_{-0.06}$ & 2.7$^{+0.6}_{-0.4}$
$\times$ 10$^{-3}$& 0.34$^{+0.06}_{-0.05}$&$-$&$-$&$-$&178 &202 \\

1307$+$085 & $4.2^{+0.2}_{-0.3}\times10^{-4}$ & $1.46^{+0.04}_{-0.07}$ &
$2.0^{+0.4}_{-0.4}\times10^{-3}$ & $0.32^{+0.08}_{-0.04}$ &
$0.77^{+0.02}_{-0.03}$ & $0.5^{+0.1}_{-0.1}$ & 99.9 & 244 & 210 \\

1309$+$355&2.7$^{+0.2}_{-0.2}$ $\times$
10$^{-4}$&1.78$^{+0.04}_{-0.06}$ &9.8$^{+0.7}_{-3.2}$ $\times$
10$^{-4}$&0.28$^{+0.10}_{-0.07}$&0.75$^{+0.02}_{-0.02}$&0.5$^{+0.2}_{-0.2}$
&$>$99.99&238 &216 \\ 

1322$+$659&1.02$^{+0.07}_{-0.07}$ $\times$
10$^{-3}$&2.24$^{+0.06}_{-0.06}$&9.0$^{+0.8}_{-0.7}$ $\times$
10$^{-3}$&0.29$^{+0.02}_{-0.02}$&$-$&$-$&$-$&242&214\\ 

1352+183   &     $8.5^{+1.0}_{-1.2}\times10^{-4}$    &  $1.89^{+0.08}_{-0.11}$    &
$4.7^{+0.3}_{-0.3}\times10^{-3}$       &        $0.45^{+0.04}_{-0.03}$       &
$0.71^{+0.02}_{-0.03}$ & $0.26^{+0.07}_{-0.06}$ & $>$99.9  & 263 &  244 \\

1402+261  &   $1.22^{+0.12}_{-0.15}\times10^{-3}$  &   $2.14^{+0.07}_{-0.09}$   &
$7.8^{+0.5}_{-0.6}\times10^{-3}$     &    $0.37^{+0.04}_{-0.03}$     &
$0.75^{+0.04}_{-0.04}$ & $0.17^{+0.07}_{-0.06}$ & $>$99.9 & 211 & 215 \\

1407$+$265&1.78$^{+0.09}_{-0.09}$ $\times$
10$^{-3}$&2.20$^{+0.03}_{-0.03}$
&1.1$^{+0.5}_{-0.4}$$\times$10$^{-3}$&0.5$^{+0.2}_{-0.1}$&$-$&$-$&$-$&224&244\\


1415$+$451 & $5.5^{+0.5}_{-0.5}\times10^{-4}$ & $2.03^{+0.07}_{-0.07}$ &
$2.8^{+0.3}_{-0.2}\times10^{-3}$ & $0.36^{+0.04}_{-0.04}$ &
$0.73^{+0.05}_{-0.05}$& $0.15^{+0.07}_{-0.07}$ & 99.9 & 179 & 228 \\

1427$+$480 &  $5.9^{+0.3}_{-0.7}\times10^{-4}$ & $1.94^{+0.04}_{-0.09}$
& $2.06^{+0.16}_{-0.14}\times10^{-3}$ & $0.43^{+0.05}_{-0.03}$ &
$0.75^{+0.04}_{-0.04}$ & $0.12^{+0.07}_{-0.04}$ & $>$99.9 & 257 & 279 \\


1444$+$407 &  $5.2^{+0.4}_{-0.8}\times10^{-4}$ &
$2.23^{+0.06}_{-0.14}$ & $3.4^{+0.4}_{-0.4}\times10^{-3}$ &
$0.34^{+0.06}_{-0.06}$ & $0.72^{+0.04}_{-0.03}$&
$0.21^{+0.11}_{-0.07}$ & 99.9 & 191 & 164 \\

1501+106 &  $5.8^{+0.2}_{-0.2}\times10^{-3}$ & $1.91^{+0.03}_{-0.03}$
& $3.4^{+0.1}_{-0.1}\times10^{-2}$ & $0.34^{+0.02}_{-0.02}$ &
$0.69^{+0.02}_{-0.02}$ & $0.17^{+0.04}_{-0.03}$ &  $>$99.9 & 259 & 221 \\

1512$+$370&1.05$^{+0.08}_{-0.07}$ $\times$ 10$^{-3}$&
1.84$^{+0.05}_{-0.05}$ & 2.9$^{+0.5}_{-0.3}$ $\times$ 10$^{-3}$&
0.41$^{+0.04}_{-0.06}$&$-$&$-$&$-$&239 &255 \\

1613$+$658&1.38$^{+0.08}_{-0.09}$ $\times$ 10$^{-3}$&
1.8$^{+0.1}_{-0.1}$ & 5$^{+1}_{-1}$ $\times$ 10$^{-3}$&
0.33$^{+0.09}_{-0.07}$&0.75$^{+0.02}_{-0.03}$&0.4$^{+0.1}_{-0.1}$&99.9&154 &180 \\

1626$+$554&1.67$^{+0.08}_{-0.10}$ $\times$
10$^{-3}$&2.04$^{+0.06}_{-0.05}$ &5.2$^{+2.2}_{-1.1}$ $\times$
10$^{-3}$&0.24$^{+0.05}_{-0.05}$&$-$&$-$&$-$&267 &268 \\

1630$+$377&3.9$^{+0.7}_{-0.7}$ $\times$
10$^{-4}$&2.0$^{+0.1}_{-0.2}$&2$^{+6}_{-1}$ $\times$
10$^{-3}$&0.40$^{+0.4}_{-0.2}$&$-$&$-$&$-$&55&59\\ 

1634+706 &$2.1^{+0.4}_{-0.4}\times10^{-3}$ & $1.95^{+0.08}_{-0.11}$ &$2.9^{+0.6}_{-0.6}\times10^{-3}$ & $1.0^{+0.2}_{-0.2}$ & $-$ & $-$ & $-$ &203 & 208 \\ 

2302+029 & $4.7^{+0.7}_{-0.8}\times10^{-4}$ &
$2.03^{+0.06}_{-0.11}$ & $2.8^{+4}_{-1.3}\times10^{-3}$ & $0.37^{+0.13}_{-0.11}$
& $-$ & $-$ & $-$ & 58 & 66 \\ \hline
\end{tabular}
\end{center}
($^{\dag}$)~Flux at 1 keV in units of photons/keV/cm$^{2}$/s.
\end{table*}

\begin{figure*}
\begin{center}
\epsfig{figure=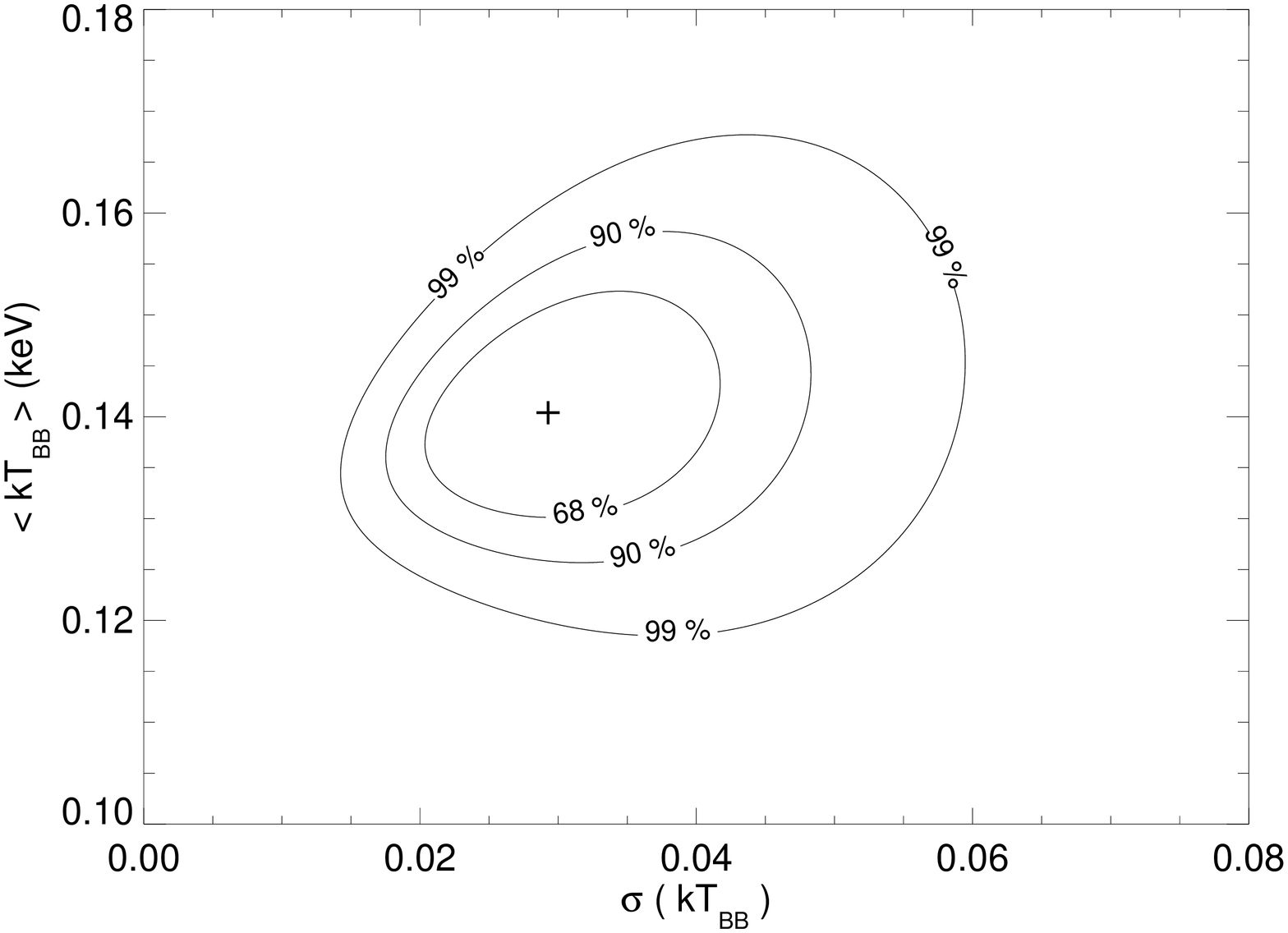,width=4.3cm,angle=90}\hspace{0.6cm}\epsfig{figure=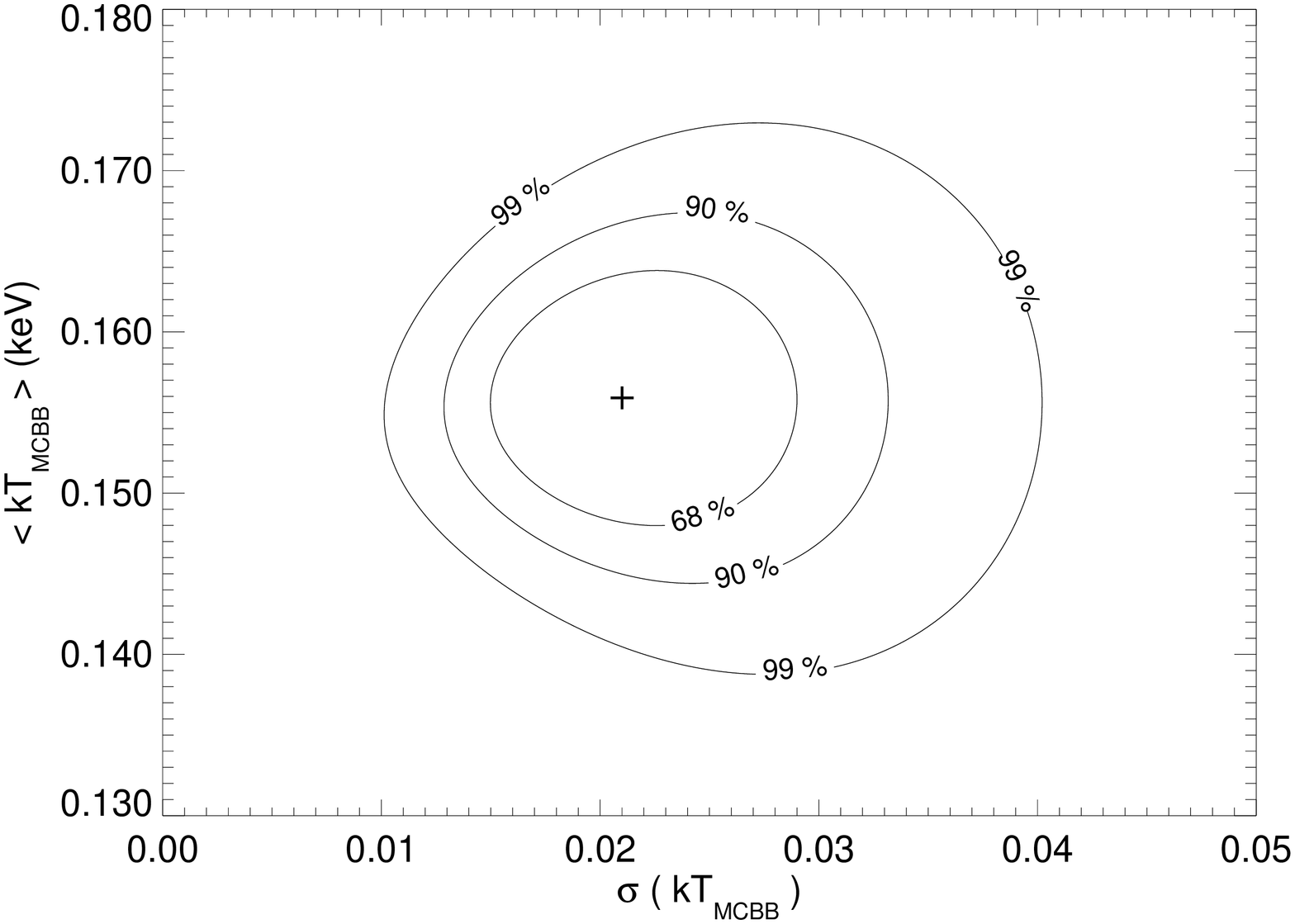,width=4.3cm,angle=90}
\epsfig{figure=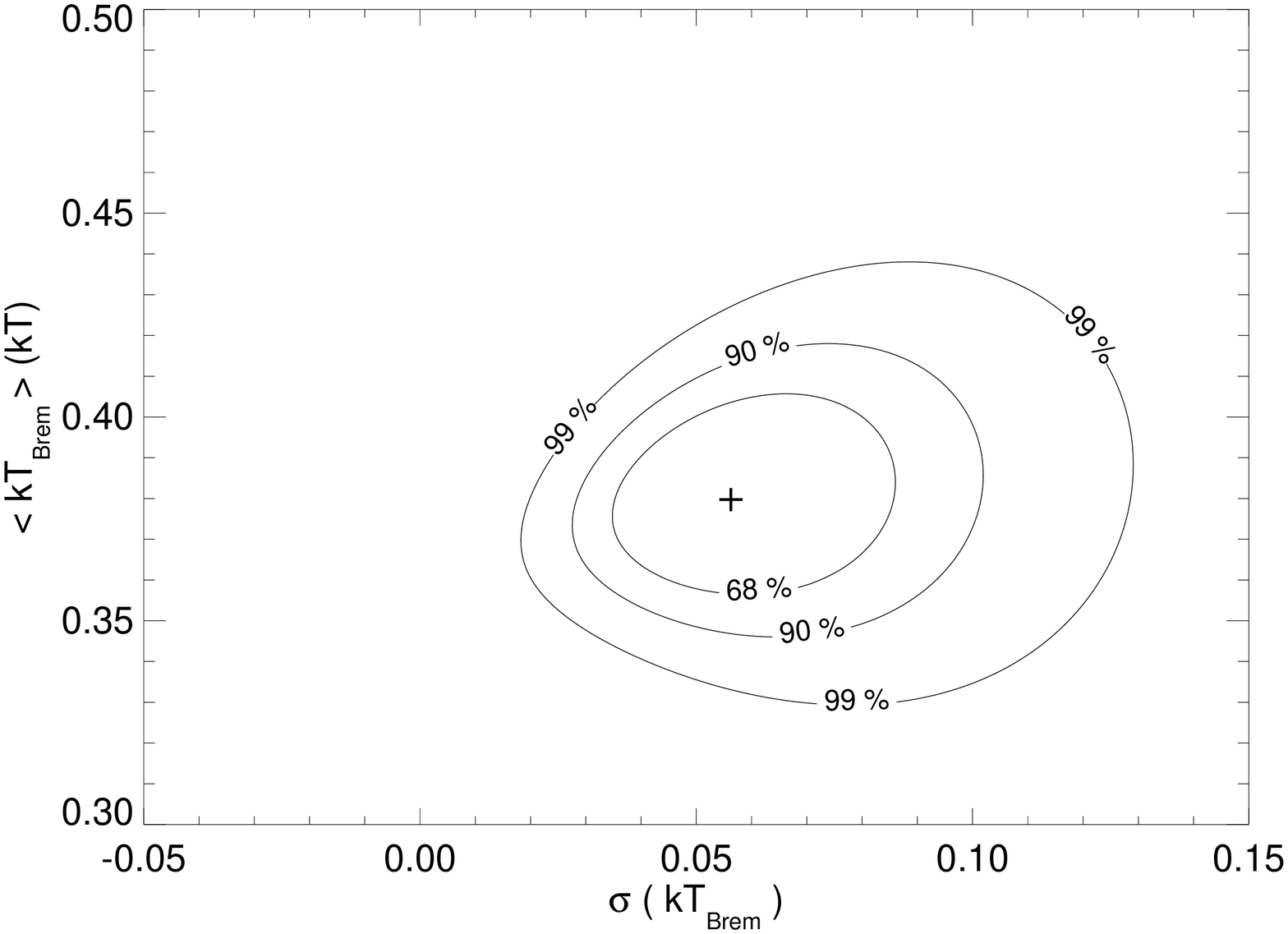,width=4.3cm,angle=90}\hspace{0.6cm}\epsfig{figure=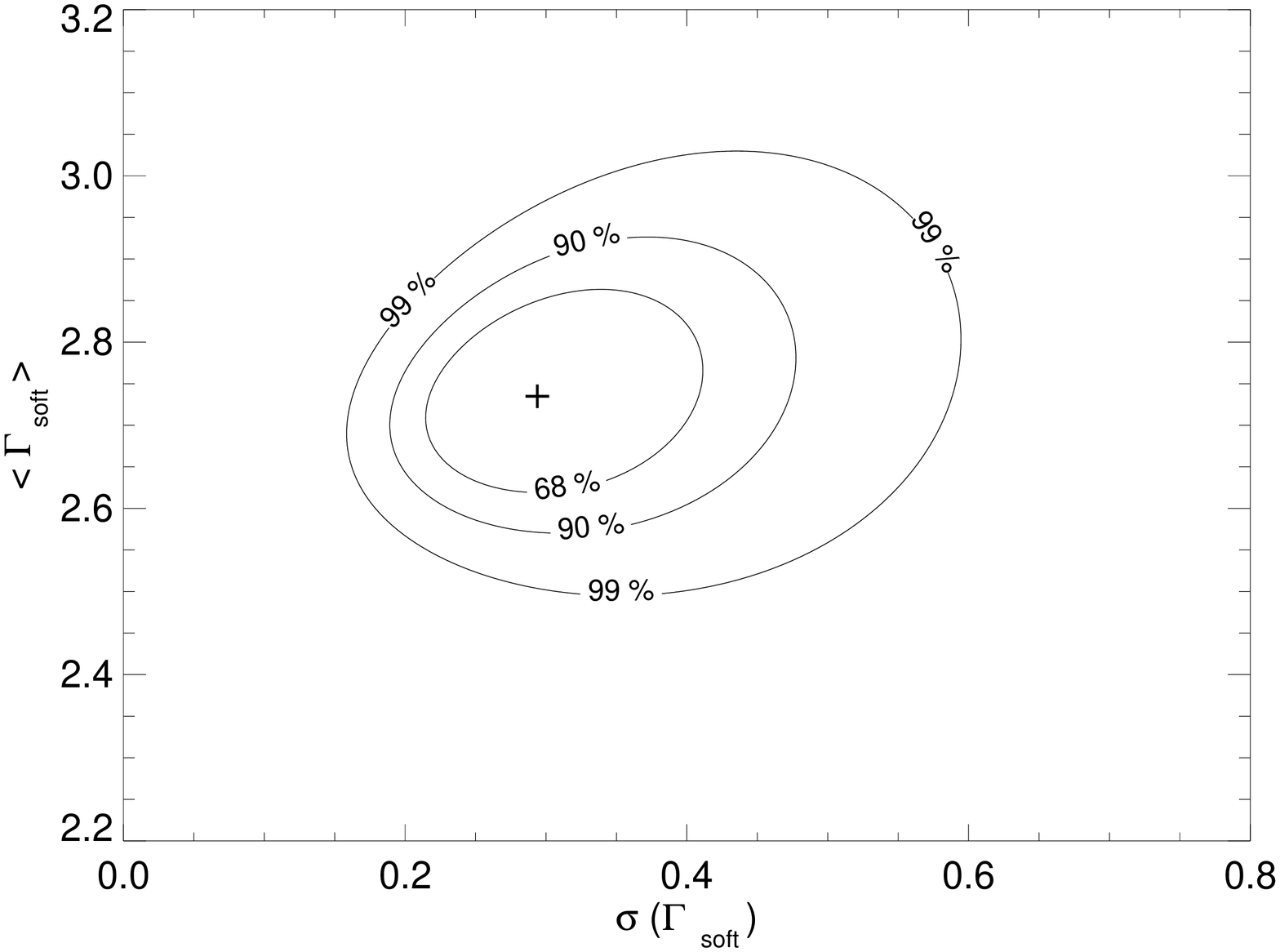,width=4.3cm,angle=90}
\caption{Mean value and intrinsic dispersion of the relevant parameters of soft excess in models A ($\langle$kT$_{BB}\rangle$), B ($\langle$kT$_{MB}\rangle$), C ( $\langle$kT$_{BS}\rangle$)and D ($\langle\Gamma_{\rm soft}\rangle$) together with 68\%, 90\% and 99\% confidence contours derived with the maximum likelihood method.}
\label{fig:means}
\end{center}
\end{figure*}

All  objects for  which  we  tested  these models  showed  a
statistically significant improvement in  the goodness of fit upon the
addition of a spectral component accounting for the soft excess. 
Using the  maximum likelihood  technique, 
we derived the best  simultaneous   estimate  of   the  mean value  and   the  intrinsic
dispersion of the relevant parameter (i.e. $kT_{BB}$, $kT_{MB}$, $kT_{BS}$ and $\Gamma_{\rm soft}$) of each tested model.
Figure~\ref{fig:means} shows such best fit  values as  well as the corresponding 68\%, 90\%  and 99\%  
confidence contour levels. Finally, in Table~\ref{tab:means_laor} are reported the same values with the relative
uncertainties given at the 68\% confidence level for two interesting parameters.

\begin{table*}
\caption{Spectral Fitting results. VI: Double Power law (Model D).}
\label{tab:D}
\begin{center}
\begin{tabular}{c c c c c c c c c c}
 \hline \\ {\bf PG Name} &  A$^{\dag}_{hard}$ & $\Gamma_{\rm{hard}}$ &
A$^{\dag}_{soft}$ &$\Gamma_{\rm{soft}}$  & Edge &$\tau$&$P_{F-test}$ &
$\chi^2_{\nu}$ & d.o.f. \\ 
&     & & & &(keV)& &(\%) & &\\ \hline\hline\\

0007$+$106&9$^{+3}_{-6}$ $\times$
10$^{-5}$&0.6$^{+0.7}_{-0.4}$ & 2.2$^{+0.1}_{-0.3}$ $\times$
10$^{-3}$&1.95$^{+0.10}_{-0.05}$&$-$&$-$&$-$&329 &305 \\ 

0157$+$001& $5^{+2}_{-4}\times10^{-5}$ & $1.1^{+0.7}_{-0.8}$ &
$7.9^{+0.5}_{-1.0}\times10^{-4}$ & $2.66^{+0.20}_{-0.09}$& $-$ & $-$ & $-$ &
151 & 166 \\ 

0804$+$761&7$^{+25}_{-6}$ $\times$
10$^{-4}$&1.3$^{+0.6}_{-0.8}$&6.5$^{+0.9}_{-2.6}$ $\times$
10$^{-3}$&2.5$^{+0.3}_{-0.1}$&$-$&$-$&$-$&294 &264\\ 

0844$+$349 &
2.9$^{+0.3}_{-0.4}$ $\times$ 10$^{-3}$ & 2.13$^{+0.06}_{-0.09}$ &
$0.9^{+0.5}_{-0.3}\times10^{-4}$ &4.4$^{+0.7}_{-0.6}$ & $-$
&$-$ & $-$ & 184 & 164\\

0947$+$396 &  $1.19^{+0.10}_{-0.06}\times10^{-3}$ &
$1.1^{+0.2}_{-0.3}$ & $1.19^{+0.07}_{-0.10}\times10^{-3}$ &
$2.58^{+0.09}_{-0.07}$ & $-$ & $-$ & $-$ & 239 & 263 \\

1048$+$342              &1.1$^{+0.7}_{-0.5}$               $\times$
10$^{-4}$&1.1$^{+0.2}_{-0.2}$&7.1$^{+0.4}_{-0.7}$       $\times$
10$^{-4}$ &2.56$^{+0.10}_{-0.08}$&$-$&$-$&$-$&273&273\\

1100$+$772&3$^{+3}_{-2}$ $\times$ 10$^{-4}$&
1.2$^{+0.2}_{-0.3}$ & 1.8$^{+0.2}_{-0.3}$ $\times$ 10$^{-3}$&
2.3$^{+0.1}_{-0.1}$&$-$&$-$&$-$&374 &346 \\

1115$+$407&1.30$^{+0.08}_{-0.08}$ $\times$
10$^{-3}$&1.3$^{+0.1}_{-0.4}$&1.2$^{+0.9}_{-0.6}$ $\times$
10$^{-4}$&2.98$^{+0.07}_{-0.08}$&8.1$^{+0.3}_{-0.3}$&0.5$^{+0.4}_{-0.2}$&99&239
&230 \\ 

1116+215  &  $6^{+5}_{-3}\times10^{-4}$   & $1.6^{+0.2}_{-0.3}$      &
$2.7^{+0.4}_{-0.5}\times10^{-3}$         &        $2.9^{+0.2}_{-0.1}$&
0.94$^{+0.08}_{-0.08}$ & 0.07$^{+0.05}_{-0.04}$ & 95 & 209 & 230 \\

1202+281  &     
$6^{+2}_{-2}\times10^{-4}$           &           1.42$^{+0.1}_{-0.2}$&
$1.15^{+0.3}_{-0.17}\times10^{-3}$   & $2.68^{+0.12}_{-0.16}$   & $-$ & $-$ & $-$ & 331 & 321 \\

1216$+$069&4$^{+1}_{-2}$ $\times$ 10$^{-4}$&
1.6$^{+0.1}_{-0.3}$ & 5.9$^{+2.1}_{-1.5}$ $\times$ 10$^{-4}$&
2.9$^{+0.3}_{-0.3}$&$-$&$-$&$-$&175 &202 \\

1309$+$355&2.0$^{+0.5}_{-0.9}$ $\times$
10$^{-4}$&1.7$^{+0.1}_{-0.2}$ &1.3$^{+1.1}_{-0.6}$ $\times$
10$^{-4}$&3.1$^{+0.6}_{-0.5}$&0.76$^{+0.02}_{-0.02}$
&0.6$^{+0.1}_{-0.1}$ &$>$99.9 &238 &216 \\ 

1352$+$183       &        $7^{+8}_{-4}\times10^{-5}$       &
0.8$^{+0.4}_{-0.4}$  &   1.52$^{+0.10}_{+0.10}$  $\times$  10$^{-3}$   &
$2.68^{+0.07}_{-0.05}$         &  0.73$^{+0.05}_{-0.05}$ & 0.09$^{+0.05}_{-0.05}$&
96 & 278 & 244 \\

1402$+$261 & $1.7^{+1.2}_{-0.7}\times10^{-4}$  &$1.3^{+0.3}_{-0.2}$
&$1.85^{+0.09}_{-0.10}\times10^{-3}$  &$2.95^{+0.08}_{-0.06}$  & $-$ & $-$ & $-$ & 216 & 217 \\

1407$+$265&1.1$^{+0.5}_{-0.8}$ $\times$
10$^{-3}$&2.1$^{+1.2}_{-0.4}$&9$^{+7}_{-7}$ $\times$
10$^{-4}$&2.7$^{+0.5}_{-1.0}$&$-$&$-$&$-$&227 &244 \\

1415+451 & 
$1.6^{+0.9}_{-0.6}\times10^{-4}$ & 1.5$^{+0.2}_{-0.2}$ & $6.5^{+0.6}_{-0.1}\times10^{-4}$ & $2.92^{+0.1}_{-0.09}$ & $-$ & $-$ & $-$ &
199 & 230 \\ 

1427+480 &   $1.2^{+0.8}_{-0.5}\times10^{-4}$ & 1.3$^{+0.2}_{-0.2}$
& $17.6^{+0.6}_{-0.8}\times10^{-4}$ &
$2.67^{+0.10}_{-0.08}$ & $-$ & $-$ & $-$ & 277 & 281 \\ 

1440+356 &   
$3.0^{+3}_{-1.0}\times10^{-4}$ & $1.2^{+0.4}_{-0.2}$ &
$2.81^{+0.11}_{-0.3}\times10^{-3}$ &  $3.14^{+0.2}_{-0.08}$ & $-$ & $-$ &  $-$ &  191 &
158 \\

1444+407 &  $2^{+1}_{-1}\times10^{-4}$ &
$1.8^{+0.3}_{-0.3}$ & $5.7^{+0.1}_{-0.1}\times10^{-4}$ &
$3.3^{+0.2}_{-0.3}$ & $-$ & $-$ & $-$ & 196 & 166 \\

1512$+$370&5$^{+4}_{-3}$ $\times$ 10$^{-4}$&
1.6$^{+0.1}_{-0.3}$ & 9$^{+3}_{-2}$ $\times$ 10$^{-4}$&
2.8$^{+0.2}_{-0.2}$&$-$&$-$&$-$&236 &255 \\

1613$+$658&1.1$^{+0.2}_{-0.2}$ $\times$ 10$^{-3}$&
1.6$^{+0.1}_{-0.1}$ & 5$^{+2}_{-2}$ $\times$ 10$^{-4}$&
3.7$^{+0.5}_{-0.4}$&0.76$^{+0.03}_{-0.03}$&0.4$^{+0.1}_{-0.1}$&$>$99.9&159 &180 \\

1626$+$554&1.1$^{+0.4}_{-0.7}$ $\times$
10$^{-3}$&1.9$^{+0.1}_{-0.4}$ &7$^{+7}_{-4}$ $\times$
10$^{-4}$&3.1$^{+0.70}_{-0.4}$&$-$&$-$&$-$&266 &268 \\

1630$+$377&3$^{+1}_{-2}$ $\times$
10$^{-4}$&1.94$^{+0.2}_{-0.5}$&3$^{+2}_{-2}$ $\times$
10$^{-4}$&3.7$^{+1.2}_{-2.3}$&$-$&$-$&$-$&54&59\\ 

1634+706 & 
$8^{+7}_{-7}\times10^{-6}$ & $0.3^{+0.1}_{-0.1}$ &
3.7$^{+0.1}_{-0.1}$$\times$10$^{-3}$ & $2.30^{+0.09}_{-0.11}$ & $-$ & $-$ & $-$ & 203 &
208 \\ 

2302+029 & $4.0^{+1.3}_{-1.6}\times10^{-4}$ & $1.96^{+0.16}_{-0.2}$ &
3.9$^{+1.6}_{-1.6}$$\times$ 10$^{-4}$ & $4.1^{+1.4}_{-0.8}$ & $-$ & $-$ & $-$ & 60 &
66\\ \hline
\end{tabular}
\end{center}
($^{\dag}$)Flux at 1 keV in units of photons/keV/cm$^{2}$/s at 1 keV.

\end{table*}

\subsection{Spectral results for the complete subsample}\label{sec:laor}

\begin{figure*}
\begin{center}
\epsfig{figure=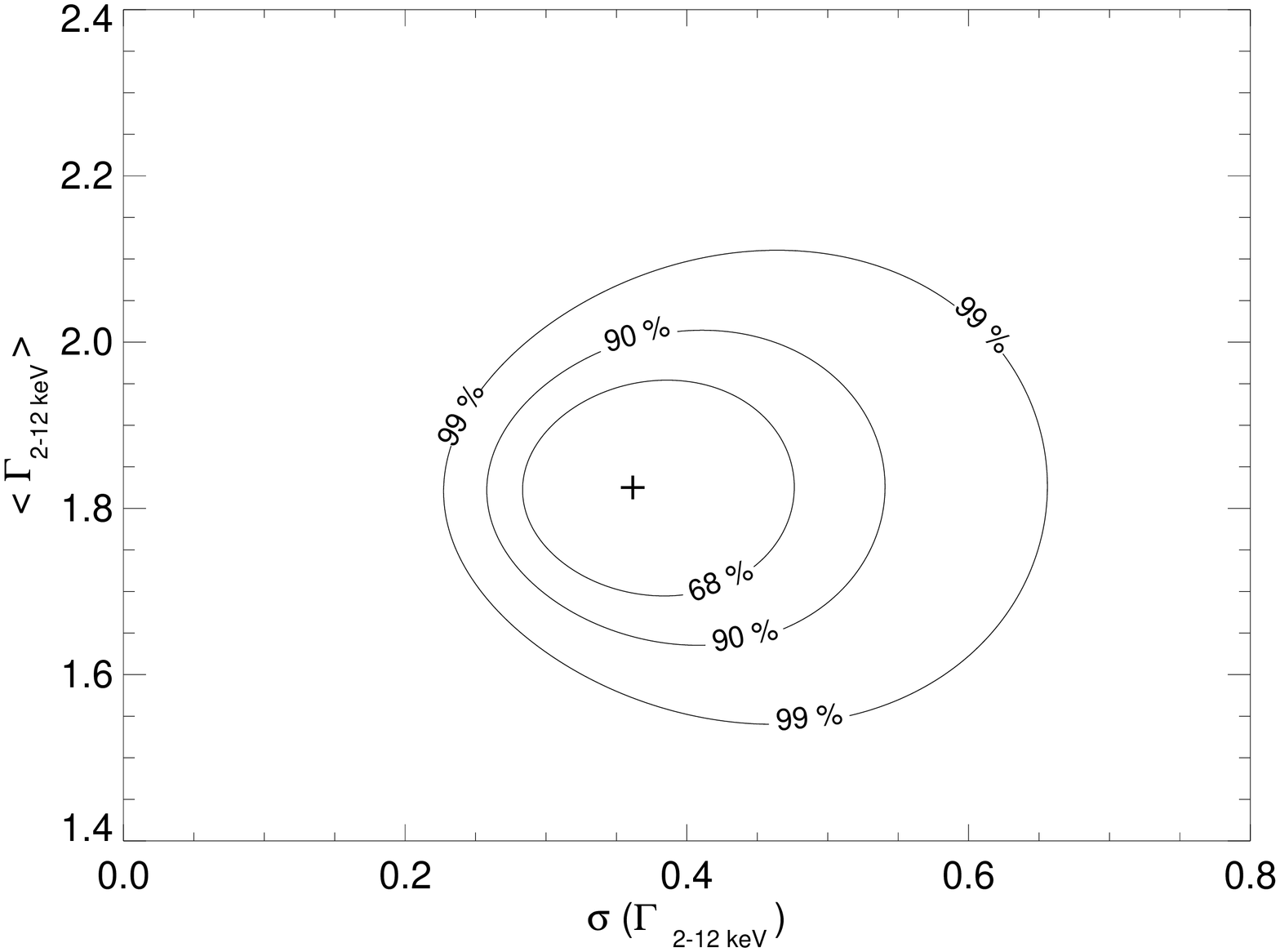,width=4.3cm,angle=90}
\epsfig{figure=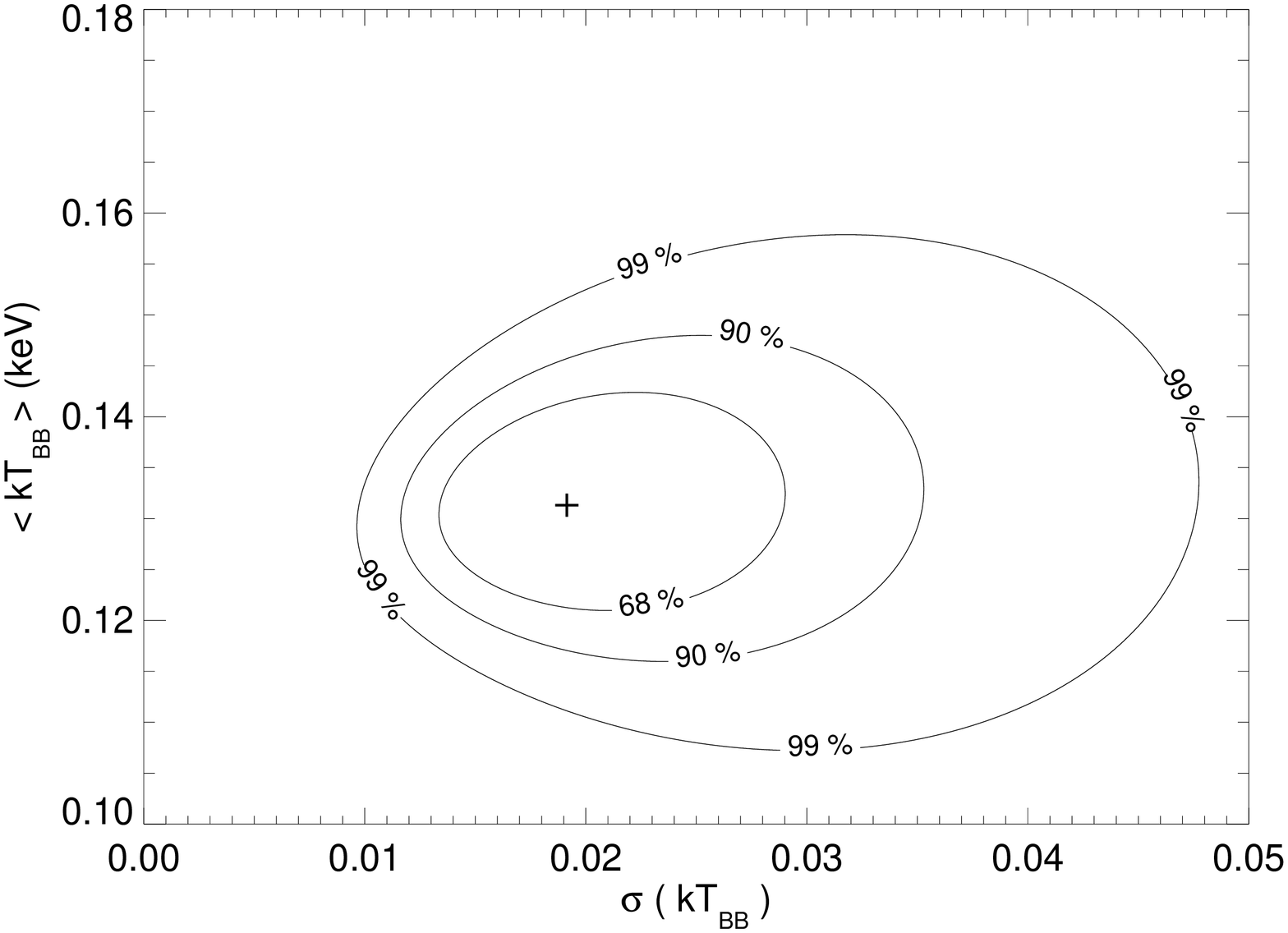,width=4.3cm,angle=90}\epsfig{figure=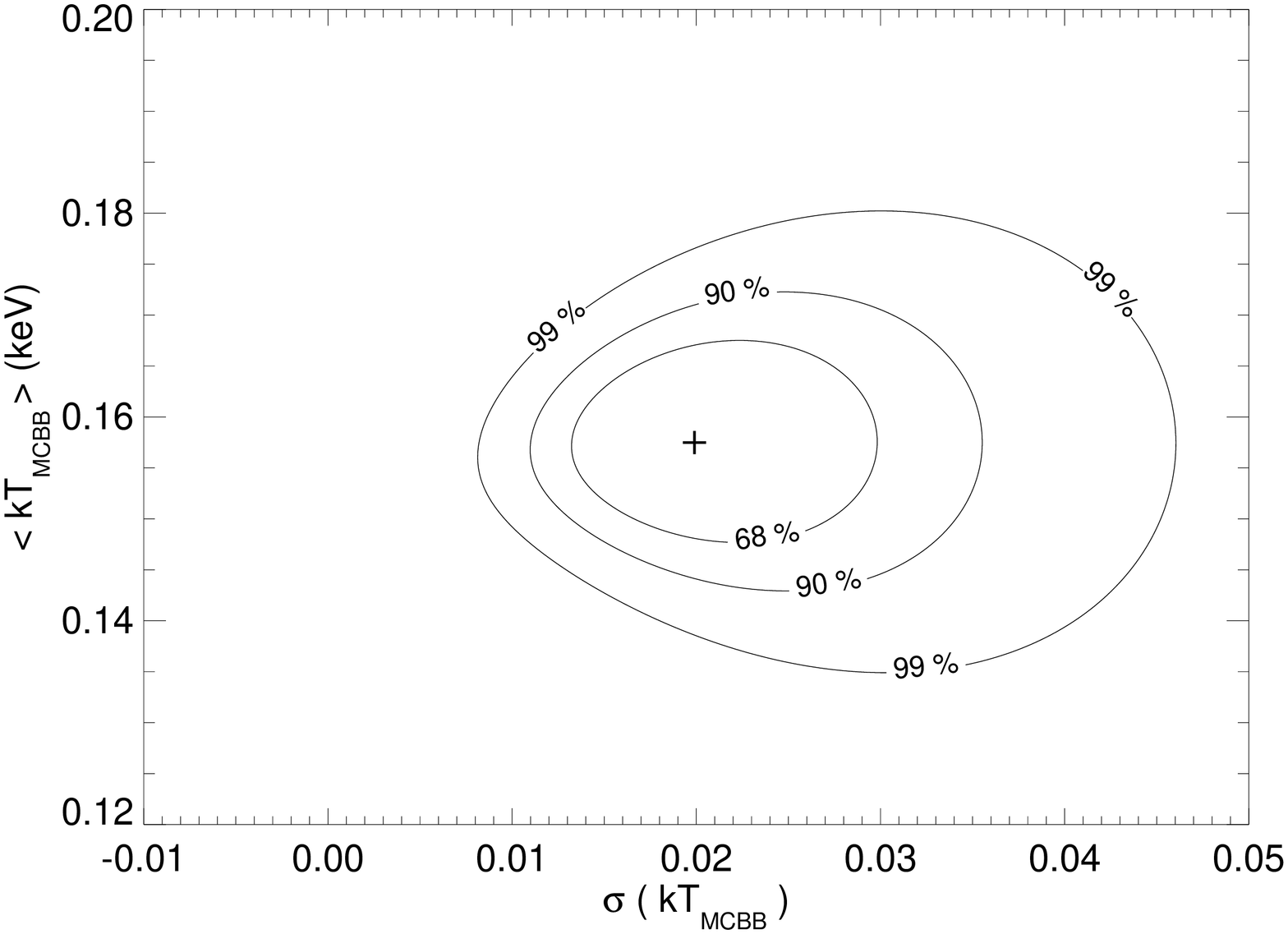,width=4.3cm,angle=90}
\epsfig{figure=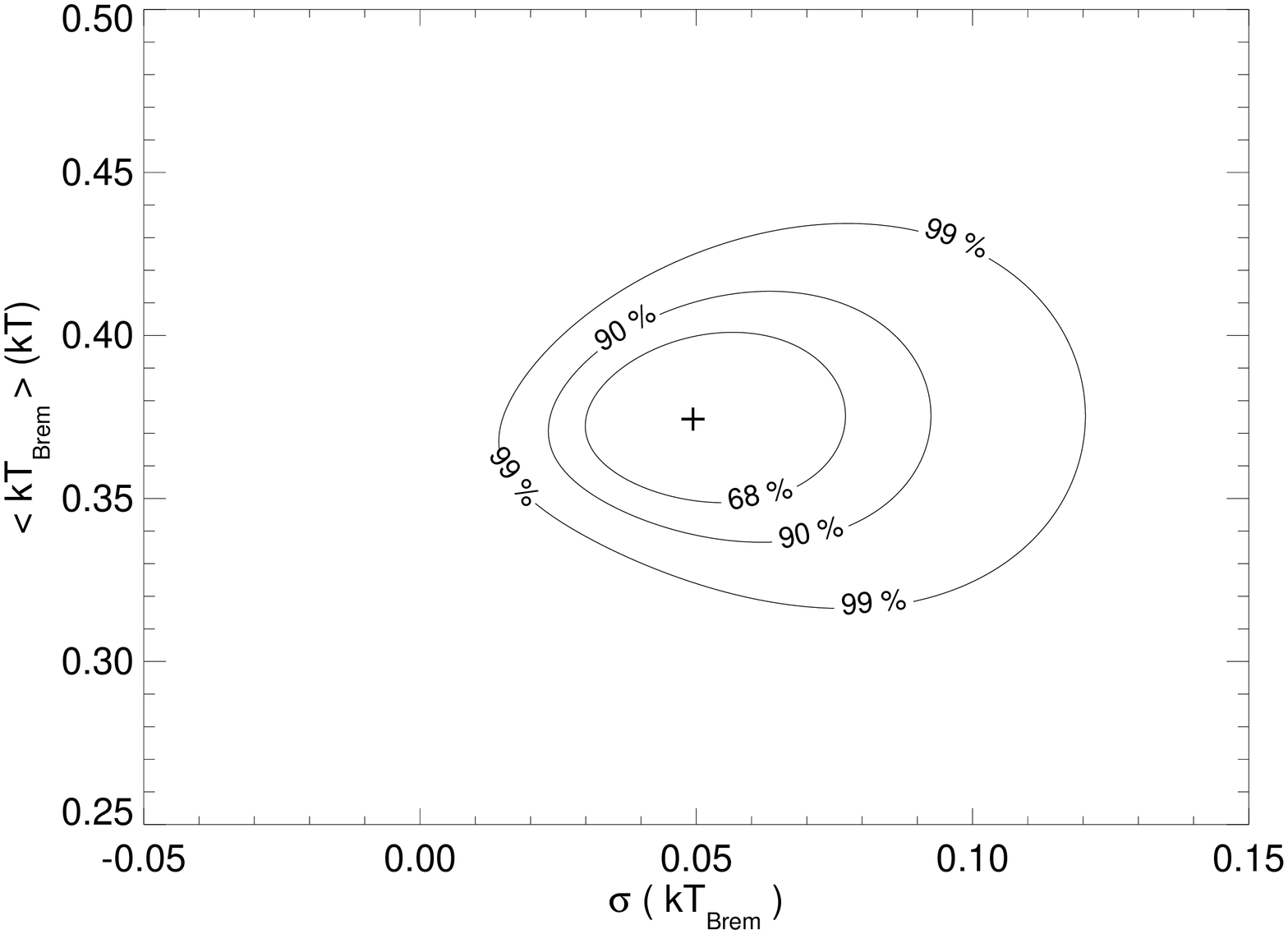,width=4.3cm,angle=90}\epsfig{figure=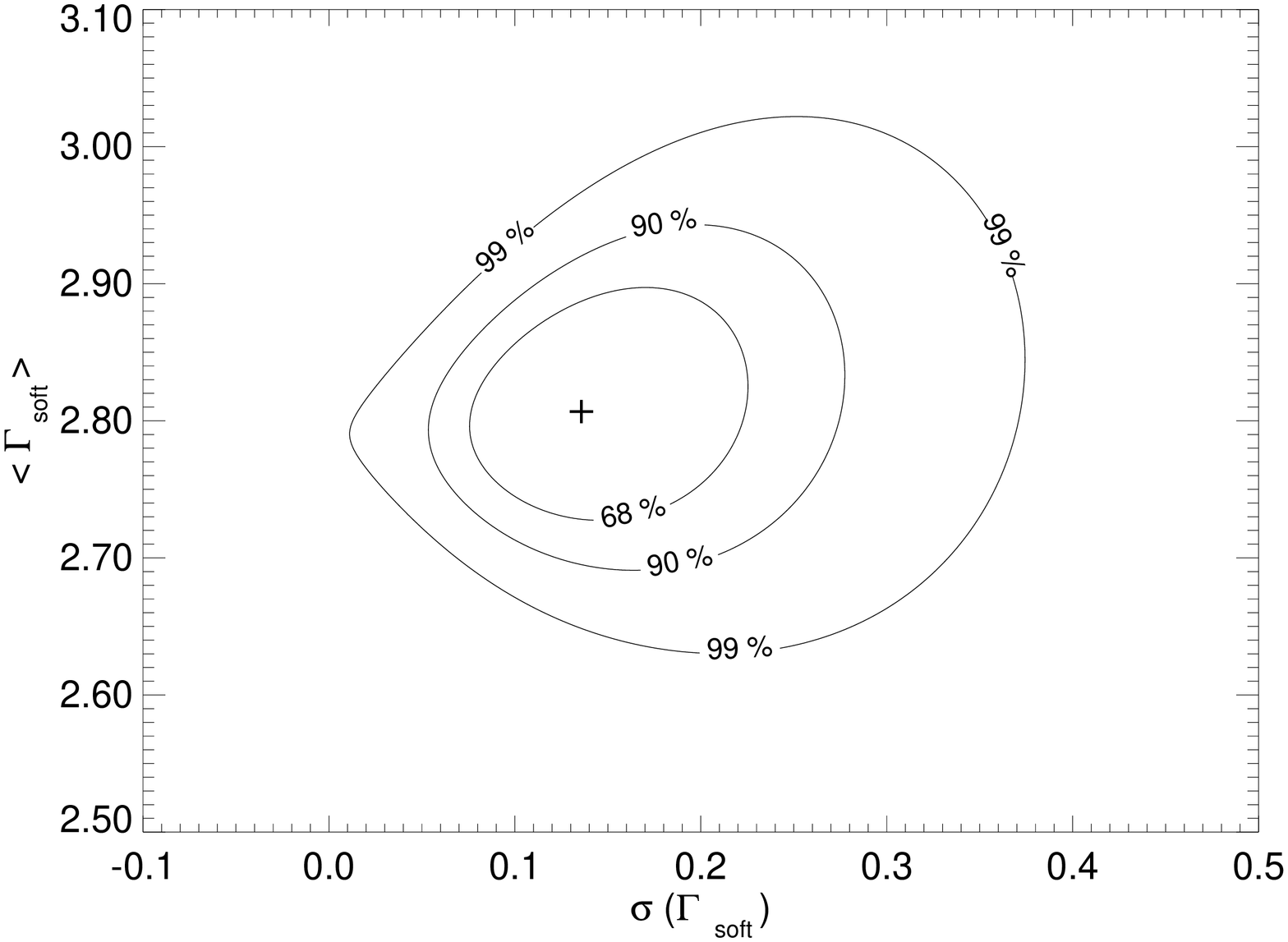,width=4.3cm,angle=90}
\caption{Mean value and intrinsic dispersion of the relevant
parameters for tested models applied to the Laor et al. (1997) quasars together with 68\%, 90\% and 99\% confidence contours derived with the maximum likelihood method.}
\label{fig:means_laor}
\end{center}
\end{figure*}

In this Section we present the results of analysis of the subsample of
objects included    in the Laor et al. (1997)   study (marked with  $^L$ in
Table~\ref{tab:sample}). Laor et al. (1997) analyzed the
\rosat~observations of 23  quasars
from  the   Bright   Quasar     Survey      with z $< 0.4$      and
$N_{H}^{Gal} < 1.9\times10^{20}$ \cm2.    We have  performed   the
analysis of 21 out of  the 23 objects  observed by Laor et al. (1997):
there are not yet public available observations in  the \xmm~archive for 1425$+$267
and 1543$+$489.  The selection criteria for the Laor et al. (1997)
sample only consider  optical properties of  the QSOs. Therefore the
complete subsample   is unbiased in terms   of X--ray properties  and it is
representative of the low--redshift, optically--selected QSO population.
\begin{table*}
\caption{ Dispersion and mean value of the relevant parameters of the
models tested for the sample and the Laor et al. (1997) subsample. See
  Sect.~\ref{sec:soft} and \ref{sec:laor} for details. The mean values of the kT in models A, B and C are given in units of keV.}
\label{tab:means_laor}
\centering
\begin{tabular}{c c c c | c c}\hline\hline\\
Model & Parameter&$\mu$&$\sigma$&$\mu_{LAOR}$&$\sigma_{LAOR}$\\
\hline\hline

Power Law 2-12 keV & $\langle\Gamma_{\rm 2-12\,keV}\rangle$ &
$1.87^{+0.09}_{-0.10} $ & 0.36$^{+0.08}_{-0.06}$ 
& $1.82\pm0.13$ & 0.36$^{+0.11}_{-0.08}$\\


A &  $\langle$kT$_{BB}\rangle$& $0.136^{+0.009}_{-0.008}$
&$0.021^{+0.010}_{-0.006}$ 
&  $0.133\pm0.010$ &$0.019^{+0.009}_{-0.005}$  \\

B & $\langle$kT$_{MB}\rangle$& $0.157\pm0.008$ &
$0.023^{+0.008}_{-0.006}$ 
& $0.160\pm0.010$ & $0.021^{+0.010}_{-0.006}$\\

C & $\langle$kT$_{BS}\rangle$& $0.38\pm0.03$ & $0.08^{+0.03}_{-0.02}$ 
& $0.38\pm0.03$ & $0.05^{+0.03}_{-0.02}$\\ 

D & $\langle\Gamma_{\rm soft}\rangle$& $2.73^{+0.12}_{-0.11}$ &
$0.28^{+0.11}_{-0.07}$ 
& $2.80\pm0.08$ & $0.13^{+0.08}_{-0.05}$\\ 
\hline
\end{tabular}
\end{table*}
Taking into account the results on the spectral analysis performed for
the whole sample of quasars, we have calculated the mean values of the
relevant parameters of the  models considered for the analysis: hard band power law and models
{\it  A}, {\it B}, {\it  C} and  {\it D}.  Table~\ref{tab:means_laor}
shows the  mean values and  the dispersion of each parameter estimated
using  the maximum  likelihood technique.  Figure~\ref{fig:means_laor}
shows  the plots for the 68\%,  90\% and 99\%  confidence contours and
the mean value  and dispersion found  for each spectral parameter. All
the mean  values  of  the  different parameters   found for the   Laor
subsample are  fully compatible within the errors  with the results of
the whole sample.
\begin{table*}
\caption{Spectral Fitting results: VII. Best fit model. A: bb; B: diskbb; C:
  brems; D: power law; E: double bb; F: absori; G: double absori $+$ bb; H: partial--covering
  (cold)$+$raymond-smith; I: partial--covering(warm)$+$emission line;
  J: absori$+$bb;  PL: simple power law. Models with $^*$ include an
  additional  cold absorption component. All but one (i.e. model PL) models
  also include a power law component. For the results concerning the Fe K$\alpha$ line see Paper II.}
\label{tab:best}
\begin{center}
\begin{scriptsize}
\begin{tabular}{c c c c c c c c c}
\hline \hline \\ PG Name &Model&$\Gamma$&kT$/\Gamma_{soft}$& kT &
 $\xi$$/$(N$_{\rm H}$)&Edge/$\tau$&$\chi^2$($\nu$)&Fe K$\alpha$\\ &
 &        & (keV)            &(keV)&   (keV)$/$10$^{22}$
 cm$^{-2}$&(keV) & &\\
\hline\\
0007$+$106 & A &1.71$^{+0.04}_{-0.04}$&0.16$^{+0.01}_{-0.01}$&$-$&$-$&$-$&312(302)&Y\\
0050$+$124 & E$^*$ & $2.31^{+0.03}_{-0.03}$ & $0.084^{+0.008}_{-0.008}$ &$0.19^{+0.03}_{-0.02}$ & 0.09$^{+0.02}_{-0.02}$ &$0.647^{+0.009}_{-0.009}/0.26^{+0.06}_{-0.06}$ &  523(383) & Y \\
0157$+$001 & A &2.24$^{+0.06}_{-0.07}$&$0.125^{+0.012}_{-0.014}$&$-$&$-$&$-$&149(166)& N\\
0804$+$761 & D &1.3$^{+0.7}_{-0.7}$&2.5$^{+0.4}_{-0.1}$&$-$&$-$&$-$&287(262)&Y\\
0844$+$349 & E & $1.96^{+0.09}_{-0.06}$ & $0.112^{+0.008}_{-0.004}$ &$0.33^{+0.03}_{-0.03}$ & $-$ &$-$ & 162(162) & N \\
0947$+$396 & C & $1.83^{+0.07}_{-0.05}$ & $0.50^{+0.05}_{-0.04}$ & $-$& $-$ & 0.79$^{+0.04}_{-0.03}/0.18^{+0.04}_{-0.04}$ & 218(259)& Y \\
0953$+$414 & E &1.97$^{+0.04}_{-0.04}$&0.119$^{+0.009}_{-0.009}$&0.25$^{+0.02}_{-0.02}$&$-$&0.73$^{+0.03}_{-0.03}$$/$0.18$^{+0.06}_{-0.09}$&308(284)&N\\ 
1001$+$054 & F &2.0$^{+0.2}_{-0.1}$&$-$&$-$&542$^{+97}_{-147}$$/$19.2$^{+11.9}_{-7.3}$&$-$&16(8)&N\\
1048$+$342 & E &1.74$^{+0.08}_{-0.08}$ &0.115$^{+0.008}_{-0.010}$&0.27$^{+0.04}_{-0.02}$&$-$&$-$&243(269)&Y\\
1100$+$772 & E &1.57$^{+0.07}_{-0.07}$&0.133$^{+0.004}_{-0.007}$&0.4$^{+0.03}_{-0.03}$&$-$&$-$&339(342) &Y\\
1114$+$445 & G &1.85$^{+0.04}_{-0.03}$&0.207$^{+0.09}_{-0.09}$&$-$&89$^{+37}_{-30}$$/$1.5$^{+0.4}_{-0.3}$&4.0$^{+0.5}_{-0.5}$$/$1.53$^{+0.12}_{-0.07}$$^a$&315(360)& Y\\
1115$+$080 &PL$^*$ & $1.85^{+0.05}_{-0.05}$ & $-$ & $-$ &$0.33^{+0.08}_{-0.07}$  &$7.1^{+0.3}_{-0.4}/0.28^{+0.13}_{-0.12}$ &214(217) & N \\& & & & & & $9.5^{+0.3}_{-0.3}/0.31^{+0.17}_{-0.15}$ & 204(215) \\
1115$+$407 & E &2.18$^{+0.08}_{-0.08}$&0.097$^{+0.006}_{-0.006}$&0.238$^{+0.003}_{-0.003}$&$-$&$-$&228(230)&Y\\
1116$+$215 & C & $2.20^{+0.06}_{-0.07}$  & $0.32^{+0.04}_{-0.04}$ &  $-$  &$-$ & $0.75^{+0.05}_{-0.06}/0.10^{+0.07}_{-0.06}$ & 197(228) & Y \\
1202$+$281 & C & $1.68^{+0.05}_{-0.05}$ &  $0.49^{+0.05}_{-0.05}$ &  $-$ &  $-$ &$0.74^{+0.03}_{-0.03}/0.20^{+0.06}_{-0.06}$ & 306(319) & N\\
1206$+$459 &PL &1.74$^{+0.09}_{-0.09}$&$-$&$-$&$-$&$-$&40(42) & N\\
1211$+$143 & E & $1.61^{+0.05}_{-0.03}$ &  $0.113^{+0.002}_{-0.004}$ &$0.24^{+0.02}_{-0.03}$ &  $-$ & $0.775^{+0.11}_{-0.05}/0.39^{+0.04}_{-0.04}$ & 402(392) & Y\\&  &  &  &   &  &$0.967^{+0.010}_{-0.009}/0.39^{+0.04}_{-0.06}$ \\&  &  &  &   &  &$7.25^{+0.14}_{-0.09}/0.66^{+0.12}_{-0.07}$ \\
1216$+$069 & E &1.67$^{+0.06}_{-0.09}$&0.123$^{+0.07}_{-0.09}$&0.349$^{+0.05}_{-0.05}$&$-$&$-$&164(200) & N\\
1226$+$023 & E &1.60$^{+0.01}_{-0.01}$&0.10$^{+0.02}_{-0.02}$&0.24$^{+0.01}_{-0.01}$&$-$&7.4$^{+0.1}_{-0.2}$$/$0.10$^{+0.01}_{-0.03}$&320(258) & N\\
1244$+$026 & A & 2.48$^{+0.03}_{-0.03}$  & $0.16^{+0.01}_{-0.01}$ &$-$ & $-$ & $-$ & 125(110) & N \\
1307$+$085 & A & 1.50$^{+0.04}_{-0.04}$ & $0.13^{+0.01}_{-0.01}$ & $-$& $-$ &0.74$^{+0.02}_{-0.02}$$/$0.54$^{+0.10}_{-0.14}$ & 232(210) & N \\
1309$+$355 & A &1.81$^{+0.04}_{-0.04}$&0.103$^{+0.011}_{-0.007}$&$-$&$-$&0.75$^{+0.03}_{-0.03}$/0.4$^{+0.2}_{-0.1}$&225(214) & Y\\
1322$+$659 & C &2.25$^{+0.06}_{-0.08}$&0.29$^{+0.01}_{-0.02}$&$-$&$-$&$-$&235(212)&Y\\
1352$+$183 & E & $1.84^{+0.14}_{-0.10}$ &  $0.112^{+0.012}_{-0.011}$ & 0.26$^{+0.03}_{-0.03}$ &  $-$ &$ 0.69^{+0.03}_{-0.04}/0.21^{+0.09}_{-0.11}$ & 247(240) & Y \\
1402$+$261 & C &  $2.15^{+0.08}_{-0.07}$ & $0.37^{+0.03}_{-0.03}$ & $-$ & $-$ &$0.75^{+0.04}_{-0.04}/0.17^{+0.07}_{-0.06}$ & 204(213) & Y\\
1404$+$226 & J &2.3$^{+0.4}_{-0.4}$&0.114$^{+0.002}_{-0.003}$&$-$&44$^{+123}_{-26}$$/$1.4$^{+0.6}_{-0.3}$&$-$&60(75)& N\\
1407$+$265 & A &2.21$^{+0.03}_{-0.03}$&0.20$^{+0.03}_{-0.03}$&$-$&$-$&$-$&223(244)&N\\
1411$+$442 & H &2.3$^{+0.2}_{-0.2}$&$\equiv\Gamma$&0.15$^{+0.03}_{-0.03}$&23.2$^{+3.8}_{-3.8}$$^c$&$-$&82(102)& Y\\
1415$+$451 & E & $2.00^{+0.07}_{-0.10}$ & $0.099^{+0.006}_{-0.006}$ &$0.24^{+0.03}_{-0.03}$ & $-$ & $-$ & 169(126) & Y\\
1427$+$480 & C & $1.93^{+0.06}_{-0.03}$ & $0.44^{+0.04}_{-0.04}$ & $-$ & $-$ &$0.76^{+0.03}_{-0.04}/0.12^{+0.05}_{-0.04}$ & 251(277) & Y\\
1440$+$356 & D & $1.2^{+0.4}_{-0.2}$ & $3.14^{+0.2}_{-0.08}$ & $-$ & $-$& $-$ &  191(158) & N\\
1444$+$407 & D &$1.8^{+0.3}_{-0.3}$  &$3.3^{+0.2}_{-0.3}$ & $-$ &$-$ & $-$ & 196(166) & N\\
1501$+$106 & E & $1.84^{+0.02}_{-0.02}$ & $0.0937^{+0.0014}_{-0.002}$ &$0.235^{+0.005}_{-0.007}$ & $-$ & $-$ & 240(221) & N\\
1512$+$370 & D &1.51$^{+0.09}_{-0.09}$&2.71$^{+0.22}_{-0.15}$&$-$&$-$&$-$&226(253)&Y\\
1613$+$658 & A &1.80$^{+0.07}_{-0.03}$&0.13$^{+0.01}_{-0.01}$&$-$&$-$&0.73$^{+0.02}_{-0.02}$$/$0.5$^{+0.1}_{-0.2}$&148(180)& N\\
1626$+$554 & D &1.9$^{+0.1}_{-0.4}$&3.1$^{+0.7}_{-0.4}$&$-$&$-$&$-$&266(268)&N\\
1630$+$377 & D &1.9$^{+0.3}_{-0.3}$&3.4$^{+2.4}_{-1.1}$&$-$&$-$&$-$&44(57)&Y\\
1634$+$706 & C & $1.95^{+0.11}_{-0.08}$ &  $1.1^{+0.2}_{-0.2}$ & $-$ & $-$ & $-$& 203(208) & N\\
2214$+$139 & I &1.88$^{+0.04}_{-0.04}$&$\equiv\Gamma$&0.57$^{+0.02}_{-0.01}$$^b$&89$^{+14}_{-12}$$/$8.9$^{+0.5}_{-0.5}$&6$^{+3}_{-1}$$/$1.7$^{+0.2}_{-0.4}$$^a$&335(358)& Y\\ 
2302$+$029 & C & $2.03^{+0.06}_{-0.11}$ & $0.37^{+0.13}_{-0.11}$ & $-$ & $-$ & $-$ & 58(66) & N \\
\hline\\

\end{tabular} 

$^a$ values of $\xi$ and N$_{\rm H}$ of the second warm absorber
component in erg$/$cm$^{2}$$/$s and  10$^{22}$ cm$^{-2}$,
respectively. $^c$ value of N$_{\rm H}$ $\times$ 10$^{22}$ cm$^{-2}$.
$^b$ energy (in keV) of the emission line (see text for details).
$^c$ value of N$_{\rm H}$ $\times$ 10$^{22}$ cm$^{-2}$.
\end{scriptsize}
\end{center}
\end{table*}

\subsection{Best fit models}
\label{sec:complex}
The  models  tested   to account for   the   soft excess  can  explain
successfully the spectra  of 30 out of the  40 sources. The spectra of
the remaining nine sources have been analyzed separately (see  below) while
none of the  two component models improved the quality of the fit for 1206$+$459 likely because of
the low signal--to--noise  ratio of its spectrum. 
Even though
the four  models  considered to explain  the  soft excess provided good
fits ($\chi^2_\nu<1.2$) for the majority of the sources, we found
that for half of the sources, more complex parameterizations were required.  Table~\ref{tab:best}  lists   the best fit model  
 as well as the resulting values of the spectral parameters for each
quasar in the sample.\\

In the following we report on those sources for which 
the best fit model to the X--ray emission is more complex than a two-component parameterization.
A consistent fraction of  the  QSOs in the sample, i.e. 13 out of 40,   
required two  thermal components to match their broad soft excess.
Such sources have been fitted by the models labeled as E and H  in
Table~\ref{tab:best}.   
In the case of model E, both thermal components
can be modeled with a blackbody model.   The mean temperatures obtained
through       the          maximum    likelihood       method      are
$\langle$kT$_{BB}\rangle=0.108\pm0.007$ keV for the   lower    blackbody temperature
and $\langle$kT$_{BB}\rangle=0.27^{+0.03}_{-0.02}$ keV for
the  blackbody with higher temperature, while    the   dispersion   is
$\sigma=0.010^{+0.007}_{-0.004}$ and  $\sigma=0.041^{+0.030}_{-0.017}$,
respectively.   
On the other hand, only one source, i.e. 1411$+$442, required a blackbody plus a
Raymond-Smith plasma (i.e. Model H in Table~\ref{tab:best}) to account for its soft excess (see below for further
details on this source).\\

Spectral features due to absorption edges  were reported in 16 quasars.
The majority of them were found in  the range 0.7--0.8 keV: most these
features are probably due to He--like oxygen and/or a blend of M-shell
iron inner   shell  transition (the  so--called  unresolved transition
array   or UTA; Behar,   Sako  \& Kahn  2001).    In three  QSOs, i.e.
1115$+$080,  1211$+$143,   1226$+$023,  an  edge   was detected around
7.1--7.4  keV,  implying   reprocessing  in  ionized  iron   material.
Interestingly,  the   spectrum of  1211$+$143  shows  three absorption
edges, which are  located, respectively, at $0.775^{+0.011}_{-0.005}$,
$0.967^{+0.010}_{-0.009}$   and   $7.25^{+0.14}_{-0.09}$  keV.  Since
these  absorption features are  indicative of  the  presence of a warm
absorber along the line--of--sight,  we attempted a further fit  using
the model {\tt ABSORI} in XSPEC (Done et al.  1992) to account for the
ionized  absorbing plasma in  all the  sources showing absorption edges.
However, such parameterization   did  not  significantly improve   the
goodness--of--fit statistic of  any  source expect for  1001$+$054, 1114$+$445, 1404$+$226 and
2214$+$139 (see below for more details on  these sources), leaving the
resulting gas parameters (i.e.  T,
\nh~and $\xi$\footnote{The ionization   parameter $\xi$ is  defined as
$\xi$ = $L$/$n$$r^{2}$, where $L$  is the isotropic luminosity of  the
ionizing source in  the interval 5  eV to 300 keV,  $n$ is  the number
density of the warm plasma and $r$ is the distance between this latter
and  the   central source.}) basically   unconstrained.   The most
likely  explanation for this result is  the relatively limited photon
statistics affecting the
\epic~spectrum   of these QSOs, which did   not  allow a more detailed
description of  the  warm  absorber than  in  terms  of one (or  more)
edge(s).\\

Finally, in the following we report more details on the spectral
analysis of  the eight most peculiar  sources in our sample, for which
model   A, B, C, and D  gave an  associate \xnu~$\geq$1.2 (see
Sect.~\ref{sec:soft}).

{\it 0050$+$124 (I Zw 1)}. The \xmm~spectrum shows a complex soft X--ray
emission dominated by cold absorption. The observed data are best
explained by a power law, $\Gamma=2.31\pm0.03$, accounting for the
hard band and a double black body component, $kT_1=0.084\pm0.008$~keV
and $kT_2=0.19^{+0.03}_{-0.02}$~keV. Our analysis reveals the presence of
a cold absorption modeled by the  {\tt ZWABS} model in XSPEC,
\nh~= $9^{+2}_{-1} \times 10^{20}$ \cm2, and an absorption edge
located at $0.65^{+0.01}_{-0.02}$~keV. 
These results are in agreement with a previous
\xmm~data analysis by Gallo et al. (2004).

{\it 1001$+$054}. We found that a power law modified by a  warm absorber
provides the best fit  for this soft X--ray weak QSO. The resulting photon index
is $\Gamma \sim$ 2, while we derived a column density \nh~$\sim$ 2 $\times$ 10$^{23}$
\cm2~and an ionization parameter
$\xi \sim$ 500 \cgs~for the absorbing gas. The spectrum of this QSO
is discussed in detail in a separate paper (Schartel et al. 2004).

\begin{figure*}
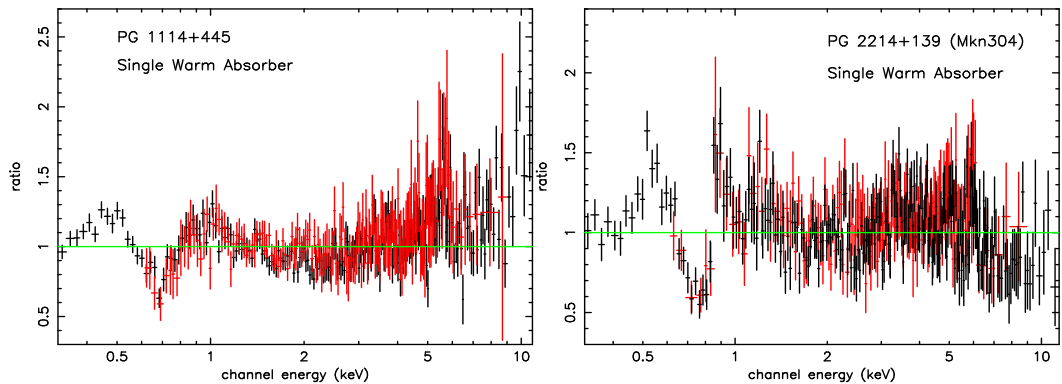

\begin{center}
\epsfig{figure=ep1621_f5a.ps,height=7cm,width=5.0cm,angle=-90}\epsfig{figure=ep1621_f5b.ps,height=7cm,width=5.0cm,angle=-90}
\caption{The ratios between the \epic~data (\pn~data in black, \mos~in red) and a fit with single warm absorber model for 1114+445 ({\it left}) and 2214+139 ({\it right}):
the large residuals below 1 keV strongly suggest the presence of a complex ``multi-component'' warm absorber. Note also the residuals around 5-6 keV which indicate Fe K$\alpha$ fluorescence emission (e.g. Paper II).}
\label{fig:1absori}
\end{center}
\end{figure*}

{\it 1114$+$445}. A {\it HST}  spectrum of this quasar revealed strong
UV  absorption lines  (Mathur et  al.  1998).  In the X--ray  band the
presence of  optically  thin partially  ionized material  was detected
(Laor et al. 1997; George et al. 1998). By applying  a power law model
to the 2--12 keV data we obtained a flat photon index ($\Gamma_{2-12}
\approx$ 1.4). Extrapolation down to lower energies shows a broad
absorbing feature in the range 0.5--1 keV.  The broad band X--ray
continuum of this QSO is well described by a power law with $\Gamma
\approx$ 1.85 modified by two ionized absorption components ({\tt
ABSORI} in XSPEC) plus a blackbody component accounting for the soft
excess (see Table~\ref{tab:best} for the best fit parameters). The
presence of the second ``warm'' component yielded an improvement in
the fit at $>$99.9\% confidence level (e.g. Fig.~\ref{fig:1absori}).
 \xmm~has therefore clearly revealed, for the first time in this
QSO, a multi-zone ionized absorber  similarly to what we found for
2214$+$139 (see below; Piconcelli et al. 2004).  We also reported
the presence of a significant ($P_F >$ 99.9\%) fluorescence iron
emission line with a narrow profile at 6.45$^{+0.02}_{-0.08}$ keV
(e.g. Paper II).  On the other hand, we did not find
any evidence of a Fe K--shell absorption edge as detected at $\approx$
7.25 keV with a $\tau$ = 0.35$^{+0.38}_{-0.29}$ in the \asca~spectrum
by George et al. (1998).  Our results agree with those recently
published by Ashton et al. (2004) for an independent analysis of the
same \xmm~data. They also detected a two--phase warm absorber even if
they found a best fit value of the ionization parameter for the
hotter--phase component of the warm absorber larger than ours (and,
consequently, also a larger column density).

{\it  1115$+$080}. \xmm~data of this BAL  quasar have been also analyzed by
Chartas et al. (2003).  This study   revealed  an absorbed
soft energy spectrum. Even though
Chartas et al. (2003) suggested that  the soft spectrum is best fitted by an
ionized  absorber, our  study  found that  a neutral absorption model
cannot be ruled out. Our analysis indicates that the
best fit model for the observed spectrum consists of a power
law modified by cold absorption and two absorption edges at $7.1^{+0.3}_{-0.4}$~keV
and   $9.5\pm0.3$~keV     (see  Table~\ref{tab:best}    for    further
details). Taking into account Chartas et al. (2003) results, we have tried to
explain the  soft energy absorption by an   ionized absorber using the
{\tt  ABSORI}  model in  XSPEC. The  value   found for  the  ionization
parameter ($\xi$) is extremely low and therefore the model does not
differ from the neutral absorption one.

{\it 1226$+$023 (3C~273)}  The main feature in the \pn~spectrum of
this well--known radio--loud QSO  is the broad soft excess.  A simple
two-component model indeed fails to adequately describe its 0.3--12
keV emission (see Sect~\ref{sec:soft}  and Fig.~\ref{fig:se}).  Our
best fit includes two blackbody components (with kT$_{BB,1} \approx$
0.1 and  kT$_{BB,2} \approx$ 0.24 keV, respectively) and a power law
with a quite flat photon index ($\Gamma$ = 1.60$\pm$0.01).  This
result confirms the finding by Page et al. (2004a), based on the same
\xmm~observation.  We also detected the significant (at $>$ 99.9\%
confidence level) presence of an absorption edge at
7.4$^{+0.1}_{-0.2}$ keV and with an optical depth of $\tau$ =
0.09$^{+0.02}_{-0.03}$. Such an energy centroid suggests that the
absorbing material is weakly ionized (Fe V--XV). The most likely
origin for this edge feature is via reflection in optically thick
matter as an origin in a line--of--sight plasma is not supported by
the {\it RGS} data analysis results published in Page et al. (2004a),
which did not report any absorption features apart from a OVII
He$\alpha$ due to warm gas in the local intergalactic medium.
Consequently, we also tried to fit the data with a model including a
Compton reflection component ({\tt PEXRAV} in XSPEC), yielding,
however, no statistical improvement respect to the best fit model
listed in Table~\ref{tab:best}. The upper limit for the covering
factor of the material irradiated by the X--ray source is $R
=\Omega/2\pi =$ 0.4.  On the other hand, no Fe K$\alpha$ emission line
(i.e. another hallmark of reprocessing in an accretion disk) was
detected,  and the upper limit on the equivalent width for a narrow
line at 6.4 keV resulted to be 6 eV.  Finally, Page et al. (2004a)
reported the evidence for a weak broad Fe line using ten co-added
\epic~observations which is, however, not detected in our data even
when the absorption edge has been removed from the fitting model.

{\it 1404$+$226}. This narrow line QSO was observed by {\it ROSAT} and {\it ASCA}. Ulrich et al. (1999)
reported a very steep soft X-ray continuum, and
evidence of complex ionized absorber. In particular, a peculiar absorption
feature around 1 keV led these authors to claim for deviations from
solar abundances in this source.
Our results confirm the spectral complexity of 1404$+$226: we have found a 
best fit model consisting of a blackbody with kT
= 0.144$^{+0.002}_{-0.003}$ keV and a steep ($\Gamma$ =
2.34$^{+0.35}_{-0.37}$) power law modified by ionized absorption. For
the gas we derive a column density \nh~$\approx$ 1.4 $\times$ 10$^{22}$ \cm2~and
an ionization parameter $\xi$ = 44$^{+123}_{-26}$ \cgs. No evidence of
elemental abundances different from solar values have been found.
Finally, no significant fluorescence iron emission line has been detected.

{\it 1411$+$442}. This BAL  QSO is X--ray faint ($F_{0.3-12} \approx$7
$\times$  10$^{-13}$ \cgs).  The 2--10  keV spectrum is extremely flat
with a $\Gamma_{2-12} \approx$  0.4 and a  clear excess around  6 keV.
The addition of a line with  a narrow gaussian profile to parameterize
the  Fe  K$\alpha$ emission  feature  produced a  very significant (at
99.8\%  confidence level) improvement    in  the fit.   The   centroid
($E_{K\alpha}$  = 6.43$^{+0.05}_{-0.13}$   keV) and narrowness  of the
line suggest a likely origin from a neutral  material located far away
from the innermost regions of the accretion disk.

The best fit model requires a power law with partial--covering
({\tt ZPCF} in XSPEC) and a Raymond--Smith plasma component (this
latter significant at $>$99.9\% confidence level). For the absorbing gas 
we obtained a column density \nh~$\approx$ 2.3 $\times$ 10$^{23}$ \cm2
and a covering fraction of 96\%.
The plasma temperature and the photon index of the power law resulted to be kT =0.15$\pm$0.03
keV and $\Gamma$ = 2.3$\pm$0.2, respectively.

The limited statistics did not allow us to test more
complex fitting models aimed, for instance, at checking the presence of a
warm absorber in this source as observed in other AGNs with prominent
absorption lines in the UV spectrum (e.g. Crenshaw et al. 2003; Monier et al. 2001;
Piconcelli et al. 2004).  This \epic~observation has been also
analyzed by Brinkmann et al. (2004a) which reported results consistent
with ours.

{\it 2214$+$139 (MKN~304)}. Our analysis matches well with previous results
based on {\it Einstein} and {\it ROSAT} observations,
which reported a very flat continuum for this source, and suggested
the likely existence of heavy absorption.
The \xmm~data reveal a complex spectrum, dominated by strong
obscuration due to ionized gas  (see Fig.~\ref{fig:1absori}). With a \nh~$\approx$ 10$^{23}$ \cm2,
it is one of the ionized absorber with the highest column density seen so
far by \xmm~and \chandra. 
A two--component warm gas provides an excellent description
of this absorbing plasma.  From the spectral analysis we derived $\xi^{T5}$ =
89.3$^{+13.9}_{-12.0}$ \cgs~and $\xi^{T4}$ = 5.9$^{+2.4}_{-0.9}$
\cgs~for the hot (i.e. with T = 1.5 $\times$ 10$^{5}$ K) and the cold (i.e. with T
= 3 $\times$ 10$^{4}$ K) component, respectively. 
In addition to a narrow Fe K$\alpha$
emission line at $\sim$ 6.4 keV, another emission line feature was
significantly detected  at 0.57 keV, likely due to the helium--like
oxygen triplet (maybe originating in the warm absorber itself). We also reported
the presence of a weak soft excess component which can be interpreted
as partial covering or scattered emission from the ionized outflowing plasma.
For the complete and detailed presentation and discussion of these results
see Piconcelli et al. (2004).

\subsection{Fluxes and luminosities}
\label{sub:flux}

Flux in the soft (0.5--2 keV) and hard (2--10 keV)
energy band assuming the  best fit model for each quasar as in Table~\ref{tab:best} are listed in
Table~\ref{tab:flux}, together with the corresponding luminosities
corrected for both the Galactic absorption and (if present) the additional intrinsic warm/neutral
absorber column densities. The hard (soft) X--ray fluxes range from $\sim$ 0.1(0.1) to $\sim$ 80(45) $\times$
10$^{-12}$ \cgs. With a 0.5--10 keV X--ray luminosity of $\sim$ 10$^{43}$ \ergs, 1001$+$054 is the faintest
object in the sample. The most luminous quasar in the hard (soft) band
is 1226$+$023 (1407$+$265) with a luminosity
of $\sim$ 50 $\times$ 10$^{44}$ \ergs. 
In the two last columns of Table~\ref{tab:flux}, the ratio between the strength of the soft
excess  and the high energy power law component in the 0.5--2 and
0.5--10 keV band are listed.

\begin{table*}
\caption{Fluxes and absorption-corrected luminosities calculated assuming the best fit
model. The last two columns list the ratio between the strength of the soft
excess  and the high energy power law component in the 0.5--2 and
0.5--10 keV band, respectively }
\label{tab:flux}
\begin{center}
\begin{tabular}{c c c c c c c }
\hline \hline \\ {\bf PG Name}  & {\bf
F$_{0.5-2\,keV}$} & {\bf F$_{2-10\,keV}$} &{\bf L$_{0.5-2\,keV}$} &
{\bf L$_{2-10\,keV}$}&$R^{0.5-2}_{S/P}$& $R^{0.5-10}_{S/P}$ \\ 
 & (10$^{-12}$ erg cm$^{2}$ s$^{-1}$)&(10$^{-12}$ erg cm$^{2}$ s$^{-1}$) &(10$^{44}$ erg s$^{-1}$) &(10$^{44}$ erg s$^{-1}$) & & \\ \hline \hline\\
0007$+$106	&3.7    &7.2    &0.87    &1.4&           0.160&         0.072\\    
0050$+$124 	& 9.15 & 8.44 & 1.39 & 0.78& 		  0.360&         0.212\\
0157$+$001 	& 1.33 & 0.93 & 1.05 & 0.71&		  0.169	&       0.126\\
0804$+$761 	&11.6    &11.12 &3.43    &2.86&    	  0.489	 &     0.238\\
0844$+$349 	&     6.9 &5.5 & 0.81 & 0.55& 		  0.954	 &     0.444\\
0947$+$396 	& 1.79 & 1.87 & 2.55 & 2.26 &		  0.814	  &    0.336\\
0953$+$414 	&3.79&3.21&7.77&5.41		&	  0.836	&      0.379\\
1001$+$054 	&0.03    &0.12    &0.10    & 0.13  & 	  $-$    &       $-$    \\
1048$+$342 	&1.24&1.46&1.07&1.10			&  0.712&	       0.277\\
1100$+$772 	&2.48    &3.83    &8.97    &11.19   & 	  0.650    &    0.318\\
1114$+$445 	&0.62&2.3    &2.65    &1.45    	&  1.75        &0.724\\
1115$+$080 	& 0.23  & 0.36 & 42.1 & 65.3& 		  $-$  &         $-$\\    
1115$+$407 	&2.13    &1.27    &1.67&0.85 &    	  0.787	&      0.427\\
1116$+$215 	& 5.2 & 3.3 & 5.0 & 3.1  	&	  0.473	&      0.258\\
1202$+$281	& 2.68 & 3.72 & 2.22 & 2.68& 	 	  0.646	&      0.226\\
1206$+$459      &0.14& 0.24 & 8.55 & 14.9 	&	  $-$    &       $-$    \\
1211$+$143 	& 3.0 & 3.1 & 0.61 &0.50 		&  1.825&	       0.572\\
1216$+$069 	&1.08    &1.38    &4.69    &4.8 	 & 0.948&	       0.351\\
1226$+$023	&43.81    &78.74    &31.75    &51.05     & 0.310 &       0.103\\
1244$+$026 	& 5.87 & 2.53 & 0.36 & 0.14 & 		  0.470  &      0.297\\
1307$+$085	& 0.94 & 2.01 & 0.63 &1.19  	&	  0.443	&      0.118\\
1309$+$355 	&0.45    &0.73    &0.46    &0.67 &   	  0.175	&     0.077\\
1322$+$659 	&2.25&1.33&2.21&1.08			&  0.650&	       0.367\\
1352$+$183 	& 2.30 & 2.00 & 1.85 & 1.36& 		  1.121	&      0.464\\
1402$+$261 	& 3.0 & 1.9 & 2.7 & 1.4&		  0.851	&      0.427\\
1404$+$226 	&0.51    &0.10    &2.31    &2.87&    	  0.423	&      0.246\\
1407$+$265 	&0.94    &0.80    &53.31    &41.20&       0.105	&      0.058\\
1411$+$442 	&0.09 &0.46    &0.35    &0.25&    	  0.061  &     0.062\\
1415$+$451 	& 1.4 & 1.1 & 0.5 & 0.4& 		  0.696	&      0.331\\
1427$+$480 	& 1.2 	& 1.1 & 2.0 & 1.6& 		  0.634	&      0.284\\
1440$+$356 	& 5.9 & 3.9 & 1.0 & 0.58 & 		  2.432	&      1.076\\
1444$+$407 	& 0.9 & 0.6 & 2.6 & 1.3 &		  3.009	&      1.420 \\
1501$+$106	&17.2 &16.2    & 0.57    & 0.49&     	  0.761	&      0.311\\
1512$+$370	&1.56 &1.97    &8.51    &8.84&     	  2.050  &      0.859\\
1613$+$658	&2.77    &4.15    &1.40 &1.78&     	  0.327	&      0.121\\
1626$+$554 	&3.04    &3.08    &1.57    &1.46&    	  0.708	&      0.354\\
1630$+$377 	&0.14&0.15&31.1&20.9&			  1.570  &      0.774\\
1634$+$706 	&	 1.0 &1.16 & 141.6 & 127.9& 	  0.591  &      0.335\\
2214$+$139 	&0.31    &3.26    &0.39&0.48  &   	  0.168	&      0.111\\
2302$+$029 	& 0.24 & 0.27 & 25.9 & 16.2 &   	  0.649 &      0.336\\
\hline
\end{tabular}\end{center}
\end{table*}
\begin{table*}
\caption{Spearman--rank correlations among spectral parameters of the
soft excess tested models and X--ray parameters (luminosity and $\Gamma_{2-12}$). The displayed values are
(r$_s$, $P_s$), where r$_s$ is the rank correlation coefficient and $P_s$ is the
significance probability. The significant (i.e. $>$~95\%) correlations
are marked in boldface (see text for details).}
\label{tab:correlations}
\begin{center}
\begin{tabular}{c c c c c c}
\hline\\ 
	                           &&kT$_{BB}$	&	kT$_{DiskBB}$&	kT$_{Brem}$& $\Gamma_{Soft}$\\
\hline\hline\\        
L$_{0.5-2\,keV}$	&&	(0.46,{\bf 97.9})&	($-$0.26,82.4)&	(0.41,{\bf 96.8})&	($-$0.15,52.1)\\

L$_{2-10\, keV}$	&&	(0.48,{\bf 98.5})&	($-$0.22,73.1)&	(0.44,{\bf 98.2})&	($-$0.23,73.7)\\

$R_{S/P}^{0.5-2}$	&&	(0.04,13.3)&		($-$0.09,35.1)&	($-$0.07,29.0)&		(0.39,94.7)\\
$R_{S/P}^{0.5-10}$	&&	(0.12,41.2)&		($-$0.11,42.3)&	($-$0.10,39.9)&		(0.45,{\bf 97.4})\\
$\Gamma_{2-12}$         && (0.26,78.1)               &   ($-$0.20,70.8)                     &($-$0.07,27.6)               &(0.33,89.4)      \\
 \hline
\end{tabular}
\end{center}
\end{table*}

\subsection{Correlations among spectral parameters and luminosity}
\label{sec:correlations}

We performed  Spearman--rank correlation tests.   This test calculates
how well  a linear  equation  describes  the relationship between  two
variables by means of the rank correlation coefficient (r$_s$) and the
probability ($P_s$).  Positive(negative) values of r$_s$ indicates the
two  quantities are  (anti--)correlated,    while   $P_s$  gives   the
significance level  of such a  correlation. In  particular, we checked
the  type and strength  of the  relationship  among the representative
parameter of  each model (i.e.  kT$_{BB}$, kT$_{DiskBB}$, kT$_{Brem}$,
$\Gamma_{Soft}$;    see  Sect.~\ref{sec:soft})  and  L$_{0.5-2\,keV}$,
L$_{2-10\,  keV}$,  $R_{S/P}^{0.5-2}$ and  $R_{S/P}^{0.5-10}$  and $\Gamma_{2-12}$ (see
Table~\ref{tab:flux} and Table~\ref{tab:pl_212}).    The corresponding  results   are  listed  in
Table~\ref{tab:correlations}. 

This study reveals that none of the 20 pairs of parameters are
significantly (i.e. $>$~99\%) correlated.
Nevertheless, we found marginal hints of a non--trivial correlation for five of them, i.e. 
kT$_{BB}$--L$_{0.5-2\,keV}$,
kT$_{BB}$--L$_{2-10\, keV}$, kT$_{Brem}$--L$_{0.5-2\,keV}$,
kT$_{Brem}$--L$_{2-10\, keV}$ and
$\Gamma_{Soft}$--$R_{S/P}^{0.5-10}$. 
Some of these correlations appear to be driven by the higher luminosity QSOs, \lum~$>$ 10$^{45}$ \ergs.
For those relationships with $P_s >$~95\%, we also performed a linear regression
fit to the data which, contrary to the  Spearman--rank correlation test,
takes also into account the uncertainties associated to the measurements.  
The values of $\chi^{2}$ obtained by each  fit revealed that
the probability that  the two quantities are  linearly correlated
is lower than 50\%. 


\subsection{Correlations with general source properties}
\label{sec:correlations2}

\begin{table*}
\caption{Spearman--rank correlations among X--ray observables  and
the radio-loudness parameter $R_{\rm L}$, the redshift, the FWHM of the H$_\beta$
emission line, the black hole mass M$_{\rm BH}$ and the accretion rate
$\dot{m}$ respect to the Eddington rate. The displayed values are
(r$_s$, $P_s$), where r$_s$ is the rank correlation coefficient and
$P_s$ is the significance probability. The significant (i.e. $>$~95\%)
correlations are marked in boldface (see text for details).}
\label{tab:correlations2}
\begin{center}
\begin{tabular}{c c c c c c}
\hline\\  &&L$_{2-10\, keV}$& $R_{S/P}^{0.5-10}$     & $\Gamma_{2-12}$&$\Gamma_{Soft}$\\ \hline\hline\\
$R_{\rm L}$ $^{(a)}$	&& (0.12,55.2)     &($-$0.48,{\bf 99.8}) 	&($-$0.003,2.0)&($-$0.37,93.2)\\
$z$                     &&(0.76,{\bf 99.9})       &(0.24,85.2)        &(0.08,39.0)&($-$0.07,25.7) \\
FWHM(H$_\beta$)$^{(b)}$	&&(0.50,{\bf 99.7}) 	&($-$0.12,51.1) 	&($-$0.62,{\bf 99.9})&($-$0.54,{\bf 98.6})\\
M$_{\rm BH}$$^{(c)}$	&&(0.67,{\bf 99.9}) 	&($-$0.03,13.1)        &($-$0.52,{\bf 99.8})&($-$0.16,51.2)\\
$\dot{m}$$^{(d)}$	&&($-$0.39,{\bf 97.5}) 	&($-$0.05,19.0)        &(0.41,{\bf 97.9})&($-$0.09,30.1)\\ \hline
\end{tabular}
\end{center}
Data taken from: $^{(a)}$ Kellermann et al. (1994); $^{(b)}$ Boroson \&
Green (1992); $^{(c)}$ Gierlinski \& Done (2004), P04 and Shields et
al. (2003); $^{(d)}$ P04, Shields et al. (2003), and Woo \& Urry (2002). 
\end{table*}

\begin{figure*}
   \centering
\epsfig{figure=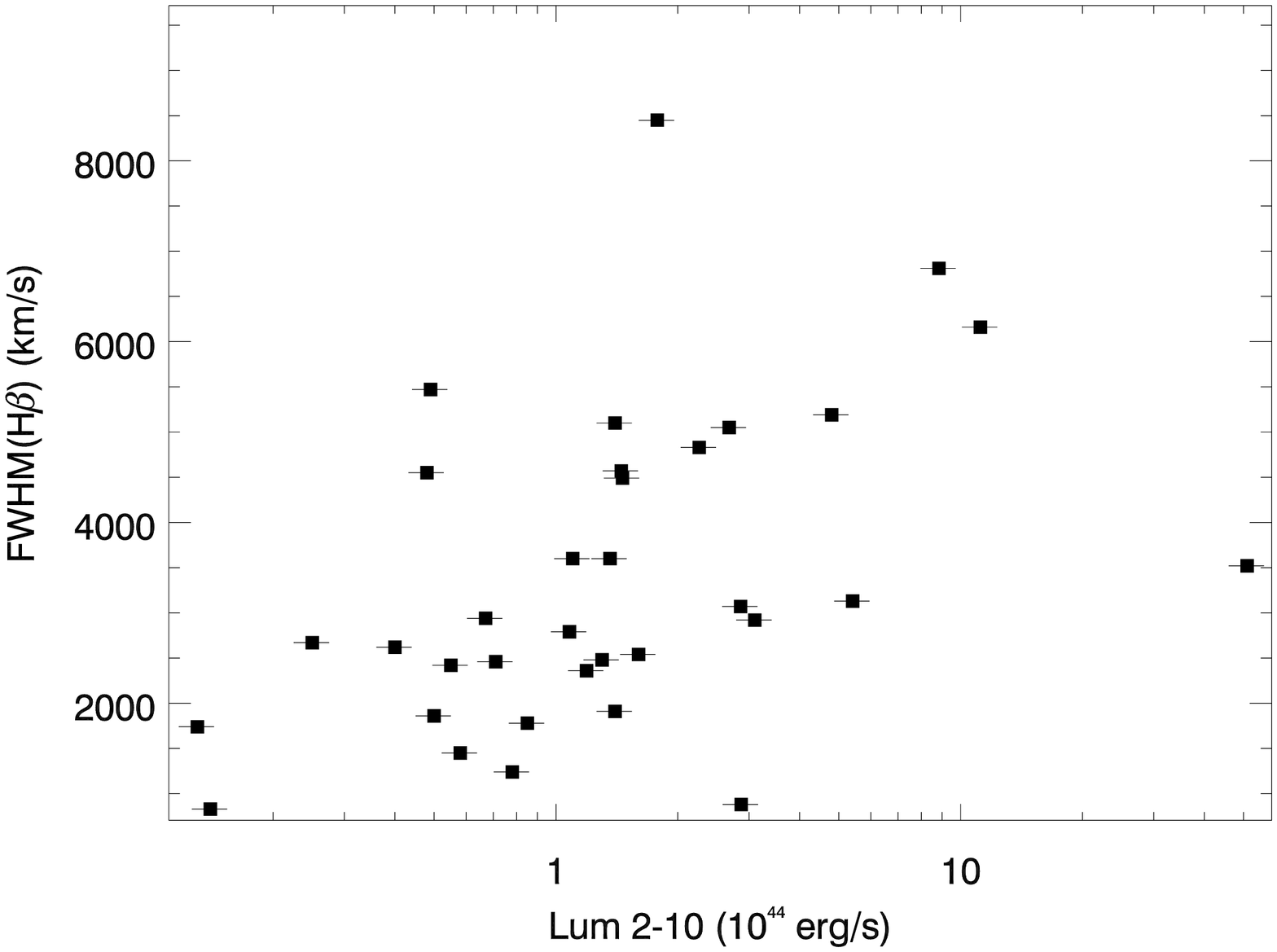,height=7.0cm,width=6.0cm,angle=90}\hspace{1cm}\epsfig{figure=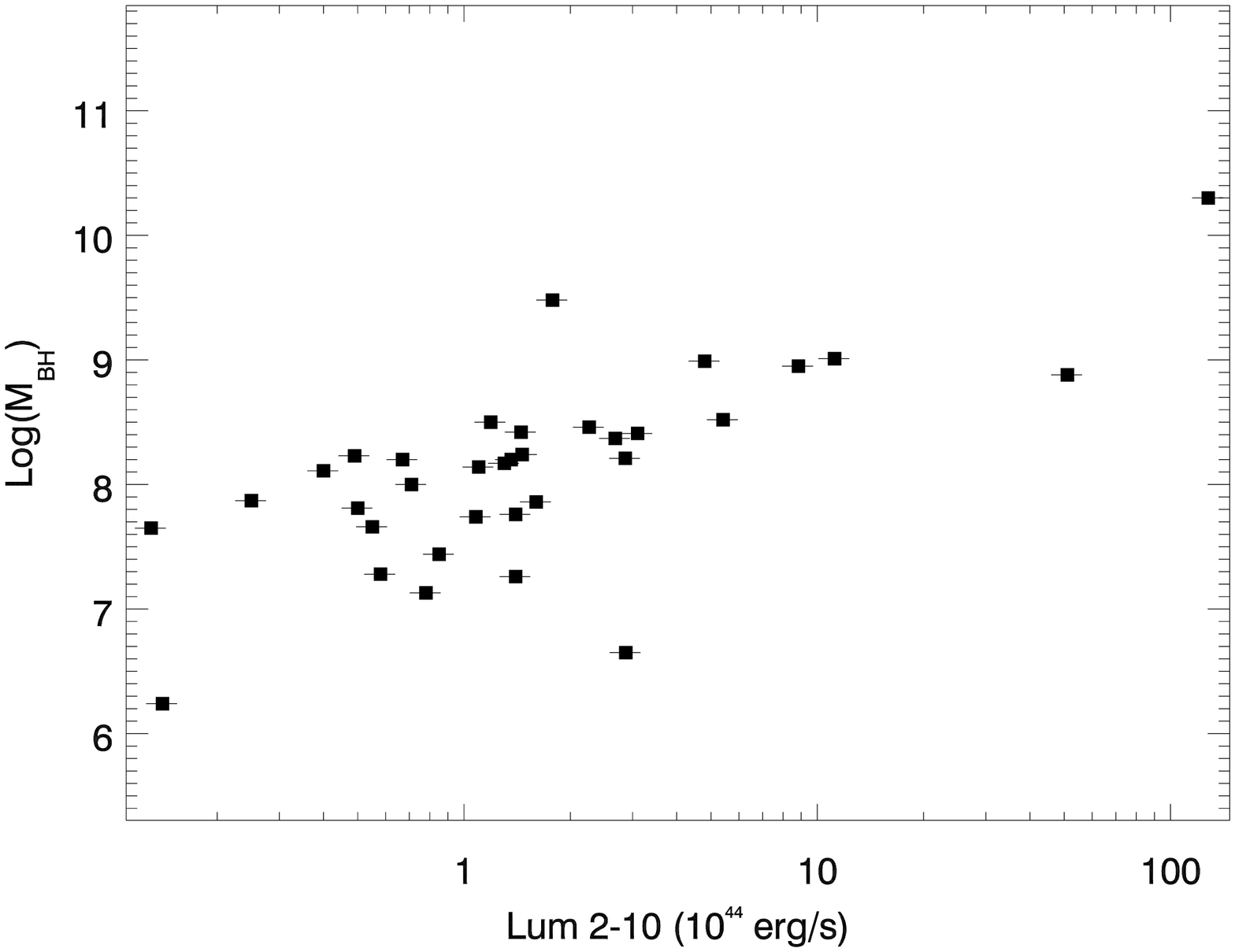,height=7.0cm,width=6.0cm,angle=90}\hspace{1cm}\epsfig{figure=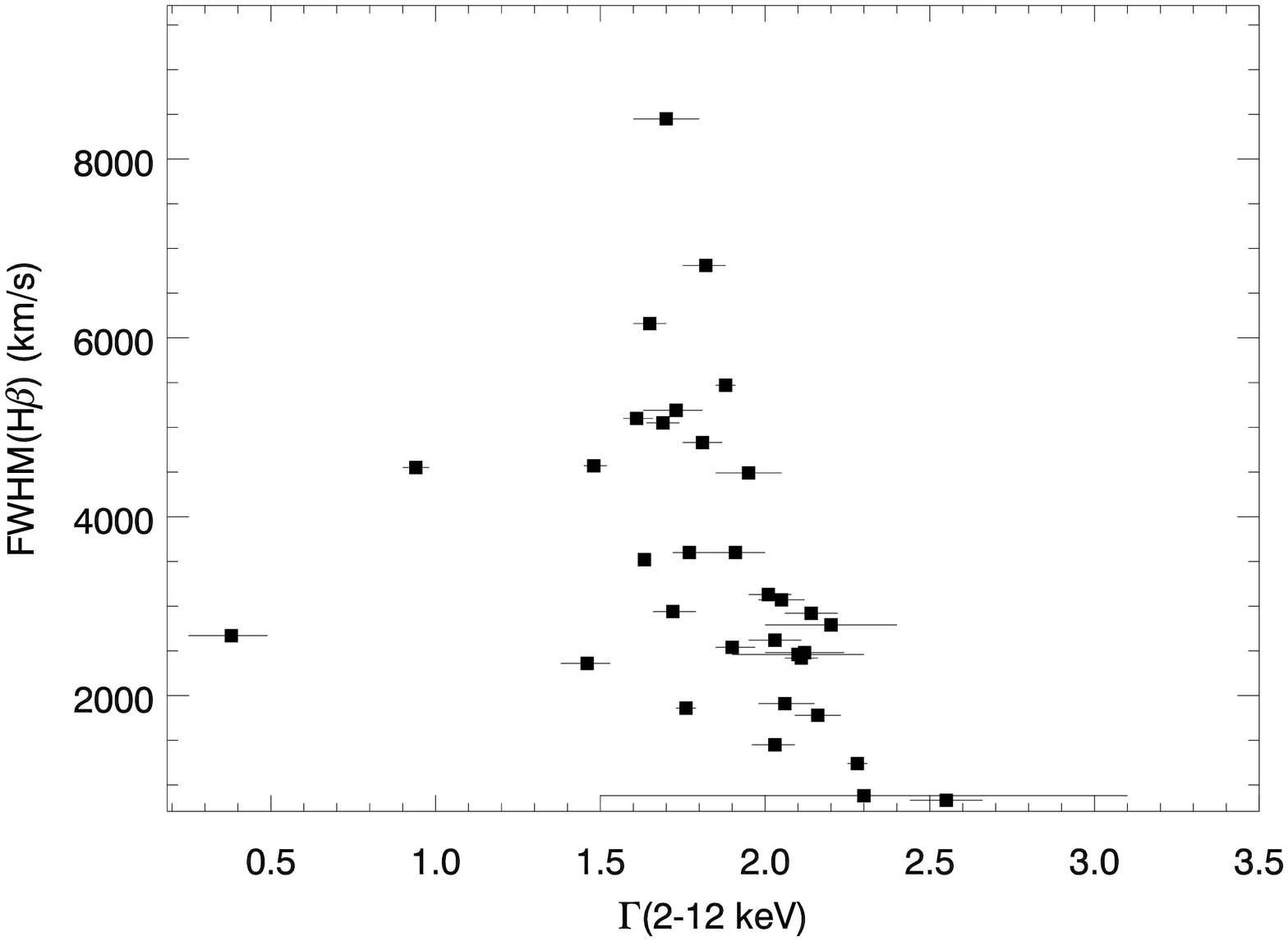,height=7.0cm,width=6.0cm,angle=90}\hspace{1cm}\epsfig{figure=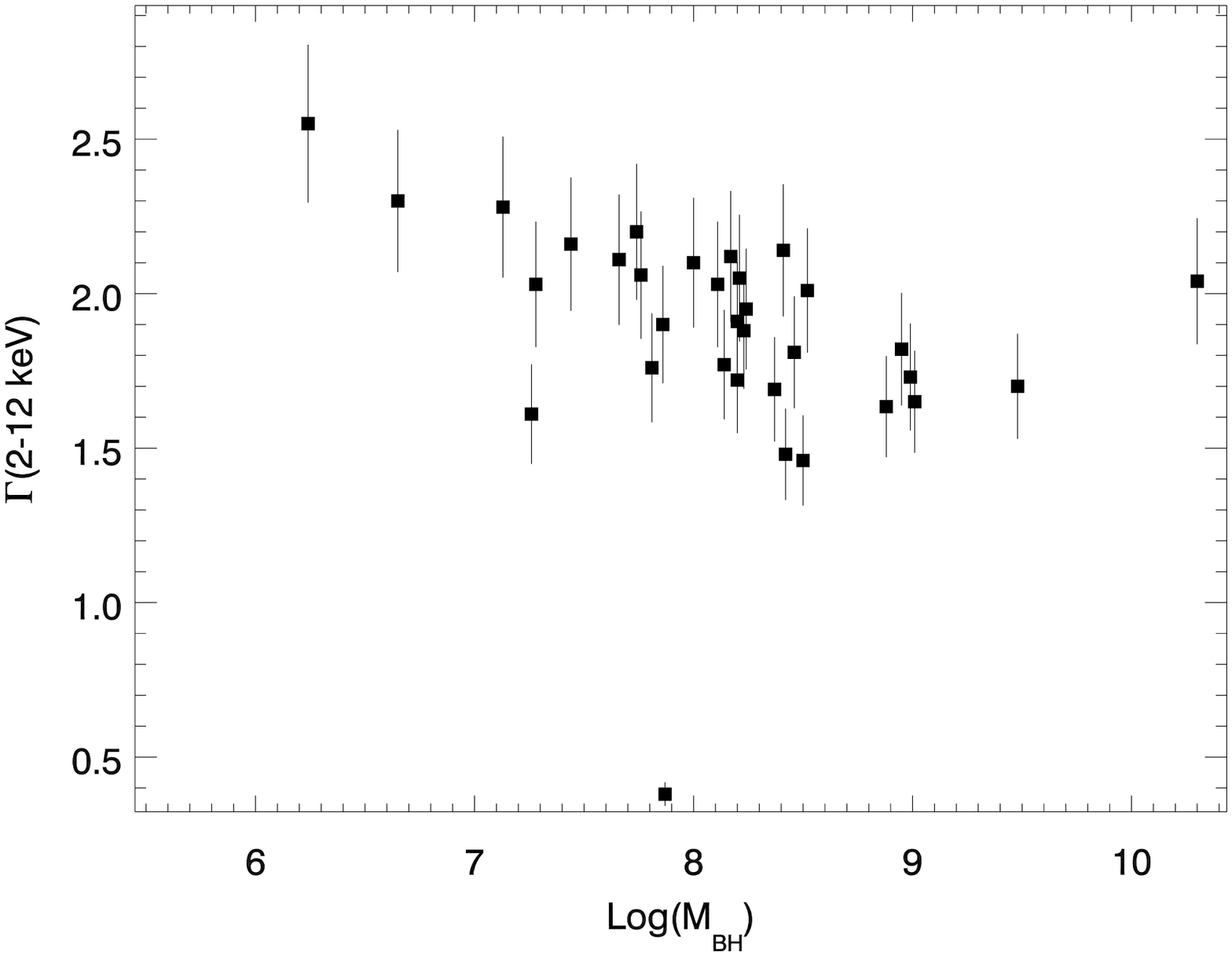,height=7.0cm,width=6.0cm,angle=90}\hspace{1cm}
\epsfig{figure=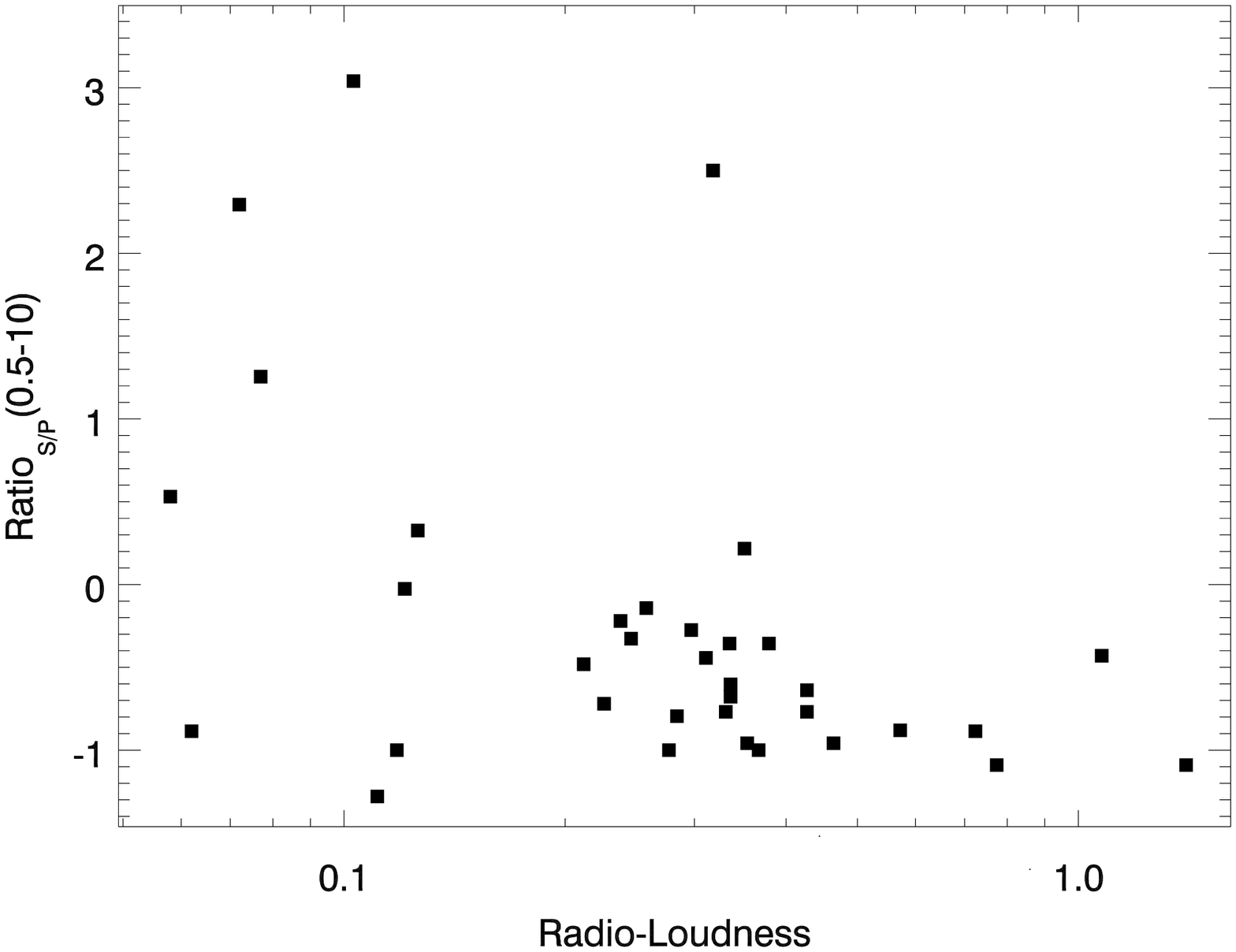,height=7.0cm,width=6.0cm,angle=90}
   \caption{Significant (i.e. at $>$99\% confidence
   level) correlations found among X--ray observables and QSO physical properties (See Table~\ref{tab:correlations2} and text for details).{\it From top-left to bottom:} (a) FWHM(H$_{\beta}$) versus L$_{2-10\, keV}$; (b) M$_{\rm BH}$ versus L$_{2-10\, keV}$ ; (c) FWHM(H$_{\beta}$) versus $\Gamma_{2-12}$; (d) $\Gamma_{2-12}$ versus M$_{\rm BH}$; (e)  $R_{S/P}^{0.5-10}$ versus $R_{\rm L}$.}
    \label{fig:correlations}
    \end{figure*}

 We also investigated by the Spearman--rank test the possible
correlations among the most representative X-ray observables inferred
by our analysis (i.e. L$_{2-10\, keV}$, $R_{S/P}^{0.5-10}$, and
$\Gamma_{2-12}$, $\Gamma_{Soft}$) and some QSO physical characteristics as the
radio-loudness parameter $R_{\rm L}$, the redshift, the FWHM of the
H$_\beta$ emission line, the black hole mass M$_{\rm BH}$ and the
accretion rate $\dot{m}$ relative to the Eddington
limit\footnote{The five  high-$z$ QSOs: 1115+080, 1206+459, 1407+265,
1630+377, 2302+029 as well as 2212+139 were excluded by the last three
correlation analysis since they have no optical data available in the
literature.  For the calculation of $\dot{m}$ we applied the formula
(1) reported in P04.} (i.e. $\dot{m}$ = $\dot{M}$/$\dot{M}_{Edd}$). The
results of the correlation analysis are presented in
Table~\ref{tab:correlations2}.  We report significant and
strong anticorrelations between the spectral indices $\Gamma_{2-12}$ and $\Gamma_{Soft}$ and
the H$_{\beta}$ FWHM confirming these well-known relationships found for
both QSO--like  (Reeves \& Turner 2000; P04; Laor et al. 1997) and Seyfert--like AGNs
(Brandt, Mathur \& Elvis 1997).  As already obtained for a smaller
sample of PG QSOs by P04, the hard band slope also appears to
anticorrelate with the black hole mass and to correlate with the fractional accretion rate
$\dot{m}$.  This latter, furthermore, has been found to be inversely
linked  with the L$_{2-10\, keV}$ with a significance of $P_s$ =
97.5\%, i.e. larger than in P04.  
PG QSOs with small M$_{\rm BH}$ tend to be less X--ray luminous. Laor (2000) pointed out
that nearly all the PG RLQs with M$_{\rm BH}$ $>$ 10$^{9}$ M$_\odot$ are radio-loud and
this fact could be have represented a bias in our correlation analysis. 
The (anti-)correlations between ($\Gamma_{2-12}$)L$_{2-10\, keV}$ and M$_{\rm BH}$ still holds
when RLQs are excluded.
An interesting discovery yielded
by the present analysis is the very significant ($P_s$ $>$99\%)
anticorrelation between the radio-loudness $R_{\rm L}$ and
$R_{S/P}^{0.5-10}$, i.e. the ratio between the strength  of the soft
excess  and the high energy power law component in the 0.5--10 keV
band. This finding seems to indicate that the soft excess emission in
RLQs is less prominent than in RQQs. However, a larger number of
objects  with $R_{\rm L}$ $>$ 10 (i.e. radio-loud) is needed to
confirm the robustness of this relationship.  Finally, no correlation
among the three X-ray observables and redshift was detected apart of
the trivial case (L$_{2-10\, keV}$, $z$) which is due to selection
effects.  Fig.~\ref{fig:correlations} shows the plots corresponding to the very significant (i.e. at $>$99\% confidence
   level) non-trivial correlations found in our analysis.

\section{Discussion}\label{sec:discussion}

\subsection{The Hard X--ray continuum}

In Section~\ref{sec:pwlw}  we presented the results for  the  power law fit in
the high energy (2-12~keV) band for all the QSO. The $\chi^2$ analysis
shows that the fits are acceptable,  i.e. $\chi_\nu^2<1.2$, in all but four
cases,       i.e.    1411$+$442     ($\chi_\nu^2=1.66$);    1440$+$356
($\chi_\nu^2=1.66$);    1630$+$377   ($\chi_\nu^2=1.70$);   2214$+$139
($\chi_\nu^2=1.83$). Except for 1630$+$377, absorption was 
significantly  detected in  all these  QSOs, being the most probable
cause for the poor quality of the fit. In the case of 1630$+$377, the
presence of an intense iron line  could be responsible for the large $\chi_\nu^2$.

The  mean  value   found  for   the  index   of   the power   law  was
$\Gamma_{2-12\,keV}=1.87^{+0.09}_{-0.10}$   and shows a significant
spread, $\sigma=0.36^{+0.08}_{-0.06}$.   The mean photon  index is  in
good   agreement   with  the results found from observations with
previous satellites.  Reeves \& Turner (2000)  and George et
al. (2000) reported  an average $\Gamma_{2-10\,keV}$ = 1.89$\pm$0.05
and $\Gamma_{2-10\,keV}$ = 1.97$^{+0.08}_{-0.09}$  observing,
respectively,  27 and 26 RQQs   with  \asca~(we compare
$\Gamma_{2-12\,keV}$ with results from RQQ samples since RQQs
represents $\sim$ 90\% of our objects).  A higher (but consistent with
ours within the statistical uncertainties) value       was    derived
from  \beppo~data  of  10  QSOs, i.e. $\Gamma_{2-10\,keV}=2.0\pm0.3$
(Mineo et al. 2000).  It has  been  observed in several studies
(Zamorani et al. 1981; Reeves \& Turner 2000) that RLQs have  flatter
spectral indices that the  RQQs. Our results  confirm this trend with
these  mean   values for the photon indices:   $\Gamma_{\rm{RLQ}}$ =
1.63$^{+0.02}_{-0.01}$  and $\Gamma_{\rm{RQQ}} = 1.89\pm0.11$.   The
possible flatness of the spectral index for RLQs  is consistent with a
different process being  responsible for the  emission in the two type
of QSO. While in RQQs the X-ray  emission is thought to be associated
to   Compton up-scattering of soft photons originated in  the
accretion disk   by  hot electrons  with a  thermal distribution
probably located in  a corona above the accretion disk (Haardt \&
Maraschi  1993; Reeves \&  Turner 2000), in RLQs the effect of the
relativistic jet could play an important role in the X-ray emission
(Ghisellini et  al. 1985; Reeves  \&   Turner 2000).   Even though
our results favor the  fact that the spectral index  in  RLQs is
flatter than in RQQs, a study of a large sample of RLQs observed with
\asca~(Sambruna et al. 1999) showed that  comparing RLQs and RQQs
which match in X--ray luminosity, there is no clear  evidence of a
difference in the spectral slope of these two types of QSOs.

We note that our mean $\Gamma_{soft}$ = 2.73$^{+0.12}_{-0.11}$ (see
Table~\ref{tab:means_laor}) is similar to that found by Laor et
al. (1997), i.e. $\Gamma_{soft}$ = 2.62$\pm$0.09. Therefore
\xmm~\epic~results   on the QSO overall spectral shape appear to be
consistent with those from {\em ROSAT} PSPC and \asca~GIS/SIS
observations.\\

Finally, a very significant ($P_s >$ 99\%) (anti-) correlation between
($\Gamma_{2-12\,keV}$)L$_{2-10\,keV}$ and the black hole mass  was
found  (e.g. P04 and Sect.~\ref{sec:correlations2}). Moreover, both
these relationships still hold  if only RQQs are considered. This
points to the possibility that the black hole mass could play a major
role in the definition of the hard X-ray properties in a
QSO. Interestingly, a similar suggestion is reported by Laor (2000)
who found a clear  bimodality  in the radio--loudness distribution as
a function of the M$_{BH}$ in the PG sample.  These results can
provide new insights on the investigations of the ultimate driver of
the AGN activity.

\subsection{The Soft excess}

The systematic modeling of the 0.3--12 keV emission by different
two--component continua presented in Sect.~\ref{sec:soft} allows us to
infer useful insights on the physical nature of the soft excess.

It is worth to note the ubiquity of such a spectral component: we
detected a low--energy X--ray turn--up in 37 out of 40
QSOs\footnote{For objects obscured by heavy absorption as 1411$+$442
and 2214$+$139, however, the presence of the additional soft power law
component can be  more properly explained in terms of scattered
emission from the absorbing medium and/or partial covering than a
truly soft excess `emission'.}.  Previous studies based on {\it ASCA}
(George et al. 2000; Reeves \& Turner 2000) and {\it BeppoSAX} (Mineo
et al. 2000)  QSO samples reported the detection of an  excess of
emission with respect of the high  energy power law below 1--1.5 keV
in $\sim$ 50--60\% of the objects. However, as pointed out by these
authors, it should be borne in mind the limited statistics (in the
case of  some {\it BeppoSAX} exposures) and the limited bandpass
(0.6--10 keV in the case of {\it ASCA}) affecting these observations.

All the models applied to account for the  soft excess component (i.e.
blackbody,  multicolor   blackbody, bremsstrahlung   and a  power
law) provided a   statistically acceptable description  of the
broad--band X--ray spectrum for the majority of QSOs.   In  the
following   we discuss our results in the light of the physical
scenario assumed in each tested model.

In first  order approximation, the  emission  due to a standard
optically--thick  accretion disk can be modeled by a blackbody or a
multicolor blackbody components. The $\chi^2_\nu$~values obtained   by
the application   of these  two models   reveal that  these
parameterizations  provide   a fair representation of the \epic~QSO
spectra (e.g. Tables~\ref{tab:A} and \ref{tab:B}). Nevertheless, the
observed    values of kT$_{BB}$ and kT$_{MB}$ $\sim$ 0.15 keV exceed
the maximal temperature for an accretion disk (e.g.  Frank, King \&
Raine 1992 for a review) which is kT$_{Max} \sim$ 20--40 eV for a
black hole mass of   10$^{7}$--10$^{8}$  M$_\odot$ accreting   at the
Eddington rate. Furthermore,   kT$_{Max}$ is    an  upper  limit  to
the   disk temperature since  the accretion could   be sub--Eddington
and/or  outer parts  of the disk  would likely be cooler. Hence, the
mean values showed in Table~\ref{tab:means_laor}   for the
temperature of        blackbody (kT$_{BB}$)   and multicolor
blackbody (kT$_{MB}$) argue   that these models  must  be  considered
just  as  phenomenological and  not  as physically consistent
models. Such  an ``optically--thick''  scenario is even  more critical
in those quasars with double blackbody component  as the best fit for
the soft excess. In fact, they show a soft energy feature too  broad
to   be  fitted by  a  single temperature  blackbody   and,
consequently,  the   second  blackbody component usually has an
unrealistic kT$_{BB}$~\simgt~0.25 keV.\\

 The soft X--ray emission of starburst regions (SBRs) is known to be
dominated by the emission  from a one-- or multi--component  hot
diffuse  plasma at  temperatures in the range of kT$\sim$0.1-1~keV
(Ptak et al. 1999; Read \& Ponman 2001; Strickland \&  Stevens 2000).
This thermal emission can  be  represented by  a bremsstrahlung model.
We obtained  that this model is  an appropriate  fit
($\chi^2_\nu<1.2$) to the soft excess emission of an important
fraction of the  objects in the sample, (i.  e.  28 out of 37 QSOs)
and is the best fit model  for 8 sources.  The mean temperature  for
the bremsstrahlung model obtained for the 28  objects   of  our
sample, kT$_{BS}=0.38\pm0.03$~keV,    is compatible   with  typical
values in SBRs.  The diffuse gas  associated to the soft X--ray
emission of the  SBRs have typically abundances in the range
0.05-0.3$Z$\sun   (Ptak  et al.   1999;  Read   \& Ponman  2001).
However, the bremsstrahlung model  does not include any emission
lines and therefore  it is not possible to  measure the abundances of
the diffuse gas and  test if they are compatible  with the typical
ones measured in SBRs. In order to determine these abundances,  we
fitted  the spectra of the eight sources for which  the bremsstrahlung
model  is the best fit model   for their soft excess emission with
the  {\tt MEKAL} model (Mewe et  al.  1986) leaving the  abundances as
a free parameter.  The  resulting values of the abundances were in all
cases consistent with zero with upper limits in the range
0.011--0.15$Z$\sun.  Except for two objects, 1634$+$706 with $Z \leq
0.15Z$\sun~ and  2303$+$029 with $Z \leq 0.1Z$\sun,   the abundances
are therefore lower than  the typical values observed in SBRs.
Moreover, in the soft band, the X--ray  luminosities of SBRs  vary in
the range of  $10^{38-41}$~\ergs~(Ptak et al.  1999;  Read \& Ponman
2001) while  the luminosities associated to the optically thin soft
X--ray component for  these 8 QSOs are in all  cases above
$8\times10^{43}$~\ergs, and, therefore,  two  orders  of magnitude
larger than the typical SBRs values even considering extreme cases as
Ultra--Luminous Infrared Galaxies (Franceschini et  al. 2003).  Hence,
on the basis of the abundances and luminosities inferred for our QSOs,
it is very  unlikely that the  bremsstrahlung emission  could be
associated to  powerful emission from a starburst region.

Even if (at the moment)  quite speculative, another possible  physical
interpretation  for the  bremsstrahlung  model is the  emission from a
transition layer  between  the   accretion  disk and  the   corona, as
proposed by  Nayakshin et al.  (2000). These authors  indeed suggested
that in the case of sources with SED dominated  by the UV thermal disk
emission (as the present PG  QSOs) bremsstrahlung emission from such a
layer could significantly contribute  to the soft X--ray emission in
AGNs  and, also,   plasma temperature  of $\sim$   0.4  keV should  be
observed.
 
Alternatively,   the  bremsstrahlung model can    be interpreted, in a
similar way of the blackbody and multicolor blackbody  models, as a
purely mathematical parametrization  of the QSO continuum (e.g. Fiore
et al. 1995).\\

The power  law is also, obviously, just  a functional form to describe
the  true soft   excess  emission,  without  any  underlying  specific
physical scenario.  For  six  QSOs  in the sample the power  law
model resulted  the  best  fit.   Notably, two   QSOs (i.e.
0804$+$761  and 1440$+$356) show a slope  of  the high  energy power
law   particularly flat ($\Gamma_{hard}$ = 1.3$\pm$0.7 and
$\Gamma_{hard}$ = 1.2$^{+0.4}_{-0.2}$,    respectively,   but    note
the      large errors). Furthermore both $\Gamma_{hard}$ are very
marginally (or not) consistent  with    the $\Gamma_{2-12}$   measured
for  these  QSOs, suggesting so   the  presence  of a  hard   tail
instead  of  a  soft excess. Such  a hard X--ray feature could  be
also related to a strong reflection component (or/and  a peculiar
variability behavior) which   cannot   be  constrained  due    to
limited  \epic~bandpass.  However, such a scenario seems to be
 unlikely for these QSOs because of the lack of a strong iron line in
their spectra (not detected in 1440$+$356 and with an EW$\sim$100 eV
in 0804$+$76; see Paper II).  Interestingly both these  QSOs are
classified  as Narrow Line  Quasar (Osterbrock  \& Pogge  1985), a
class  of  objects which usually shows extreme  flux  variations,
steep X--ray  slopes and  very  strong soft excess extending up  to
$\sim$ 3 keV   in some cases (e.g.  O'Brien et al. 2001). This   seems
also the  case  for  1440$+$356  where  such a spectral  component
dominates the broad band X--ray luminosity (e.g.
Table~\ref{tab:flux}).\\

Comptonization of thermal disk emission has been suggest as a likely
physical origin of the soft  excess in the X--ray spectra of Type 1
AGNs. The electron population responsible of the up-scattering is
thought to have a temperature cooler (typically from 0.1 to few keVs)
than the population producing the observed hard X--ray emission
(approximatively a power law in the 2--10 keV band) with a k$T_e$
\simgt~80 keV  (Perola et al. 2002). The exact location of such an
electron population is still  an open issue: a warm skin on the disk
surface (Rozanska 1999), a transition  between a cold and a hot disk
(Magdziarz et al. 1998), or a single hybrid non--thermal/thermal
plasma (Vaughan et al. 2002 and reference therein) are the most viable
candidates.  Interestingly, this soft X--ray component would contain
the bulk of the source bolometric luminosity.
 
According to this scenario, recent works (P04; Gierli{\' n}ski \& Done
2004) indeed modeled the soft X--ray excess of PG QSOs by a `cool'
Compton--scattering medium.  This fit successfully reproduces the QSO
spectral shape.  However, the typical value found for the electron
temperature is k$T_e$ $\sim$ few hundred eVs, with a very small  range
of variation      through the broad distribution in luminosity and
black hole mass within the sample.
As pointed out by different authors (Brinkmann et al. 2004b; Page et
al. 2004b; P04) the major problem with the Comptonization fit is the
limited bandpass of \xmm~that does not allow to adequately constrain
neither the blackbody emission from the disk nor the exponential
cut--off at high energy. So a plausible explanation for the observed
narrow range of  k$T_e$ values is an observational bias due to the
lack of sensitivity below 0.3 keV in the \epic~data.  Furthermore, as
pointed out by P04, the inferred values of k$T_e$ imply an extreme
compactness of the Comptonizing region, i.e. 0.1 gravitational radii
($\sim$ 10$^{12}$ cm).  This size is difficult to explain even in  a
scenario where the hard X--ray continuum is thought to be produced in
discrete magnetic flares (instead of a extended corona) above the
accretion disk (e.g.  Merloni \& Fabian  2002 for a review).
Nevertheless, Comptonization seems the most realistic interpretation
for the nature of the soft excess  in QSOs. Unfortunately, the
spectral capabilities of \chandra~and \xmm~are not sufficient enough
to disentangle the well--known degeneracy between  the  k$T_e$ and the
optical depth of the scattering plasma  (e.g. Vaughan et al. 2002;
Brinkmann et al. 2004b). EUV to very hard X--rays (i.e. few hundred
keV) broad band observations can give the key contribution to solve
this issue.\\

Gierli{\' n}ski \& Done (2004) suggested to interpret the soft excess
 in RQQs as an artifact of unmodeled  relativistically smeared
 partially ionized absorption according to the fact that strong O and
 Fe absorption features at 0.7--0.9 keV can lead to an apparent upward
 curvature below these energies mimicking the presence of a true soft
 excess component.  This scenario predicts a very steep ($\Gamma \gg$
 2) intrinsic continuum slope in the hard X--ray band (i.e. $>$ 10
 keV) for RQQs. However, such large values of the intrinsic
 (i.e. corrected both by reflection and absorption) photon index have
 not been observed neither by {\it BeppoSAX} or {\it Ginga}
 observations (e.g. Mineo et al. 2000 and Lawson \& Turner 1997,
 respectively). Nonetheless, even if not in such a dramatic way, broad
 ``warm'' absorption features due to atomic transitions could impact
 at a minor  level on the estimate of the real slope of the primary
 continuum emission.  More sensitive high--energy spectroscopic
 studies carried out in the next future with {\it Astro-E2} or {\it
 Constellation--X} will be able to definitively test this hypothesis.\\

On the basis of {\it RGS} observations Branduardi--Raymond et
al. (2001) proposed to explain the observed soft excess in the Seyfert
galaxies MKN~766 and MCG--6-30-15 in terms of strong relativistically
broadened recombination H--like emission lines of O, N and C. However,
as pointed out by Pounds \& Reeves (2002), all the objects for which
this interpretation seems to be feasible show low X--ray luminosity
(i.e. $L_X \sim$ 10$^{42}$ \ergs) and their soft excess component
emerges very sharply below $\sim$ 0.7 keV from the extrapolation of
the 2--10 keV power law.  On the contrary, all the soft excesses
observed in PG QSOs turn up smoothly below $\sim$ 2 keV.  So this soft
X--ray broad emission lines scenario seems to fail for the sources
(all with $L_X >$ 10$^{43}$ \ergs) in our sample.\\

Finally, the finding of an anticorrelation between $R_L$ and
$R_{S/P}^{0.5-10}$ (see Table~\ref{tab:correlations2}) seems to 
support the hypothesis that the soft excess in radio--loud QSOs is
due to an X--ray emission component originating in the jets.

\subsection{The intrinsic absorption}
\subsubsection{Cold absorber}

In order to study  the presence of a cold  absorber  in QSOs, we  have
added a neutral absorption component  to the different models  tested.
The  analysis  reveals that the majority of the spectra  does not
require this component, and that the equivalent hydrogen column  is
negligible in comparison with  the Galactic one.  Only  three
sources, all  of  them RQQs,   show  the presence of a   cold
absorber,    i.e.  0050$+$124, 1115$+$080     and 1411$+$442.  The
two former ones have columns of \nh~$\leq 10^{21}$~cm$^{-2}$ likely
associated to their host galaxies.   The strongest    absorption
was       found      for    1411$+$442, \nh~= 2.3$\pm$0.4 $\times$
10$^{22}$ \cm2,  although  due  the low statistic of the data it is
not possible to confirm if  this feature is originated by a cold
absorption  or is  indeed related  to a warm  absorber.  The rarity of
intrinsic neutral absorption observed  here is in agreement with
previous studies of  optically--selected QSOs (Laor et al. 1997; Mineo
et al. 2000)   and, in general, with    the findings of   recent
wide--angle deep \xmm~surveys (e.g. Barcons et al. 2002; Piconcelli et
al. 2003;  Caccianiga et al. 2004) which  reported a small fraction of
broad line AGNs with a significant cold absorption component in excess
to the Galactic value.

Therefore, we  can conclude that for PG QSOs there is a good
correlation between optical spectral type and X--ray absorption
properties in agreement with the predictions of the AGN Unification
Models.

\subsubsection{Warm absorber}
{\it Warm} (i.e. partially ionized) {\it absorber} in active galaxies
are revealed by the presence of absorption edges in their soft X--ray
spectra  (Pan, Stewart \& Pounds 1990). In particular, OVII (0.739
keV) and OVIII (0.871 keV) absorption edges were the typical
signatures of such ionized gas detected in  {\it ROSAT} and {\it ASCA}
low--resolution  observations.  Notably, Reynolds (1997) found a warm
absorber in $\approx$ 50\% of the well studied Seyfert galaxies.

The study of this spectral component took a remarkable step forward
exploiting the higher resolution ($R>$ 100) of the grating
spectrometers on--board \xmm~and \chandra.  H-- and He--like
absorption lines of cosmically abundant elements such as C, O, Ne, Mg
and Si  are the most numerous and strongest  features observed in the
soft X--ray spectra of bright Seyfert 1 galaxies. The presence between
0.73 and 0.77 keV of the unresolved transition array (UTA, e.g. Behar,
Sako \& Kahn 2001) of iron M--shell ions is also particularly common.
The X--ray absorption lines are typically blueshifted by  a few
hundreds km/s and show complex profile due to a wealth of kinematic
components.  The ionized material is therefore in an outflow and it
generally shows multiple ionization states (Kaastra et al. 2000;
Collinge et al. 2001; Piconcelli et al. 2004).

On the contrary, the properties of the warm absorber in QSOs are
poorly known so far. Nonetheless, it is widely accepted that the
presence of ionized absorbing gas  in such high luminosity AGNs is not
as common as in Seyfert galaxies. Laor et al. (1997) indeed detected
it in $\leq$ 5\% of the optically--selected QSO population, and, on
the basis of large {\it ASCA} sample, Reeves \& Turner (2000) and
George et al. (2000) both confirmed the rare occurrence in their
spectra of opacity due to ionized gas along line of sight.   Bearing
in mind the limited statistics, George et al. (2000) did not find,
however, significant differences in the physical parameters of the
warm absorber (i.e. ionization state and column density) in low and
high luminosity AGNs.\\

The scenario emerging by our analysis is completely different.  In
fact, we significantly detect absorption features due to ionized gas
in the spectrum of 18 out 40 QSOs, i.e. the 45\% of the sample.  This
fraction is similar to what found for Seyfert galaxies.  The detection
rate of warm absorber appears therefore to be independent  from the
X--ray luminosity. This finding confirms the recent result of P04
based on a smaller sample of 21 QSOs observed by \xmm, and put it on a
sounder statistical ground.

As reported in Sect.~\ref{sec:complex} most of these absorption
features are observed in the range 0.7--0.8 keV: this fact suggests
they are likely due to the He--like O edge and/or the blend of
M--shell Fe lines (UTA).  In the case of the edges found in the
high--energy portion of the spectrum around 7 keV (as in 1115$+$080,
1211$+$143 and 1226$+$023), the energy of these features is still
consistent with an origin in "neutral" material (i.e. FeI--V). So an
alternative explanation in terms of reflection from the optically
thick  accretion disk cannot be ruled out.  The edge detected at
$\sim$ 9 keV in 1115$+$08 would imply an ionization state higher than
FeXXVI, so the possible origin of this feature in an ionized outflow
(with a velocity of 0.34$c$ to account for the observed  blueshift) as
proposed by Chartas et al. (2003) seems to be the most reasonable.

Due to the strength of the absorption features in their spectra,  for
4 objects  (i.e. 1001$+$054, 1114$+$445, 1404$+$226 and 2214$+$139) we
could apply the {\tt ABSORI} model in the fit and, consequently,
derive some physical parameters of the warm absorber in these sources
(see Table~\ref{tab:best}). For 2214$+$139 and the BAL QSO 1001$+$054
the inferred column densities of the ionized gas are larger than
10$^{23}$ \cm2, i.e.  among the highest seen by \xmm~and \chandra~so
far.  Even more interesting are the results we report for 1114$+$445
and 2214$+$139.  Their \xmm~spectra indeed show for the first time the
evidence for a multi-zone warm absorber in both QSOs.

These findings agree with the recent results obtained for Seyfert
galaxies which claim for a multiphase outflowing plasma as origin of
the warm absorber. Unfortunately, due to the \epic~spectral
resolution, we cannot draw any conclusion on the ionization structure
(i.e. discrete or continuous) of the ionized wind.\\

All the QSOs in our sample classified as ``Soft X--ray Weak'' QSOs
(SXWQs) by  Brandt, Laor \& Wills (2000) (namely 1001$+$054,
1411$+$442 and 2214$+$139)  have been found to be heavily absorbed.
This finding supports the hypothesis that the observed soft X--ray
``weakness'' of SXWQs is likely due to the presence of absorbing
matter along line of sight instead of  an ``intrinsic'' weakness due
to a different SED and/or emission/accretion mechanisms at work in
SXWQs with respect to ``normal'' QSOs.  Furthermore,  all SXWQs
analyzed here also show prominent absorption lines in the UV and in 2
out of 3 SXWQs the absorption occurs in an ionized gas\footnote{There
is also a strong suggestion for the presence of a warm absorber also
in 1411$+$442 (e.g. Sect.~\ref{sec:complex}, Brinkmann et al. 2004a)}.
These facts support the idea of a possible physical connection between
UV and X--ray absorbers (Crenshaw, Kramer \& George 2003).

\section{Conclusions}\label{sec:conclusions}

We have presented the results of the spectral analysis of 40 QSOs
\epic~spectra observed by \xmm.  These objects represents $\sim$ 45\%
of the complete PG QSOs sample. The main findings of our study are:\\

\begin{itemize}

\item The hard X--ray continuum slope resulted $\Gamma =$
1.89$\pm$0.11 and $\Gamma =$ 1.63$^{+0.02}_{-0.01}$ for the
radio--quiet and radio--loud QSOs in the sample, respectively.  These
values are consistent with previous X--ray observations of these
objects and, in particular, we confirm a flatter slope in RLQs than
RQQs. Such a difference could suggest the presence of an
extra--continuum contribution from the self--synchrotron Compton jet
emission (Zamorani et al. 1981).\\

\item The soft excess is an ubiquitous feature of the low energy
X--ray spectrum of the optically--selected QSOs.  We systematically
fitted the spectrum of each object with four different two--component
models.  None of the two-component models tested  provides an
satisfactory description for all the QSO spectra. This result suggests
that the shape of the soft excess is not an universal QSO
property. Spectral parameters inferred by these fits strongly indicate
that all the applied models should be interpreted as phenomenological
(and not physical) parameterizations of the observed spectral shape.
Direct thermal emission from the accretion disk is ruled out on the
basis of  the very high temperature derived for this
optically--thick component. Emission from a thermal plasma in a
starburst is also ruled out because of the very large X--ray
luminosities observed in our QSOs.  A double power law model provides
as well a good description of some QSO spectra and corresponds to a
double Comptonization scenario for the X--ray emission. However, as
also suggested by other authors (e.g. Gierli{\' n}ski \& Done 2004,
Brinkmann et al.   2004b) current X--ray observations cannot
effectively constrain the physical parameters of the two plasma due to
the lack of sensitivity in the EUV and at few hundred keVs where the
emission peaks of the two component are expected.\\

\item Significant evidence of cold absorption  was detected in only
three  objects.  Of them,  two show column density (\nh~$\sim$
10$^{21}$ \cm2) consistent with the contribution from the host
galaxy. For 1411$+$442 (with \nh~$\sim$ 10$^{23}$ \cm2), there are
strong indications that the obscuration occurs in an ionized medium
(Brinkmann et al. 2004a).  Our result therefore agrees with the
prediction of the  AGN Unified Models (as  well as recent
observational results from wide--angle X--ray  surveys) on the
relationship between optical classification and X--ray absorption
properties.  \\

\item About 50\% of the QSOs show  warm absorber features in their
spectra. This finding is in contrast with  previous studies  based on
{\it  ASCA} or  {\it BeppoSAX}  data, where warm  absorbers appeared
to   be  rare. The   fraction we  observed is similar to  that
inferred for Seyfert  1  galaxies.  The existence of ionized gas
along  the line of  sight  is independent on the X--ray luminosity.
For  two  QSOs with the highest  S/N ratio (i.e. 1114$+$445 and
2214$+$139)  we also significantly  detected a  complex (two--phase)
warm absorber (see Piconcelli et al. 2004).\\

\item  All the SXWQs in the sample show evidence of obscuration
  (occurring in an ionized medium in most of the cases). This finding
  suggests that their `weakness' is just a matter of absorption
  instead of  an unusual shape of the spectral energy distribution of
  such a class of QSOs.\\

\item We performed correlation analysis between the spectral
parameters of the soft excess tested models and the X--ray
luminosities  (Sect.~\ref{sec:correlations}) as well as between the
hard X--ray observables (i.e.  L$_{2-10\, keV}$, $R_{S/P}^{0.5-10}$
and $\Gamma_{2-12}$,  $\Gamma_{Soft}$) and some physical parameters (i.e. $R_L$, $z$,
FWHM(H$_\beta$) and $\dot{m}$) (see Sect.~\ref{sec:correlations2}).
We confirmed the anticorrelations between both photon indices and
FWHM(H$_\beta$) as reported in previous papers (P04, Reeves \& Turner
2000).  We also found that $\dot{m}$ is inversely correlated  with the
L$_{2-10\, keV}$ and that the black hole mass (anti-)correlates with
($\Gamma_{2-12}$)$L_{2-10\, keV}$.  Furthermore, we discovered a very
significant anti-correlation between $R_L$ and  $R_{S/P}^{0.5-10}$,
i.e. the ratio  between the strength  of the soft excess and the high
energy power law component in the 0.5--10 keV band.
\end{itemize}

The \xmm~view of optically--selected QSOs  suggests that there are no
significant changes in the spectral properties of these high
luminosity AGNs once compared with the low luminosity broad line
Seyfert galaxies. See also Paper II for the properties of the
fluorescent  iron K--shell emission lines in the same objects studied
here.

\begin{acknowledgements}
  The authors wish to thank the referee, Dr. James Reeves, for the
prompt and encouraging report and the useful comments, which have
significantly improved the paper.  We also would to thank the \xmm~SOC
science support team members at ESAC.  This paper is based on
observations obtained with \xmm, an ESA science mission with
instruments and contributions  directly funded by ESA Member States
and the USA (NASA).  This research has made use of the NASA$/$IPAC
Extragalactic Database (NED) which is operated by the Jet Propulsion
Laboratory,  California Institute of Technology, under contract with
the National Aeronautics and Space Administration.
\end{acknowledgements}

\end{document}